\documentclass[11pt,a4paper]{article} 
\pdfoutput=1
\usepackage{jheppub}
\usepackage{graphicx, color}
\usepackage{amsmath, amssymb, amscd, latexsym, bm, braket}
\usepackage[normalem]{ulem}

\begin{document}
\title{Exact WKB analysis of the vacuum pair production by time-dependent electric fields}

\author[a]{Hidetoshi~Taya}
\author[a]{Toshiaki~Fujimori}
\author[b,a]{Tatsuhiro~Misumi}
\author[a]{Muneto~Nitta}
\author[a]{Norisuke~Sakai}

\emailAdd{h\_taya@keio.jp}
\emailAdd{toshiaki.fujimori018@gmail.com}
\emailAdd{misumi@phys.akita-u.ac.jp}
\emailAdd{nitta@phys-h.keio.ac.jp}
\emailAdd{norisuke.sakai@gmail.com}

\affiliation[a]{Department of Physics, and Research and Education Center for Natural Sciences, Keio University, 4-1-1 Hiyoshi, Yokohama, Kanagawa 223-8521, Japan}
\affiliation[b]{Department of Mathematical Science, Akita University, Akita 010-8502, Japan}

\abstract{We study the vacuum pair production by a time-dependent strong electric field based on the exact WKB analysis.  We identify the generic structure of a Stokes graph for systems with the vacuum pair production and show that the number of produced pairs is given by a product of connection matrices for Stokes segments connecting pairs of turning points.  We derive an explicit formula for the number of produced pairs, assuming the semi-classical limit.  The obtained formula can be understood as a generalization of the divergent asymptotic series method by Berry, and is consistent with other semi-classical methods such as the worldline instanton method and the steepest descent evaluation of the Bogoliubov coefficients done by Brezin and Izykson.  We also use the formula to discuss effects of time-dependence of the applied strong electric field including the interplay between the perturbative multi-photon pair production and non-peturbative Schwinger mechanism, and the dynamically assisted Schwinger mechanism.  }


\maketitle

\section{Introduction}

Our vacuum is not a vacant space, and there are full of quantum fluctuations popping in and out of existence.  When the vacuum is exposed to a strong field, those quantum fluctuations can interact with the field, leading to non-trivial responses of the vacuum.  A prominent example of such a response is pair production of charged particles from the vacuum by a strong electric field (for review, see Refs.~\cite{gel15, ruf10, dun05}).  Since the first proposal by Sauter in 1931 \cite{sau31}, the vacuum pair production has been under intensive investigation not only as a fundamental prediction of quantum-field theory but also to understand actual physics processes under extreme conditions (e.g., the early-stage dynamics of heavy-ion collisions \cite{tan09, low75, nus75, gle83, kaj85, gat87, tay17}, the vacuum decay induced by a superheavy nucleus \cite{dar28, gor28, pom45, zel71, mul72}, the magnetogenesis in the early Universe \cite{kob14, sor19, sob18, sha17, sta18, kob19, dom20}, and possibility of prohibition of anisotropic inflation \cite{Kitamoto:2020tjm}) as well as some analogous phenomena with different kinds of fields/forces (e.g., the Hawking radiation by a strong gravitational field \cite{haw74, haw75}, the dynamical Casimir effect by time-varying boundaries \cite{yab89, sch92, moo70, for82, dod93, dod96, mun98, tay20b}, and axion production by a time-dependent condensate \cite{gri02, gri99, hua20}).  The vacuum pair production has never been verified by experiments yet, as it requires an extremely strong electric field of the order of $eE_{\rm cr} = m_{\rm e}^2 \sim \sqrt{10^{29}{\rm W}/{\rm cm}^2}$, with $m_{\rm e}=511\;{\rm keV}$ being electron's mass (cf. the current world record is $eE \sim \sqrt{10^{22}\;{\rm W}/{\rm cm}^2}$, achieved by HERCULES laser \cite{yan08}).  Nevertheless, the up-coming intense laser facilities such as Extreme Light Infrastructure (ELI) may reach or go beyond the critical field strength $eE_{\rm cr}$, providing the very first opportunity to observe the vacuum pair production in laboratory experiments \cite{pia12}.

Quantitative features of the vacuum pair production drastically change depending on the time-dependence of an applied strong electric field \cite{kel65, bre70, pop71, tay14}.  If the electric field is fast (i.e., the typical frequency $\Omega$ is large), the field interacts with the vacuum fluctuations {\it incoherently}.  That is, the electric field behaves like a dynamical photon $\gamma$ to produce particles via perturbative multi-photon processes $n \gamma \to e^+e^-$, with $n$ being the number of photons involved.  This is an analog of the photo-absorption effect in materials.  The production number $N_{e^+e^-}$ becomes proportional to powers of the coupling constant $e$ as
\begin{align}
	\lim_{\Omega \to \infty} N_{e^+e^-} \propto e^{2n}.  \label{EQ1}
\end{align}
Note that we have a factor of $2$ in the exponent because the production number $N_{e^+e^-}$ is the square of the amplitude $\propto e^n$.  On the other hand, for small $\Omega$, the field interacts with the fluctuations {\it coherently}, rather than incoherently, and produces particles via quantum tunneling.  The production number $N_{e^+e^-}$ becomes purely non-perturbative with respect to the coupling constant $e$ as \cite{sau31, hei36, sch51}
\begin{align}
	\lim_{\Omega \to 0} N_{e^+e^-} \propto {\rm e}^{- {\rm const.} \times e^{-1} }.  \label{EQ2}
\end{align}
This production mechanism is called the {\it Schwinger mechanism}, named after Schwinger who first derived the exponential formula (\ref{EQ2}) in a fully quantum-field theoretical manner \cite{sch51}, and can be understood as an analog of the electrical breakdown of materials.  In the above, we have implicitly assumed that the applied electric field is dominated by a single frequency mode $\Omega$.  For a field with several frequency modes $\Omega_1, \Omega_2, \cdots$, the production mechanisms with different frequency modes interfere with each other.  The interference leads to, for example, substantial enhancement in the production number (the dynamically assisted Schwinger mechanism \cite{sch08, piz09, dun09, mon10a, mon10b}), characteristic momentum signatures in the spectrum \cite{heb09, ort11, chr12, pan15, ces10, ces11, fk, fk2, tay20, gre17, gre19}, and spin-dependences \cite{fk2, Huang:2019szw}.  Those time-dependent effects are utilized to ``optimize" electromagnetic field profiles, so that the signatures of the vacuum pair production become the most manifest in intense laser experiments \cite{fra17, lin15, heb-14, don20, koh13, abd13}.  Thus, getting a deeper understanding of the time-dependent effects is important, so as not only to understand actual physics phenomena correctly but also to design a better experimental setup.

The time-dependent effects have been studied with various analytical approaches such as semi-classical methods (e.g., the worldline instanton method \cite{dun05a, dun06a, dun06b}, the imaginary-time method \cite{pop05}, and the steepest descent evaluation of the Bogoliubov coefficients by Brezin and Izykson \cite{bre70}), the standard perturbative calculation \cite{tay14}, and the perturbation theory in the Furry picture \cite{fk, fk2, tay20, gre17, gre19, gre18}.  Those approaches have different regimes of applicability.  For example, the semi-classical methods are justified only in the semi-classical limit, in which one takes a formal limit of $\hbar \ll 1$, while the standard perturbative calculation covers the opposite regime $\hbar \gg 1$.  The perturbation theory in the Furry picture can be applied to arbitrary values of $\hbar$, but it is valid only if one can clearly separate a given electromagnetic field into a strong field and perturbations on top of it.  It is, therefore, desirable to develop a novel method for the vacuum pair production in order to get a deeper understanding of the time-dependent effects and to cover a wider parameter regime for the production.

The purpose of this paper is to use the exact Wentzel-Kramers-Brillouin (WKB) analysis as a first step toward developing a novel method for the vacuum pair production by a time-dependent electric field.  The exact WKB analysis, which has been developed mainly in mathematics since the pioneering work by Voros in 1983 \cite{vor83}, is a powerful tool to analyze Stokes phenomena of Schr\"{o}dinger-type differential equations.  We apply the exact WKB analysis to derive a production number formula, using the fact that the vacuum pair production can be regarded as a Stokes phenomenon of a given field equation.  By clarifying the generic structure of a Stokes graph for systems with the vacuum pair production, we show that the production number $N_{e^+e^-}$ is given by a product of connection matrices for Stokes segments connecting pairs of turning points.  The connection matrices are then evaluated using the semi-classical approximation.  We show that our formula gives a systematic improvement of the divergent asymptotic series method proposed by Berry \cite{ber89} (see also Dingle's book \cite{din73}) and agrees with other semi-classical approaches such as the worldline instanton method \cite{dun05a, dun06a, dun06b} and Brezin-Izykson's steepest descent evaluation \cite{bre70}.  We also use the formula to discuss the time-dependent effects including the interplay between the perturbative multi-photon pair production and non-perturbative Schwinger mechanism and their interference effects.

This paper is organized as follows: In Sec.~\ref{sec2}, we review the basics of the exact WKB analysis.  Section~\ref{sec3} is the main section of the paper, in which we apply the exact WKB analysis to formulate the vacuum pair production by a time-dependent electric field.  Section~\ref{sec4} is devoted to summary and discussion.  In Appendix~\ref{sec2.2}, we discuss the Airy equation by means of the exact WKB analysis.  We briefly review other computational methods for the vacuum pair production such as the standard perturbation theory and the perturbation theory in the Furry picture in Appendices~\ref{appa4} and \ref{appa-4}, respectively.  We also present a detailed analysis of a Sauter-type electric field in Appendix~\ref{appa-5} and of Stokes graphs in the dynamically assisted Schwinger mechanism in Appendix~\ref{appe}.

\section{Preliminaries: Exact WKB analysis}\label{sec2}

To be self-contained, we here review the basics of the exact WKB analysis.  The exact WKB analysis, originally proposed by Voros \cite{vor83} and developed, e.g., by Pham and his collaborators \cite{CNP, DP, DDP1, DDP2} and by Aoki, Koike, and Takei \cite{AKT1, AKT2, AKT3}, is an extension of the conventional (J)WKB method, proposed independently by Jeffreys, Wentzel, Kramers, and Brillouin \cite{jef24, wen26, kra26, bri26}.  The central idea of the exact WKB analysis is to apply the Borel resummation technique \cite{eca81} to the conventional WKB method, which does not only make the conventional WKB method mathematically well-defined but also provides a powerful tool to investigate Stokes phenomena of WKB solutions.  The exact WKB analysis can also be regarded as a generalization and a mathematically rigorous justification of the divergent asymptotic series method proposed by Berry \cite{ber89}.

The conventional WKB method is a method to analyze differential equations with a small parameter.  To be precise, let us consider a second order differential equation of the form,
\begin{align}
	0 = \left[  \epsilon^2 \frac{{\rm d}^2}{{\rm d}t^2} + Q(t) \right] \phi(t), \label{eqa106}
\end{align}
where $t \in {\mathbb R}$ is a real variable and $\epsilon>0$ is some small parameter (e.g., the Planck constant $\hbar$).  One may understand that the parameter $\epsilon$ controls fastness/slowness of the potential $Q$.  Indeed, by introducing $\tau \equiv t /\epsilon$, one can rewrite the differential equation (\ref{eqa106}) as $0 = \left[ \partial_\tau^2 + Q(\epsilon \tau) \right] \phi$.  Thus, the potential $Q$ effectively becomes slow and fast in the limit of $\epsilon \to 0$ and $\infty$, respectively.  The starting point of the WKB method is to make the so-called WKB ansatz, 
\begin{align}
	\phi_{\pm} (t; \epsilon) 
		\equiv \frac{1}{\sqrt{2\Omega(t)}} \exp\left[ \mp \frac{{\rm i}}{\epsilon} \int^t_{t_0} {\rm d}t' \Omega(t') \right],    \label{eqa1}
\end{align}
where $t_0$ is an arbitrary point on ${\mathbb R}$ and the WKB ansatz is normalized as 
\begin{align}
	1 = +{\rm i}\epsilon\,  \phi_{-} \overset{\leftrightarrow}{\partial}_t \phi_{+} , \nonumber\\
	0 = +{\rm i}\epsilon\,  \phi_{\pm} \overset{\leftrightarrow}{\partial}_t \phi_{\pm} .   \label{eq2.3}
\end{align}
Note that $\phi_+$ ($\phi_-$) describes an out-going (in-coming) wave propagating in the positive (negative) time direction, and hence we added the subscript $\pm$.  By substituting the WKB ansatz (\ref{eqa1}) into the differential equation (\ref{eqa106}), one obtains an equation for $\Omega$ as
\begin{align}
	0 
		&= Q - \Omega^2 + \epsilon^2 \left[  \frac{3}{4} \left( \frac{\Omega'}{\Omega} \right)^2  - \frac{1}{2} \frac{\Omega''}{\Omega} \right] . \label{eqa3} 
\end{align}
Assuming $\epsilon \ll 1$, one may expand $\Omega$ as
\begin{align}
	\Omega \equiv \sum_{n=0}^{\infty} \epsilon^n \Omega_n , \label{eeeq85}
\end{align}
with which one may iteratively solve $\Omega_n$ as 
\begin{align}
\label{eqa110}
	&\Omega_0 = Q^{1/2}, \nonumber\\   
	&\Omega_2	= \frac{5 {Q'}^2-4 Q Q''}{32 Q^{5/2}} , \nonumber\\ 
	&\Omega_4	= \frac{64Q^3Q''''-448Q^2Q'Q'''-304Q^2Q''^2+1768QQ'^2Q''-1105Q'^4}{2048 Q^{11/2}},\  \cdots , 
\end{align}
and  
\begin{align}
	\Omega_{n} = 0\ \quad{\rm if}\ \quad n={\rm odd} 
\end{align}
because the differential equation (\ref{eqa3}) does not depend on the sign of $\epsilon$.  Now, it may look that the series (\ref{eeeq85}) gives a ``solution" of the differential equation (\ref{eqa106}).  Unfortunately, it turns out that the series (\ref{eeeq85}) is not necessarily convergent and therefore is in general ill-defined.  For example, see Appendix~\ref{sec2.2}, in which we explicitly show that the Airy equation gives a factorially divergent series.  This problem can be resolved by the exact WKB analysis, as we describe below.

The idea of the exact WKB analysis is to apply the Borel resummation technique to make the conventional WKB method well-defined.  To apply the Borel resummation technique, we first reexpress the WKB ansatz (\ref{eqa1}) by expanding the exponential factor as
\begin{align}
	\phi_{\pm}(z; \epsilon) 
		&\equiv \exp \left[ \mp \frac{\rm i}{\epsilon} \int^z_{t_0} {\rm d}z' \sqrt{Q(z')}  \right] \sum_{n=0}^{\infty} \psi_{\pm,n} (z) \epsilon^{n}  \nonumber\\
		&\equiv \exp \left[ \mp \frac{\rm i}{\epsilon} \int^z_{t_0} {\rm d}z' \sqrt{Q(z')}  \right]  \psi_{\pm}(z; \epsilon) , 
\label{qqq86}
\end{align}
where we have used $\Omega_0 = \sqrt{Q}$ and complexified the variable $t$ as $t \in {\mathbb R} \to z \in {\mathbb C}$.  One can explicitly compute the series coefficients $\psi_{\pm,n}$ from the iterative solution $\Omega_n$ (\ref{eqa110}) and understands that the series $\psi_{\pm}$ (\ref{qqq86}) is ill-defined in general because of the non-convergence of $\Omega$ (\ref{eeeq85}).  Now, we apply the Borel resummation technique to make the series $\psi_{\pm}$, or accordingly $\phi_{\pm}$, well-defined.  Namely, we introduce a Borel transformation $\tilde{\psi}_{\pm}$ as 
\begin{align}
	\tilde{\psi}_{\pm} (z; \eta) \equiv \sum_{n=0}^{\infty} \frac{ \psi_{\pm,n}(z) }{n!} \eta^{n}, \label{eqa113}
\end{align}
which is a well-defined object if the series $\psi_{\pm}$ is at most factorially divergent, i.e., for any $n$ there exist some constants $A, C$ such that $|\psi_{\pm,n}| < A C^n n!$.  Then, the theory of the Borel resummation guarantees that if a Borel sum $\Psi_{\pm}$ exists (Borel summable), i.e., one can perform the following Laplace transformation in a well-defined manner, 
\begin{align}
	\Psi_{\pm} (z; \epsilon) \equiv \int^{\infty}_0 \frac{{\rm d}\eta}{\epsilon} {\rm e}^{-\eta/\epsilon} \tilde{\psi}_{\pm}(z; \eta) , \label{eq1a15}
\end{align} 
then
\begin{align}
	\Phi_{\pm}(z; \epsilon) \equiv \left[ \mp \frac{\rm i}{\epsilon} \int^z_{t_0} {\rm d}z' \sqrt{Q(z')}  \right] \Psi_{\pm} (z; \epsilon)
\label{eq:borel_sum}
\end{align}
becomes a well-defined solution of the original equation (\ref{eqa106}) having the asymptotic expansion (\ref{qqq86}).  The Borel sum (\ref{eq:borel_sum}) can be analytically continued to arbitrary values of $\epsilon$, while the smallness of $\epsilon \ll 1$ is assumed in the conventional WKB method.  Note that Berry \cite{ber89} assumes some truncation of the series (\ref{qqq86}) and only considers the leading order factorial divergence of $\psi_{\pm}$ to compute (a correspondence of) the Borel transformation (\ref{eqa113}), and then further employs the saddle point method to evaluate the Borel sum (\ref{eq1a15}).  Such treatments are unneeded in the exact WKB analysis.

The Borel (non-)summability is determined by the singularity structure of the Borel transform (\ref{eqa113}) in the $\eta$-plane.  Since the Borel transform (\ref{eqa113}) is dependent not only on $\eta$ but also on the complexified variable $z$, the singularity structure in the $\eta$-plane changes as $z$ varies in general.  To proceed, let us assume for simplicity that (i) the potential $Q$ is an analytic function on the entire complex $z$-plane and (ii) its Taylor expansion around a zero always starts from a linear term as 
\begin{align}
	Q = 0 + c (z-z_{\rm t}) + {\mathcal O}(|z-z_{\rm t}|^2)\ \ (c\neq 0),  \label{eq2.12}
\end{align}
where $z_{\rm t}$ is a root of the potential such that $Q(z_{\rm t})=0$ and is called a {\it turning point} (or, specifically, a {\it simple turning point} because it is of order one) in the language of the exact WKB analysis.  Since the potential $Q$ reduces to the Airy potential $Q_{\rm Airy}=c(z-z_{\rm t})$ around the turning point $z\sim z_{\rm t}$, one can investigate the change of the singularity structure in the $\eta$-plane by analyzing the Airy equation (see Appendix~\ref{sec2.2}).  Mathematically, this procedure is justified by the so-called WKB-theoretic transformation \cite{AKT1, AKT2}.  It turns out that the singularity can hit the integration contour of the Laplace transformation (\ref{eq1a15}) when $z$ crosses the so-called {\it Stokes line} ${\mathcal C}_{z_{\rm t}}$: 
\begin{align}
	{\mathcal C}_{z_{\rm t}} \equiv \Biggl\{ z\; \Biggl| \; 0 = {\rm Im} \left[ \frac{{\rm i}}{\epsilon} \int^z_{z_{\rm t}} {\rm d}z' \sqrt{ Q(z')} \right] \Biggl\}.   \label{eq2.13}
\end{align}
Therefore, the Borel sum exists unless $z$ is on top of a Stokes line.  It should be noted that we implicitly assumed for the moment that (iii) there are no Stokes lines degenerated with other Stokes lines emanating from other turning points (such a degenerated Stokes line is called a {\it Stokes segment}), which case shall be discussed later.

The Borel summability is important not only to make the conventional WKB method well-defined but also to describe Stokes phenomena of WKB solutions.  Consider a Borel sum defined at some point in a {\it Stokes region}, which is defined as a region in the $z$-plane that is separated from other regions by some Stokes lines.  The Borel sum can be analytically continued to the entire Stokes region, since it does not hit any Stokes line within the region.  However, whenever moving to another Stokes region, $z$ must hit a Stokes line, at which the integration contour of the Laplace transformation hits singularities in the $\eta$-plane.  Then, the Borel sum experiences a sudden jump due to the integration of the singularity.  This is the Stokes phenomenon of WKB solutions.  To be precise, let us consider two Borel sums, say $\Phi_{\pm, {\rm I}}$ and $\Phi_{\pm, {\rm II}}$, defined on two neighboring Stokes regions, I and II, separated by a Stokes line ${\mathcal C}_{z_{\rm t}}$.  For an analytic potential $Q$ having the property (\ref{eq2.12}), one can explicitly evaluate the discontinuity by carrying out the integration around the corresponding singularity and finds\footnote{The exponential factors in Eq.~(\ref{e151}) appear because of the normalization of the WKB ansatz (\ref{eqa1}) and are modified by changing the lower end of the integration in the ansatz (\ref{eqa1}).  For example, if one normalizes the WKB ansatz around a turning point $z_{\rm t}$, instead of $t_0$, as $\phi_{\pm}(z) \equiv \frac{1}{\sqrt{2\Omega(z)}} \exp\left[ \mp \frac{{\rm i}}{\epsilon} \int^z_{z_{\rm t}} {\rm d}z' \Omega(z') \right]$, one should replace $\sigma_{z_{\rm t}}$ with unity in Eq.~(\ref{e151}).  For a different normalization, one also needs to multiply a proper normalization matrix in addition to the connection matrix, when computing Stokes constants with the procedure (3).  } \cite{vor83, AKT1} (see also Appendix~\ref{sec2.2})
\begin{align}
	\left\{\begin{array}{ll}
		\displaystyle \begin{pmatrix} \Phi_{+,{\rm II}} \\ \Phi_{-,{\rm II}} \end{pmatrix}
			= \begin{pmatrix} 1 & \pm {\rm i}{\rm e}^{ -\sigma_{z_{\rm t}}/\epsilon} \\ 0 & 1 \end{pmatrix}\begin{pmatrix} \Phi_{+,{\rm I}} \\ \Phi_{-,{\rm I}} \end{pmatrix} 
			&
		\displaystyle {\rm for}\ {\rm i} \int^{z\in {\mathcal C}_{z_{\rm t}}}_{z_{\rm t}} {\rm d}z' \sqrt{Q(z')} <0 \vspace*{4mm} \\
		\displaystyle \begin{pmatrix} \Phi_{+,{\rm II}} \\ \Phi_{-,{\rm II}} \end{pmatrix}
			= \begin{pmatrix} 1 & 0 \\ \pm {\rm i}  {\rm e}^{ +\sigma_{z_{\rm t}}/\epsilon }  & 1 \end{pmatrix}\begin{pmatrix} \Phi_{+,{\rm I}} \\ \Phi_{-,{\rm I}}\end{pmatrix} 
			&
		\displaystyle {\rm for}\ {\rm i} \int^{z \in {\mathcal C}_{z_{\rm t}}}_{z_{\rm t}} {\rm d}z' \sqrt{Q(z')} >0
	\end{array}\right. , \label{e151}
\end{align} 
where we have introduced
\begin{align}
	\sigma_{z_{\rm t}} \equiv +2 {\rm i} \int_{t_0}^{z_{\rm t}} {\rm d}z' \sqrt{Q(z')}  \label{eq2.14}
\end{align}
and one chooses $+$ ($-$) sign if crossing the Stokes line $ {\mathcal C}_{z_{\rm t}}$ counter-clockwise (clockwise) with respect to the turning point when moving from the region I to II.  Note that ${\rm i} \int^{z\in {\mathcal C}_{z_{\rm t}}}_{z_{{\rm t}}} {\rm d}z' \sqrt{Q} \in {\mathbb R}$ by definition (\ref{eq2.13}) and its sign does not change unless the Stokes line $ {\mathcal C}_{z_{\rm t}}$ emanating from $z_{\rm t}$ hits another turning point or a singularity, which are forbidden by the assumptions.  The connection formula (\ref{e151}) is the basis to quantify the Stokes phenomenon of WKB solutions within the exact WKB analysis: 
\begin{enumerate}
\item[(1)] Draw a Stokes graph (i.e., draw turning points and Stokes lines) for a given potential $Q$; 

\item[(2)] Draw a path that connects a WKB solution defined at some Stokes region and another one defined at a distinct Stokes region; and 

\item[(3)] The Stokes phenomenon of the two WKB solutions is quantified by successively applying the connection formula (\ref{e151}) whenever the path hits a Stokes line.   
\end{enumerate}
Remind that we have assumed the following to justify the above procedures: (i) the potential $Q$ has no singularities in the complex $z$-plane; (ii) all the turning points are simple satisfying the condition (\ref{eq2.12}); and (iii) there are no Stokes segments.  For (i), if the Stokes regions that the path traverses contain some singularities, the connection formula (\ref{e151}) receives some corrections.  Such corrections are important in, e.g., analyses of a Fuchsian differential equation \cite{AKT3}.  Nevertheless, in most physical applications, the potential $Q$ is obtained by analytically continuing some regular function defined on ${\mathbb R}$ and thus $Q$ would not have singularities near the real axis.  Therefore, very roughly speaking, as long as one considers a path that goes near the real axis (which is actually a convenient choice for the vacuum pair production; see Sec.~\ref{sec3}), singularities of $Q$ sufficiently far from the real axis might not matter.  On the other hand, one should be careful about the assumptions (ii) and (iii) if the potential $Q$ has some symmetries.  For example, potentials having a symmetry $[Q(z)]^* = Q(z^*)$ must have a Stokes segment connecting a turning point $z_{\rm t}$ and its conjugate $z_{\rm t}^*$, which is also a turning point (see Sec.~\ref{sec32} for a proof).  The existence of a Stokes segment brings some difficulties in the exact WKB analysis.  For example, in a case of the Weber potential $Q_{\rm Weber}(z) = z^2-c^2$, there appear the so-called fixed singularities in the $\eta$-plane due to a Stokes segment that connects $z=+c$ and $z=-c$ \cite{vor83, tak08, she08, AKT2, CNP, DDP1, DDP2, DP}.  The fixed singularities do not change the structure on the $z$-plane and may cover the whole right-half of the $\eta$-plane.  This implies that the integration contour of the Laplace transformation hits the singularities no matter what values of $z$, i.e., the Borel sum never exists.  Therefore, to apply the exact WKB analysis safely, one needs to avoid the existence of a Stokes segment.  In Sec.~\ref{sec33}, we consider an infinitesimally small perturbation on top of a potential $Q$ to break symmetries causing a Stokes segment and derive a connection formula for a Stokes segment by taking the vanishing limit of the perturbation after safely applying the above procedures (1)-(3).

\section{Exact WKB analysis of the vacuum pair production} \label{sec3}

We discuss the vacuum pair production on the basis of the exact WKB analysis.  We begin with clarifying our physics setup as well as working assumptions in Sec.~\ref{sec30}; in a word, we consider a scalar quantum electrodynamics (QED) in the presence of a time-dependent electric field.  After explaining in Sec.~\ref{sec31} that the vacuum pair production can be understood in terms of a Stokes phenomenon of WKB solutions, we perform the exact WKB analysis to derive the production number formula in Secs.~\ref{sec32}-\ref{sec34}, which contain the main results of the present paper.  To be specific, we first discuss the generic structure of a Stokes graph for the vacuum pair production and show the existence of Stokes segments in Sec.~\ref{sec32}.  In Sec.~\ref{sec33}, we derive a connection formula for a Stokes segment by considering an infinitesimally small perturbation and assuming the semi-classical limit, in which we take a formal limit of $\hbar \ll 1$.  Using the results obtained in Secs.~\ref{sec32} and \ref{sec33}, we explicitly derive the production number formula in Sec.~\ref{sec34} and discuss the time-dependent effects including the interplay between the perturbative multi-photon pair production and the non-perturbative Schwinger mechanism and their interference effects.  In Sec.~\ref{sec35}, we compare our exact WKB result in the semi-classical limit with the worldline instanton method \cite{dun05a, dun06a, dun06b} and Brezin-Izykson's steepest descent evaluation \cite{bre70} and show that they are equivalent up to unimportant prefactors.

\subsection{Setup}\label{sec30}

We consider a scalar QED\footnote{It is straightforward to extend our exact WKB analysis to the usual spinor QED.  Indeed, one can always reduce the Dirac equation into a Klein-Gordon type second order differential equation, to which our analysis can be applied directly.  } in the presence of a time-dependent and spatially homogeneous U(1) electric field.  The complex scalar field $\phi$ in the momentum space satisfies a Klein-Gordon equation,
\begin{align}
	0	&= \left[  \hbar^2 \partial_t^2 + m^2 + \left( {\bm p} - e {\bm A}(t)  \right)^2 \right] \phi(t; {\bm p}) \nonumber\\
		&\equiv \left[ \hbar^2  \partial_t^2 + Q(t) \right] \phi(t; {\bm p}),  \label{eq3.1}
\end{align}
where $m>0, {\bm p}, e \in {\mathbb R}$, and ${\bm A}$ are mass, (canonical) momentum, the QED coupling constant, and a U(1) gauge potential for the electric field ${\bm E} \equiv -\partial_t {\bm A}$ in the temporal gauge $A_0=0$, respectively.  Note that the U(1) gauge potential ${\bm A}$ is real-valued on $t \in {\mathbb R}$.  We can naturally identify the parameter $\epsilon$ in the exact WKB analysis (\ref{eqa106}) with the Planck constant $\hbar$ in our quantum problem.  

We assume that (i) the electric field is switched off at the infinite future and past, i.e.,
\begin{align}
	\lim_{|t| \to \infty} {\bm A} = {\rm const.} \label{eq3.3}
\end{align}
Note that ${\bm A}(+\infty) \neq {\bm A}(-\infty)$ in general.  The assumption (i) shall be used to define particle states in a well-defined manner (see Sec.~\ref{sec31}).  For simplicity, we also assume that (ii) the gauge potential ${\bm A}$, after analytically continued to the complex $z$-plane $t \in {\mathbb R} \to z \in {\mathbb C}$, is an analytic function in the entire complex plane, as we assumed in Sec.~\ref{sec2}.  To quantitatively compute the production number of the vacuum pair production (see Secs.~\ref{sec33} and \ref{sec34}), we also consider (iii) the semi-classical limit, in which we neglect higher order terms in $\hbar$ appearing in connection matrices by formally taking the limit of $\hbar \ll 1$.

\subsection{The vacuum pair production as a Stokes phenomenon}\label{sec31}

We explain how the vacuum pair production can be understood in terms of a Stokes phenomenon of WKB solutions.  Namely, based on the canonical quantization formalism of quantum-field theory, we explain that the vacuum pair production is reduced to a scattering problem and that the production number is quantified by a Bogoliubov transformation that connects solutions of a field equation at the infinite past and future (see also Refs.~\cite{bir82, tan09}).

In the standard canonical quantization formalism, one may expand the field operator $\phi$ in terms of an annihilation operator $\hat{a}$ for a particle  and a creation operator  $\hat{b}^\dagger$ for an anti-particle as
\begin{align}
	\phi(t; {\bm p}) = \hat{a}({\bm p}) \varphi_{+}(t; {\bm p}) + \hat{b}^{\dagger}({\bm p}) \varphi_{-}(t; {\bm p}),  \label{eq3.6}
\end{align}
where $\varphi_{+}$ and $\varphi_{-}$ are mode functions for positive and negative energy states, respectively, satisfying the mode equation (\ref{eq3.1}).  One may normalize the mode functions $\varphi_{\pm}$ in the same manner as the WKB ansatz (\ref{eq2.3}) as
\begin{align}
	&1 = +{\rm i}\hbar\,  \varphi_{-} \overset{\leftrightarrow}{\partial}_t \varphi_{+} , \nonumber\\
	&0 = +{\rm i}\hbar\,  \varphi_{\pm} \overset{\leftrightarrow}{\partial}_t \varphi_{\pm} . \label{eq3.7}
\end{align}
Combining with the mode expansion~(\ref{eq3.6}), one finds
\begin{align}
	&\hat{a} =  +{\rm i}\hbar\,  \varphi_{-} \overset{\leftrightarrow}{\partial}_t \phi, \nonumber\\ 
	&\hat{b}^\dagger =  -{\rm i}\hbar\,  \varphi_{+} \overset{\leftrightarrow}{\partial}_t \phi.   \label{eq3.8}
\end{align}	
The non-vanishing canonical commutation relations for the creation/annihilation operators are
\begin{align}
	\delta^3({\bm p}-{\bm p}') = [ \hat{a}({\bm p}), \hat{a}^\dagger({\bm p}') ] = [ \hat{b}({\bm p}), \hat{b}^\dagger({\bm p}') ], 
\end{align}
and the other commutations are vanishing.

One needs a special care in identifying ``positive" and ``negative" energy states in the presence of an external electric field (or a time-dependent potential, in general).  Indeed, the time-translational invariance is explicitly broken while the electric field is being turned on (which physically means that the field is supplying energy to the system).  Thus, energy is no longer a good quantum number and there is no good quantum number to characterize a ``particle."  This implies that one can define the notion of a particle in a well-defined manner only after the electric field is turned off, for which the time-translational invariance is restored.  Suppose that there are asymptotic regions where $Q(t)$ becomes constant (the electric field is turned off), which we assume to occur at the asymptotic times (\ref{eq3.3}) [assumption (i)].  Then, one can naturally identify the positive and negative energy states by plane waves in these asymptotic regions, which are eigenfunctions of the time-translation operator (or the energy operator) $+{\rm i}\hbar \partial_t$.  However, we need to distinguish two different boundary conditions, since there are two separate asymptotic regions at $t=-\infty$ and $t=+\infty$.  We, therefore, introduce two separate sets of mode functions $\varphi_{\pm,{\rm as}}$ (``as'' denotes ``in, out'') as full solutions of the field equation \eqref{eq3.1} satisfying two different boundary conditions associated to two different asymptotic regions: 
\begin{align}
	0 &\equiv \lim_{t \to -\infty} \left[ \varphi_{\pm,{\rm in}}  - \frac{1}{\sqrt{2Q^{1/2}}} \exp\left[ \mp\frac{{\rm i}}{\hbar} \int^{t}_{t_0} {\rm d}t' \, Q^{1/2} \right] \right], \nonumber\\ 
	0 &\equiv \lim_{t \to +\infty} \left[ \varphi_{\pm,{\rm out}} - \frac{1}{\sqrt{2Q^{1/2}}} \exp\left[ \mp\frac{{\rm i}}{\hbar} \int^{t}_{t_0} {\rm d}t' \, Q^{1/2} \right] \right].   \label{eq3.10}
\end{align}
One has to distinguish $\varphi_{\pm, {\rm in}}$ and $\varphi_{\pm, {\rm out}}$ because the plane waves cannot be solutions of the mode equation (\ref{eq3.1}) for $|t| < \infty$, during which the electric field is being turned on, and they mix up with each other during the time-evolution so that $\varphi_{\pm, {\rm in}} \neq \varphi_{\pm, {\rm out}}$.  Since we now have two distinct mode functions, we must distinguish the corresponding annihilation/creation operators as well.  Noting Eq.~(\ref{eq3.8}), it is legitimate to define
\begin{align}
	\hat{a}_{\rm as} &\equiv  +{\rm i}\hbar\,  \varphi_{-,{\rm as}} \overset{\leftrightarrow}{\partial}_t \phi, \nonumber\\ 
	\hat{b}^\dagger_{\rm as} &\equiv  -{\rm i}\hbar\,  \varphi_{+,{\rm as}} \overset{\leftrightarrow}{\partial}_t \phi . \label{eq3.11}
\end{align}	
Physically, $\hat{a}_{\rm as}$ and $\hat{b}_{\rm as}$ correspond to annihilation operators of a particle and an anti-particle at the corresponding asymptotic times, respectively.  Since $\varphi_{\pm, {\rm in}} \neq \varphi_{\pm, {\rm out}}$, $\hat{a}_{\rm in}, \hat{b}_{\rm in} \neq \hat{a}_{\rm out}, \hat{b}_{\rm out}$ follows, and the mismatch between $\hat{a}_{\rm in}, \hat{b}_{\rm in}$ and $\hat{a}_{\rm out}, \hat{b}_{\rm out}$ is given in terms of that between $\varphi_{\pm, {\rm in}}$ and $\varphi_{\pm, {\rm out}}$.  To quantify the mismatch, we notice that $\varphi_{\pm, {\rm in}}$ and $\varphi_{\pm, {\rm out}}$ satisfy the same second order differential equation (\ref{eq3.1}) and that a second order differential equation can have only two independent solutions.  Therefore, there exists a $2\times 2$ matrix $U \in {\rm SL}(2,{\mathbb C})$ such that
\begin{align}
	\begin{pmatrix} \varphi_{+,{\rm out}} \\  \varphi_{-,{\rm out}} \end{pmatrix}
	= U \begin{pmatrix} \varphi_{+,{\rm in}} \\  \varphi_{-,{\rm in}} \end{pmatrix} 
	\equiv \begin{pmatrix} U_{11} & U_{12} \\ U_{21} & U_{22} \end{pmatrix} \begin{pmatrix} \varphi_{+,{\rm in}} \\  \varphi_{-,{\rm in}} \end{pmatrix} .  \label{eq3.13}
\end{align}
Note that $1 = \det U$ because $\varphi_{\pm, {\rm in}}$ and $\varphi_{\pm, {\rm out}}$ satisfy the same normalization condition (\ref{eq3.7}).  Plugging this expression into Eq.~(\ref{eq3.11}), we find
\begin{align}
	\begin{pmatrix} \hat{a}_{\rm out} \\  \hat{b}^\dagger_{\rm out} \end{pmatrix}
	= [U^{-1}]^{\mathsf T}	\begin{pmatrix} \hat{a}_{\rm in} \\  \hat{b}^\dagger_{\rm in} \end{pmatrix}
	= \begin{pmatrix} U_{22} & -U_{21} \\ -U_{12} & U_{11} \end{pmatrix} \begin{pmatrix} \hat{a}_{\rm in} \\  \hat{b}^\dagger_{\rm in} \end{pmatrix} ,  \label{eq3.15}
\end{align}
which is called a {\it Bogoliubov transformation}.

The Bogoliubov transformation (\ref{eq3.15}) is the essence to describe the vacuum pair production.  It is evident from Eq.~(\ref{eq3.15}) that the vacuum states at $t = -\infty$ such that
\begin{align}
	0 = \hat{a}_{\rm in} \ket{\rm vac;in} = \hat{b}_{\rm in} \ket{\rm vac;in}
\end{align}
is no longer annihilated by the annihilation operators at $t = +\infty$ as
\begin{align}
	0 \neq \hat{a}_{\rm out} \ket{\rm vac;in}, \hat{b}_{\rm out} \ket{\rm vac;in} .  
\end{align}
This means that particles are produced from the vacuum $\ket{\rm vac;in}$, and the production number reads
\begin{align}
	\frac{{\rm d}^6N_{e^-}}{{\rm d}{\bm x}^3{\rm d}{\bm p}^3} 
		&\equiv \frac{1}{V}\frac{\braket{{\rm vac;in}| \hat{a}^\dagger_{\rm out}  \hat{a}_{\rm out}  |{\rm vac;in} }}{ \braket{{\rm vac;in}|{\rm vac;in} } } = \frac{1}{(2\pi \hbar)^3} |U_{21}|^2,\nonumber\\ 
	\frac{{\rm d}^6N_{e^+}}{{\rm d}{\bm x}^3{\rm d}{\bm p}^3}  
		&\equiv \frac{1}{V}\frac{\braket{{\rm vac;in}| \hat{b}^\dagger_{\rm out} \hat{b}_{\rm out}  |{\rm vac;in} }}{ \braket{{\rm vac;in}|{\rm vac;in} } } = \frac{1}{(2\pi \hbar)^3} |U_{12}|^2, \label{eq3.16}
\end{align}
where $V$ is the spatial volume and we have used $\delta^3({\bm p}={\bm 0}) = V/(2\pi \hbar)^3$.  Thus, analyzing the vacuum pair production is equivalent to computing the off-diagonal components of the Bogoliubov transformation (\ref{eq3.15}).

The Bogoliubov transformation (\ref{eq3.15}) is nothing but a Stokes phenomenon of WKB solutions.  Indeed, Borel sums defined around $t = -\infty$ and $+\infty$, which we write $\Phi_{\pm, {\rm in}}$ and $\Phi_{\pm,{\rm out}}$, respectively, give solutions of the mode equation (\ref{eq3.1}) that asymptote the planes waves with $t \to -\infty$ and $t \to +\infty$.  Therefore, the Borel sum $\Phi_{\pm,{\rm as}}$ surely satisfies the boundary condition (\ref{eq3.10}), and we can identify 
\begin{align}
	\Phi_{\pm,{\rm as}} = \varphi_{\pm,{\rm as}}.  
\end{align}
In general, the Borel sum $\Phi_{\pm, {\rm in}}$ cannot be analytically continued to $\Phi_{\pm,{\rm out}}$ because of a Stokes phenomenon of WKB solutions.  To be precise, consider a path in the complex $z$-plane that connects $t \to z = -\infty$ to $+\infty$ traversing $n$ Stokes regions.  We label the Stokes region containing $t \to z = -\infty$ as 1 and successively increase the number $2,3,\cdots$ as entering the neighboring regions until reaching $n$, which is the final region containing $z = +\infty$.  Whenever analytically continuing $\Phi_{\pm, {\rm in}}$ from a Stokes region $i$ to $i+1$, $\Phi_{\pm, {\rm in}}$ experiences a sudden jump, which is quantified by a $2\times 2$ matrix $T_{i} \in {\rm SL}(2,{\mathbb C})$ just like Eq.~(\ref{e151}).  Note that Stokes regions are not necessarily separated by Stokes lines emanating from a simple turning point in general; a Stokes line emanating from a turning point of order more than one or a pole, or a Stokes segment can separate Stokes regions as well.  In such cases, the connection matrix $T_i$ cannot be identified with Eq.~(\ref{e151}).  As we discuss later, the Stokes regions $1,2, \cdots, n$ are, indeed, separated by Stokes segments in the case of the vacuum pair production [or potentials $Q$ having a property $[Q(z)]^* = Q(z^*)$ in general].  For the moment, we do not have to specify the explicit form of $T_i$, but an important point here is that one may symbolically express the relationship between $\Phi_{\pm, {\rm in}}$ and $\Phi_{\pm,{\rm out}}$ in terms of $T_i$ as
\begin{align}
	\begin{pmatrix} \Phi_{+,{\rm out}} \\  \Phi_{-, {\rm out}} \end{pmatrix}
	= T_n T_{n-1} \cdots T_2 T_1 \begin{pmatrix} \Phi_{+, {\rm in}} \\  \Phi_{-, {\rm in}} \end{pmatrix} .   
\end{align}
Comparing this expression with Eq.~(\ref{eq3.13}), we find
\begin{align}
	U = T_n T_{n-1} \cdots T_2 T_1.  \label{eq3.19}
\end{align}
Thus, evaluating the off-diagonal components of the Bogoliubov transformation (\ref{eq3.15}) is reduced to evaluating the product of the connection matrices $T_i$'s.  Note that Eq.~(\ref{eq3.19}) is an exact relation for exact $T_i$'s.  Approximation enters only when one approximates $T_i$'s.

Before closing this subsection, we comment on the relationship between our use of the exact WKB analysis and another common use of it, i.e., application to bound-state problems to get exact quantization conditions (e.g., Ref.~\cite{Sueishi:2020rug}).  In this work, we apply the exact WKB analysis to a scattering problem of quantum fields, i.e., to investigate how the behavior of a wave function changes between $t=-\infty$ and $t=+\infty$, and the change is described in terms of the Bogoliubov transformation (\ref{eq3.13}).  In scattering problems, the magnitude of the Borel sums $\Phi_{\pm,{\rm as}}$ (or the mode functions $\varphi_{\pm,{\rm as}}$) is always bounded $|\Phi_{\pm,{\rm as}}|<\infty$ because the potential is positive definite $Q(t \in {\mathbb R})>0$, and accordingly the system never gets quantized.  On the other hand, the exact WKB analysis has also been applied to bound-state problems in quantum mechanical systems.  In bound-state problems, the potential $Q$ is a function of space $x$, instead of time $t$.  A crucial difference is that the potential $Q$ takes negative values at the asymptotic points $x = \pm \infty$ in bound-state problems.  Thus, either of the Borel sums $\Phi_{+}$ and $\Phi_{-}$ diverges exponentially at $x = \pm \infty$.  So as to have a normalizable wave function, the divergent Borel sum at the asymptotic points $x = \pm \infty$ must be vanishing.  The Borel sums at the asymptotic points $x = \pm \infty$ are related with each other by a Bogoliubov transformation as in the case of our scattering problem, and hence the normalization condition requires some conditions onto the Bogoliubov transformation, which eventually give exact quantization conditions of a given quantum mechanical system.  During the above procedures in bound-state problems, the exact WKB analysis is applied to get the Bogoliubov transformation.  This is completely the same usage as in our scattering problem.  Nonetheless, our analyses presented below cannot be applied directly to bound-state problems.  For example, the generic properties of Stokes graphs that we shall discuss in Sec.~\ref{sec32} [in particular, the properties (4) and (5), which assume $Q>0$] should be modified by the property $Q<0$.  Accordingly, our path (see Fig.~\ref{fig1}) to compute the Bogoliubov transformation is not necessarily suitable in bound-state problems.  It is an interesting topic to extend our analyses to bound-state problems, and we leave it as a future work.

\subsection{Generic properties of Stokes graph} \label{sec32}

As the first step toward the exact WKB analysis of the vacuum pair production, let us discuss some generic properties of a Stokes graph for the potential $Q$ (\ref{eq3.1}): 
\begin{enumerate}
\item[(1)] The Stokes graph is symmetric in the upper and lower half complex planes, i.e., symmetric with respect to ${\rm Im}\,z \leftrightarrow -{\rm Im}\,z$.  

{\it Proof}.  Since $Q$ is a real-valued function on the real axis, the Schwarz reflection principle guarantees $[Q(z)]^* = Q(z^*)$.  Therefore, if $z_{\rm t}$ is a turning point, $z_{\rm t}^*$ is also a turning point because $0 = Q(z_{\rm t}) \Rightarrow 0 = [Q(z_{\rm t})]^* = Q(z_{\rm t}^*)$.  Similarly, if a point $z$ is on a Stokes line ${\mathcal C}_{z_{\rm t}}$, $z^*$ must be on a Stokes line ${\mathcal C}_{z_{\rm t}^*}$ because $0 = {\rm Im} \left[ {\rm i} \int^z_{z_{\rm t}} {\rm d}z' \sqrt{ Q(z')} \right] \Rightarrow 0 = {\rm Im} \left[ \left( {\rm i} \int^z_{z_{\rm t}} {\rm d}z' \sqrt{ Q(z')} \right)^* \right] = - {\rm Im} \left[ {\rm i} \int^{z^*}_{z^*_{\rm t}} {\rm d}z' \sqrt{ Q(z')} \right]$.

\item[(2)] Let $z_{\rm t}$ be a turning point of the potential $Q$.  The turning point $z_{\rm t}$ is of order one if $0 \neq \left( {\bm p} - e {\bm A}(z_{\rm t}) \right) \cdot e{\bm E}(z_{\rm t})$.

{\it Proof}.  Around a turning point $z \sim z_{\rm t}$, the potential $Q$ behaves as
\begin{align}
	\lim_{z\to z_{\rm t}} Q(z) = 2\left( {\bm p} - e {\bm A}(z_{\rm t}) \right) \cdot e{\bm E}(z_{\rm t}) \times (z-z_{\rm t}),  \label{eq3.20}
\end{align}
which proves the statement.

\item[(3)] A turning point $z_{\rm t}$ and its conjugate $z_{\rm t}^*$ are always connected by a Stokes segment.  

{\it Proof}.  It is sufficient to show $0 = {\rm Im} [ +{\rm i}\int_{z_{\rm t}}^{z_{\rm t}^*}\sqrt{Q}{\rm d}z ]$.  By using $[Q(z)]^*=Q(z^*)$, one finds $[ +{\rm i}\int_{z_{\rm t}}^{z_{\rm t}^*}\sqrt{Q}{\rm d}z ]^* = +{\rm i}\int_{z_{\rm t}}^{z_{\rm t}^*}\sqrt{Q}{\rm d}z$.  Therefore, ${\rm Im}[+{\rm i}\int_{z_{\rm t}}^{z_{\rm t}^*}\sqrt{Q}{\rm d}z]=0$.  

\item[(4)] The Stokes segment connecting a pair of $z_{\rm t}$ and $z_{\rm t}^*$ crosses the real axis only once.  

{\it Proof}.  A Stokes segment connecting $z_{\rm t}$ and $z_{\rm t}^*$ should cross the real axis because of the topology.  Namely, in order to connect two points having opposite signs of the imaginary parts ${\rm Im}\,z_{\rm t} = - {\rm Im}\,z_{\rm t}^*$ via a continuous line, the line must cross ${\rm Im}\,z=0$ because of the intermediate value theorem.  This proves the first half of the statement.  To prove the second half, suppose that there are two (or more) crossings $t_1, t_2 \in {\mathbb R}$ between the Stokes segment and the real axis.  One can assume $t_1 \neq t_2$ because Stokes lines cannot cross each other except at a turning point, which is absent on the real axis since $Q(t \in {\mathbb R}) \geq m > 0$, or at a singularity, which is assumed to be absent by the assumption (ii).  Then, $0 = {\rm Im}[{\rm i}\int^{t_2}_{t_1} \sqrt{Q}{\rm d}z]$ follows from the definition of a Stokes line $0 = {\rm Im}[{\rm i}\int^{t_1}_{z_{\rm t}} \sqrt{Q}{\rm d}z] = {\rm Im}[{\rm i}\int^{t_2}_{z_{\rm t}} \sqrt{Q}{\rm d}z]$ under the assumption (ii).  However, $Q \in {\mathbb R}$ on the real axis, and thus $0 \neq {\rm Im}[{\rm i}\int^{t_2}_{t_1} \sqrt{Q}{\rm d}z]$, which contradicts with the above.  Therefore, the Stokes segment can cross the real axis only once.

\item[(5)] Stokes lines/segments emanating from a turning point $z_{\rm t}$, other than the Stokes segment connecting $z_{\rm t}$ and $z_{\rm t}^*$, cannot cross the real axis.  

{\it Proof}.  The proof is the same as that for the property (4).  Suppose there are two (or more) Stokes lines/segments emanating from the same turning point $z_{\rm t}$ crossing the real axis at, say, $t_1, t_2 \in {\mathbb R}$.  Then, one gets $0 = {\rm Im}[{\rm i}\int^{t_2}_{t_1} \sqrt{Q}{\rm d}z]$, which is a contradiction.  Therefore, only one Stokes line, which is nothing but the Stokes segment connecting $z_{\rm t}$ and $z_{\rm t}^*$, can cross the real axis.  
\end{enumerate}

\begin{figure}[!t]
\begin{center}
\includegraphics[clip, width=0.9\textwidth]{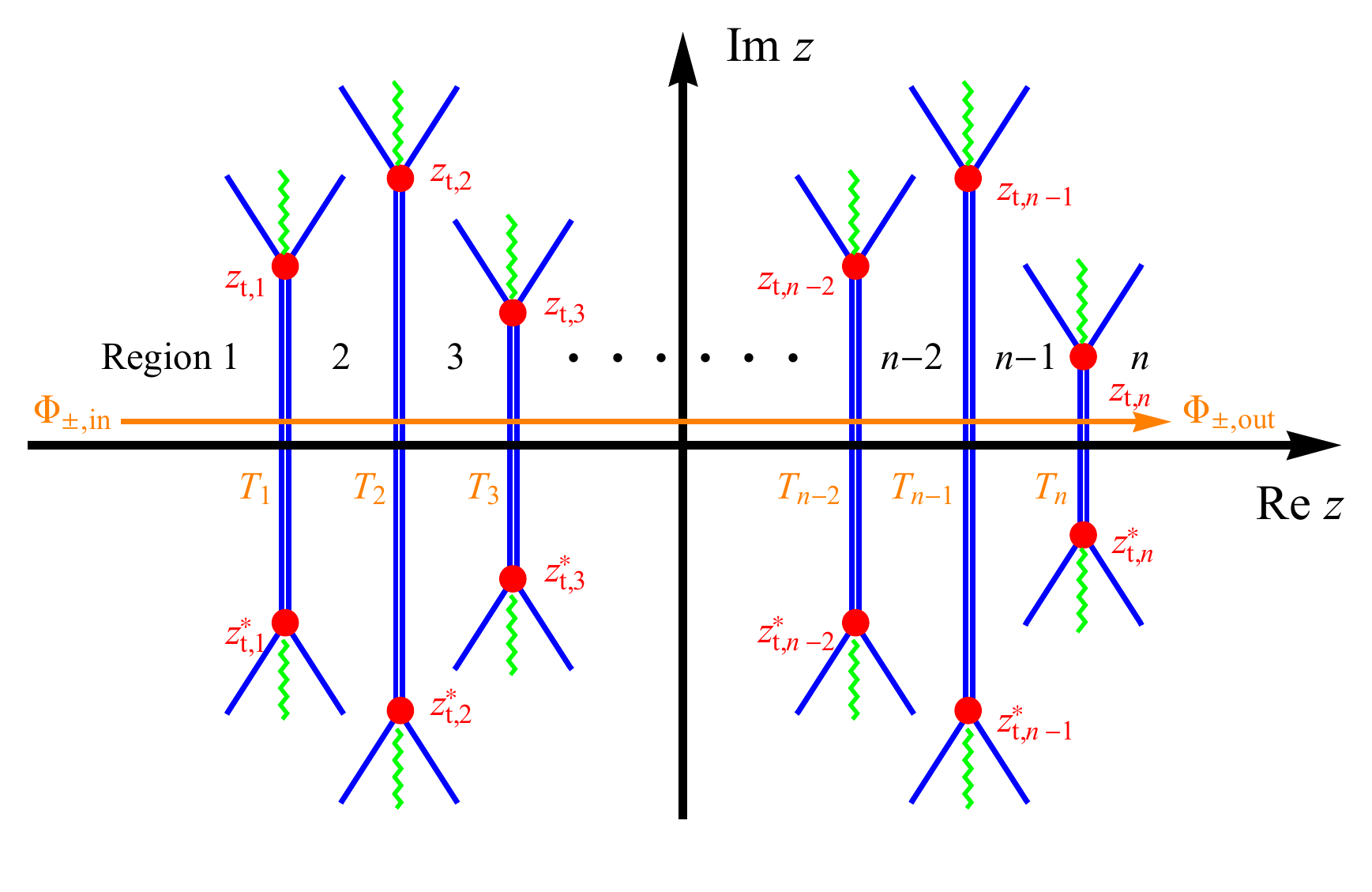}
\end{center}
\caption{\label{fig1} (color online) Generic structure of a Stokes graph for the potential $Q$ (\ref{eq3.1}).  The red points, the blue lines, and the green wavy lines are representing turning points, Stokes lines, and branch cuts, respectively.  Stokes segments connecting a pair of $(z_{{\rm t},i}, z_{{\rm t},i}^*)$ are represented by the doubled blue lines.  The orange arrow represents the path we consider to compute the Bogoliubov transformation between $\Phi_{\pm, {\rm in}}$ and $\Phi_{\pm,{\rm out}}$.  Note that the Stokes segments are depicted by straight lines.  This is just for simplification, and a Stokes segment is not necessarily a straight line in general.  Also, there could exist other Stokes segments connecting pairs on either half of the complex $z$-plane $(z_{{\rm t},i}, z_{{\rm t},j})$ or $(z_{{\rm t},i}^*, z_{{\rm t},j}^*)$.  Those Stokes segments do not cross the real axis directly (see also footnote~\ref{foot3} for appearance of multiply degenerated Stokes lines) and thus are omitted here for simplicity.  }
\end{figure}

Summarizing the above properties (1)-(5), a Stokes graph for the potential $Q$ (\ref{eq3.1}) should generically look like Fig.~\ref{fig1}.  In Fig.~\ref{fig1}, we have considered ${\bm p}$ such that $0 \neq \left( {\bm p} - e {\bm A}(z_{\rm t}) \right) \cdot e{\bm E}(z_{\rm t})$, so that any turning points $z_{{\rm t},i}$'s become of order one [property (2)].  In general, three Stokes lines emanate from a simple turning point.  Indeed, the integration of Eq.~(\ref{eq3.20}) yields
\begin{align}
	+ {\rm i} \int_{z_{\rm t}}^z \sqrt{Q(z')} {\rm d}z'
	\xrightarrow{z\sim z_{\rm t}}{} \frac{2{\rm i} }{3} \sqrt{ 2 \left( {\bm p} - e {\bm A}(z_{\rm t}) \right) \cdot e{\bm E}(z_{\rm t}) }  (z-z_{\rm t})^{3/2},  \label{eq3.21}
\end{align}
and thus Stokes lines around $z \sim z_{\rm t}$ are emanating in directions
\begin{align}
	\arg (z-z_{\rm t}) = \frac{2 \pi k}{3} - \frac{2}{3} \arg \frac{2{\rm i} }{3} \sqrt{ 2\left( {\bm p} - e {\bm A}(z_{\rm t}) \right) \cdot e{\bm E}(z_{\rm t}) },
\end{align}
where $k$ is an integer.  The integral (\ref{eq3.21}) is a multi-valued function around a turning point.  Accordingly, we inserted cuts for each turning point in such a way that the cuts do not traverse the real axis.  Then, one understands that $k$ can take three values when restricted in one Riemann sheet, and only one out of the three values corresponds to the Stokes segment that connects $z_{\rm t}$ and its conjugate $z_{\rm t}^*$.  Below, we concentrate on a Riemann sheet such that $\sqrt{Q}>0$ on the real axis.  Note that one may insert a cut in other directions, which does not alter our final results if one takes an appropriate path for the new cut.  Also, one may play with the exact WKB analysis for ${\bm p}$ such that $0 = \left( {\bm p} - e {\bm A}(z_{\rm t}) \right) \cdot e{\bm E}(z_{\rm t})$ by first considering some perturbation onto ${\bm p} \to {\bm p} + \delta{\bm p}$, for which $0 \neq \left( {\bm p} - e {\bm A}(z_{\rm t}) \right) \cdot e{\bm E}(z_{\rm t})$, and then taking $\delta {\bm p} \to 0$.

In order to compute the Bogoliubov transformation (\ref{eq3.15}) based on the exact WKB formula (\ref{eq3.19}), we consider a path from $z=-\infty$ to $+\infty$ just along the real axis as shown in Fig.~\ref{fig1}, so as to avoid the branch cuts located away from the real axis.  The path crosses all the Stokes segments that connect pairs of turning points $z_{{\rm t},i}$ and $z_{{\rm t},i}^*$, while it does not cross any other Stokes lines because they never traverse the real axis [properties (4) and (5)].  As explained in the previous subsections, the Borel sum $\Phi_{\pm, {\rm in}}$ experiences a sudden jump whenever it crosses the Stokes segments.  The jump is quantified by connection matrices $T_i$'s, whose explicit form in the semi-classical limit is obtained in the next subsection.

\subsection{Connection formula at a Stokes segment} \label{sec33}

We derive the connection matrix $T$ for a Stokes segment connecting a pair of turning points $z_{\rm t}$ and $z^*_{\rm t}$, assuming the semi-classical limit.  In the exact WKB analysis, the existence of a Stokes segment can spoil the Borel summability even inside of a Stokes region.  To circumvent this difficulty, let us recall that the existence of a Stokes segment is closely related to symmetries of the potential [e.g, $[Q(z)]^* = Q(z^*)$; see the proof for the property (3)].  This fact temps us to consider some perturbations $Q \to Q + \delta Q$ to break the symmetries, so as to de-degenerate a Stokes segment into some Stokes lines.  Then, one can safely apply the exact WKB analysis and expects that the correct connection matrix $T$ is reproduced by taking a vanishing limit of the perturbations after completing the procedures (1)-(3).  We show that this prescription works in the semi-classical limit and write down an explicit form of $T$.

\begin{figure}[!t]
\begin{center}
\includegraphics[clip, width=0.35\textwidth]{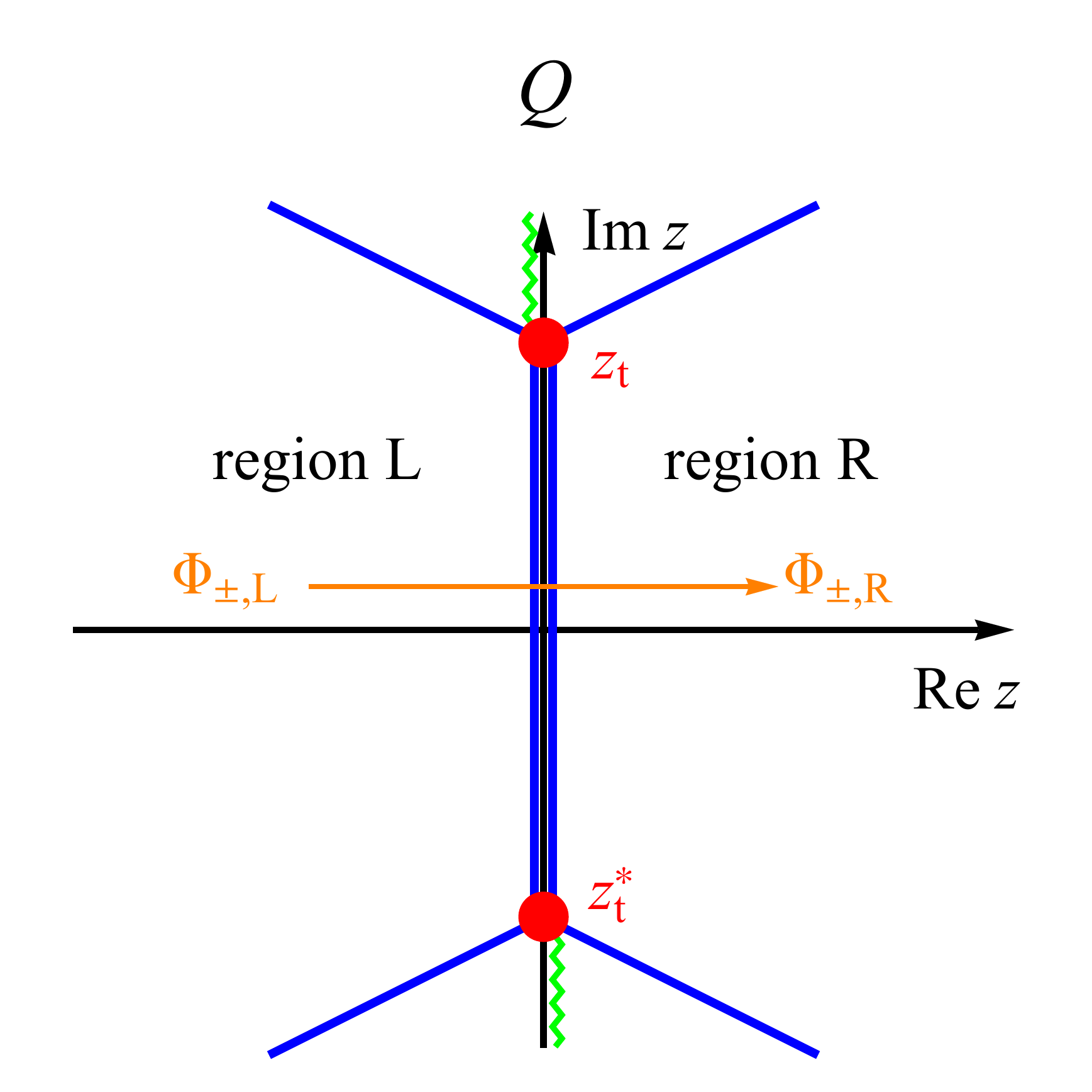} \hspace*{-8mm}
\includegraphics[clip, width=0.35\textwidth]{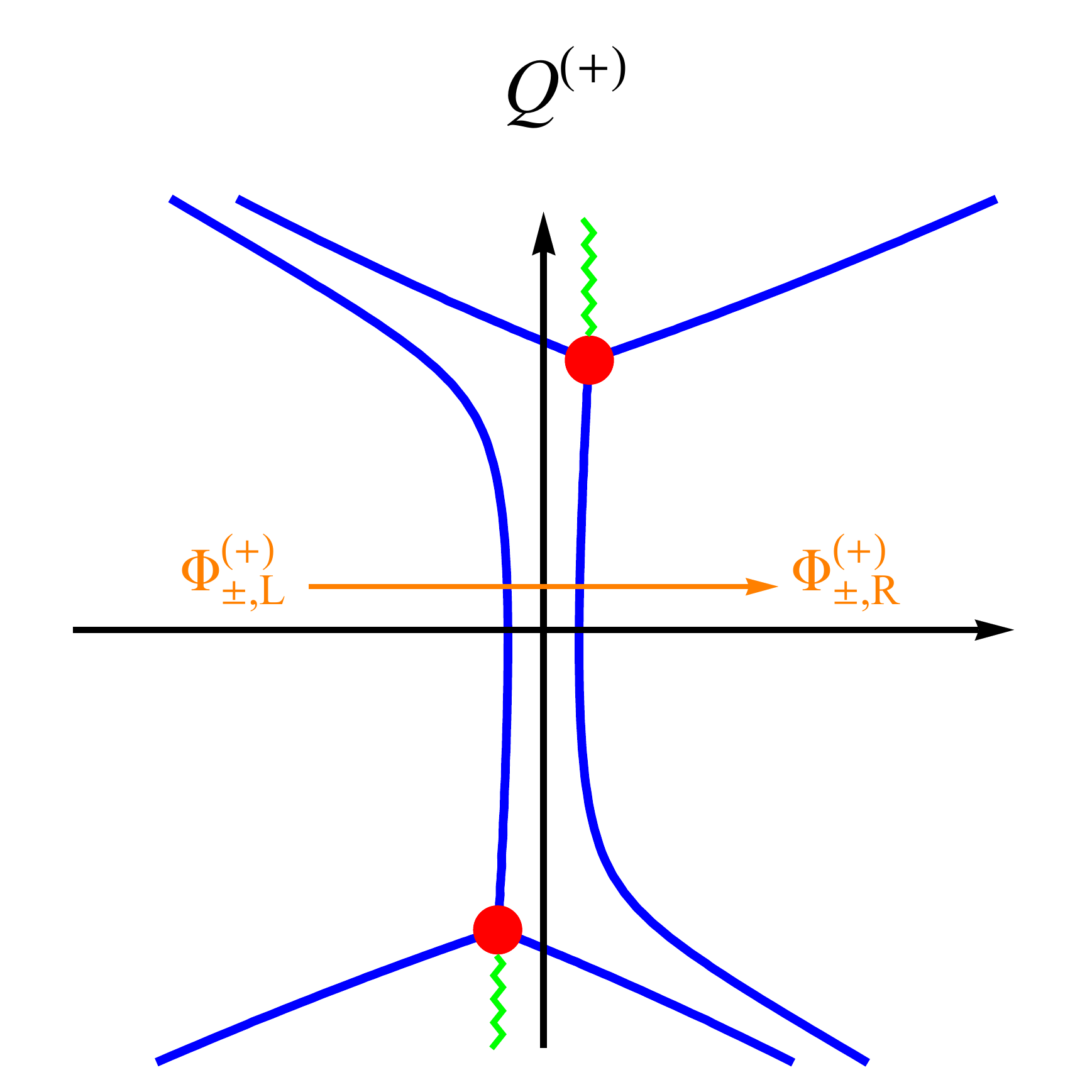}\hspace*{-7mm}
\includegraphics[clip, width=0.35\textwidth]{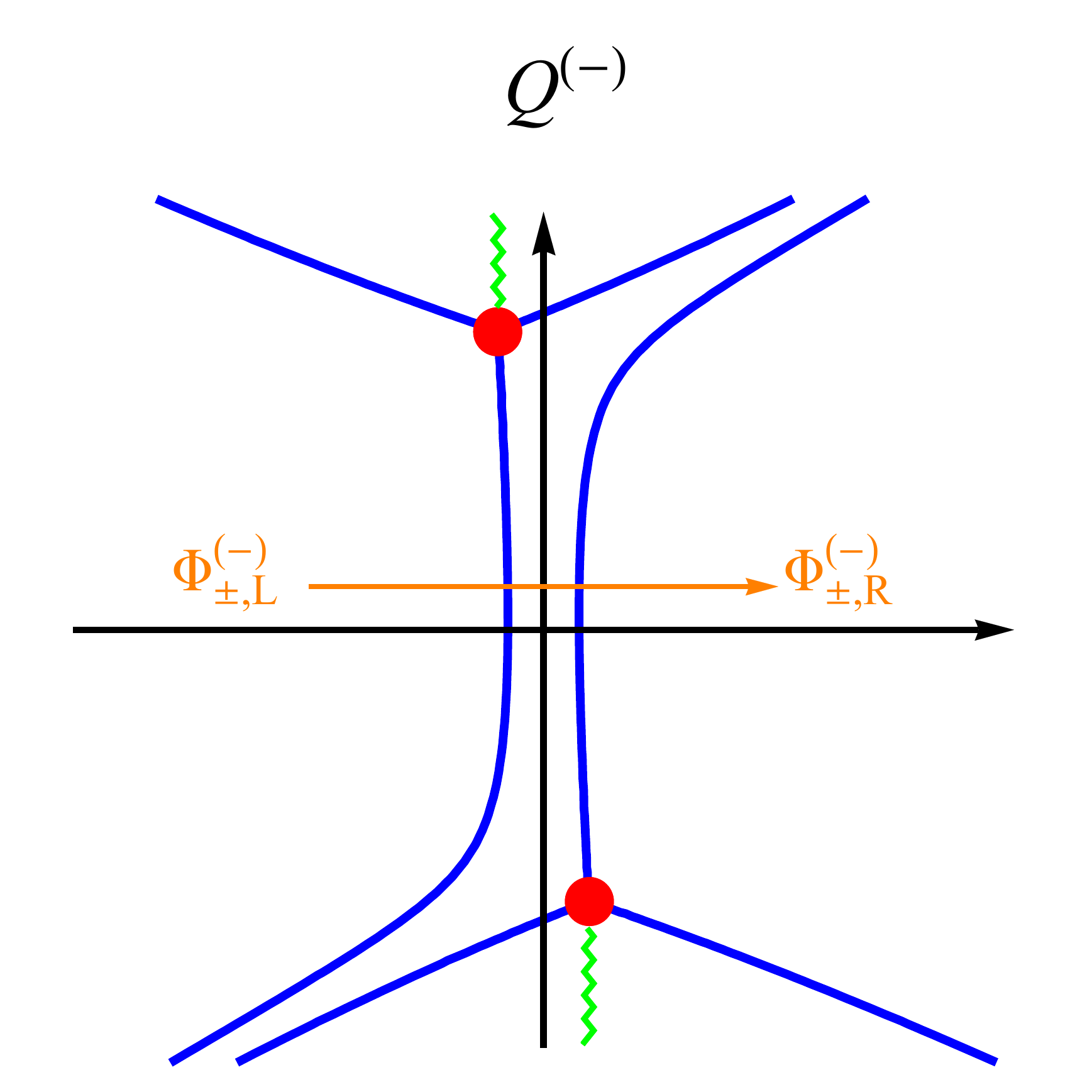}
\end{center}
\caption{\label{fig2} (color online) De-degeneration of a Stokes segment ${\mathcal C}_{z_{{\rm t}};z_{{\rm t}}^*}$ by the perturbations $Q \to Q^{(\pm)}$ (\ref{eq2.15}).  Turning points, Stokes lines, and branch cuts are represented by the red points, blue lines, and green wavy lines, respectively.  The Stokes segment ${\mathcal C}_{z_{{\rm t}};z_{{\rm t}}^*}$ is represented by a doubled blue line in the left most panel.  }
\end{figure}

To be specific, let us consider a Stokes segment ${\mathcal C}_{z_{{\rm t}};z_{{\rm t}}^*}$ into which two Stokes lines, ${\mathcal C}_{z_{{\rm t}}}$ and ${\mathcal C}_{z_{{\rm t}}^*}$, emanating from a pair of turning points, $z_{{\rm t}}$ and $z_{{\rm t}}^*$, are degenerated\footnote{\label{foot3}One can extend the discussions/results presented in this subsection to more generic situations where more than two Stokes lines are degenerated.  A Stokes segment connecting a pair of turning points $z_{{\rm t},i}$ and $z_{{\rm t},i}^*$  becomes multiply degenerated when, for example, $z_{{\rm t},i}^{(*)}$ forms a Stokes segment with a turning point on the same half of the complex $z$-plane $z_{{\rm t},i'}^{(*)}$ ($i' \neq i$).  Suppose we have such a {\it multiply degenerated Stokes segment}, into which $2n$ Stokes lines emanating from $z_{{\rm t},i}$ and $z_{{\rm t},i}^*$'s ($i=1,\cdots,n$) are degenerated [note that if a Stokes line ${\mathcal C}_{z_{{\rm t},i}}$ is degenerated with other Stokes lines, its complex pair ${\mathcal C}_{z_{{\rm t},i}^*}$ must be degenerated because of the property (1)].  Repeating the same argument, one can generalize the connection matrix for a doubly degenerated Stokes segment (\ref{eq3.32}) as
\begin{align}
	T^{(\pm)}
	\sim
	\begin{pmatrix} 
		1  
		&  -{\rm i}\sum_{i=1}^n {\rm e}^{-{\rm i}\,{\rm Im}\,\sigma_{z_{{\rm t},i}}/\hbar } {\rm e}^{-S_{z_{{\rm t},i}}/\hbar } \\ 
		+{\rm i}\sum_{i=1}^n {\rm e}^{+{\rm i}\,{\rm Im}\,\sigma_{z_{{\rm t},i}}/\hbar } {\rm e}^{-S_{z_{{\rm t},i}}/\hbar }
		& 1  	\end{pmatrix}
	\xrightarrow{\delta \to 0}{}
	T.   \label{eq-330}
\end{align}
where we have dropped ${\rm e}^{-(S_{z_{{\rm t},i}}+S_{z_{{\rm t},i'}})/\hbar}$ contributions by virtue of the semi-classical approximation.  Just for the sake of simplicity, we implicitly assume in the main text that all the Stokes segments are doubly degenerated and simply use the connection matrix (\ref{eq3.32}), instead of the generalized one (\ref{eq-330}).  One can carry out the same analysis for multiply degenerated Stokes segments in the following and can show that our main result for the production number (\ref{eq3.35}) is unchanged.  }.  Without loss of generality, we can set ${\rm Im}\,z_{{\rm t}} > 0$.  Note that ${\rm Im}\,z_{{\rm t}} = 0$ is forbidden because $Q>0$ on the real axis.  Now, consider an infinitesimally small perturbation\footnote{In principle, one can consider any perturbation here, as long as it breaks the symmetry $[Q(z)]^* \neq Q(z^*)$ so that Stokes segments are de-degenerated, and it gives the same result in the semi-classical limit.  Note that one may formally deform the Planck constant $\hbar \to \hbar {\rm e}^{\pm {\rm i}0^+}$ to break the symmetry, though it is equivalent to our perturbation (\ref{eq2.15}).  This is because, as we mentioned below Eq.~(\ref{eqa106}), changing values of $\hbar$ is equivalent to that of the argument of the potential $Q$.  Indeed, the Klein-Gordon operator $\hbar^2 \partial_z^2 + Q(z)$ is transformed under the deformation $\hbar \to \hbar {\rm e}^{\pm {\rm i}0^+}$ as $\hbar^2 \partial_z^2 + Q(z) 	\to (\hbar {\rm e}^{\pm{\rm i}0^+})^2 \partial_z^2 + Q(z)	= \hbar^2 \partial_\xi^2 + Q^{(\pm)}(\xi)$ with $\xi \equiv z{\rm e}^{\mp {\rm i}0^+}$, which is nothing but the perturbation (\ref{eq2.15}).  }, 
\begin{align}
	Q(z) \to Q^{(\pm)}(z) \equiv Q(z{\rm e}^{\pm {\rm i} \delta})\ {\rm with}\ \delta \to 0^+ , \label{eq2.15}
\end{align}
and denote the corresponding Borel sum as
\begin{align}
	\Phi_{\pm} \to \Phi_{\pm}^{(\pm)} .
\end{align}
One may interpret the perturbation (\ref{eq2.15}) as the ${\rm i}\epsilon$-prescription in quantum-field theory.  The perturbation $Q \to Q^{(\pm)}$ shifts (the real part of) the turning points $z_{{\rm t}}, z_{{\rm t}}^*$ as
\begin{align}
	{\rm Re}\,z_{{\rm t}} 
		&\to {\rm Re}\left[z_{{\rm t}} \times {\rm e}^{\mp{\rm i}\delta}\right]
		\sim {\rm Re}\,z_{{\rm t}} \pm 0^+ \times {\rm Im}\,z_{{\rm t}}, \ \nonumber\\
	{\rm Re}\,z^*_{{\rm t}} 
		&\to {\rm Re}\left[z^*_{{\rm t}} \times {\rm e}^{\mp{\rm i}\delta}\right]
		\sim {\rm Re}\,z_{{\rm t}} \mp 0^+ \times {\rm Im}\,z_{{\rm t}} \label{eq2.17}
\end{align}
because $0 = Q(z_{{\rm t}}) \Leftrightarrow 0 = Q^{(\pm)}(z_{{\rm t}} \times {\rm e}^{\mp{\rm i}\delta})$.  Equation (\ref{eq2.17}) implies that the turning points $z_{{\rm t}}$ and $z_{{\rm t}}^*$ are shifted to the right (left) and left (right), respectively, in the complex $z$-plane by the perturbation $Q \to Q^{(+)}$ ($Q \to Q^{(-)}$); see Fig.~\ref{fig2}.  The shift of the turning points $z_{{\rm t}}$ and $z_{{\rm t}}^*$ also shifts the Stokes lines ${\mathcal C}_{z_{{\rm t}}}$ and ${\mathcal C}_{z_{{\rm t}}^*}$, which in turn de-degenerate the Stokes segment ${\mathcal C}_{z_{{\rm t}};z_{{\rm t}}^*}$ as shown in Fig.~\ref{fig2}.  The two Stokes lines de-degenerated from the Stokes segment should have the following property: 
\begin{align}
	{\rm sgn}\left[ +{\rm i}\int^{z\in{\mathcal C}_{z_{{\rm t}}}}_{z_{{\rm t}}} \sqrt{Q(z')}{\rm d}z' \right] 
	= - {\rm sgn}\left[ +{\rm i}\int^{z\in{\mathcal C}_{z_{{\rm t}}^*}}_{z_{{\rm t}}^*} \sqrt{Q(z')} {\rm d}z' \right] > 0 . \label{eq2.18}
\end{align}
As a proof of Eq.~(\ref{eq2.18}), we first notice that the de-degenerated Stokes line ${\mathcal C}_{z_{{\rm t}}}$ (${\mathcal C}_{z_{{\rm t}}^*}$) crosses the real axis only once, which is a reminiscence of the property (4), and is heading to the lower (upper) half plane at the crossing because ${\rm Im}\,z_{\rm t} > 0$.  Thus, by noticing $\sqrt{Q}>0$ on the real axis, one understands that ${\rm d}z \propto -{\rm i}[\sqrt{Q}]^* \propto -{\rm i}$ (${\rm d}z \propto +{\rm i}[\sqrt{Q}]^* \propto +{\rm i}$) at the crossing.  On the other hand, the integrals in Eq.~(\ref{eq2.18}) monotonically increase or decrease as $z$ varies from their initial point to the end.  This is because if they suddenly start decreasing/increasing at some point $\bar{z}$ on the corresponding Stokes line while they were increasing/decreasing until that point, $\bar{z}$ must be a singularity or another turning point, which is assumed to be absent.  Therefore, the integral along ${\mathcal C}_{z_{{\rm t}}}$ (${\mathcal C}_{z_{{\rm t}}^*}$) is increasing $+{\rm i}\sqrt{Q} {\rm d}z > 0$ (decreasing $+{\rm i}\sqrt{Q} {\rm d}z < 0$) as $z$ varies.  This proves Eq.~(\ref{eq2.18}), as the integrals are vanishing at the initial points.  Now, we consider a path that crosses the de-degenerated Stokes lines from the left to right in the complex $z$-plane (see Fig.~\ref{fig2}).  The path first crosses the Stokes line ${\mathcal C}_{z_{\rm t}^*}$ (${\mathcal C}_{z_{\rm t}}$) in the clockwise (counter-clockwise) direction around the turning point, and then crosses ${\mathcal C}_{z_{\rm t}}$ (${\mathcal C}_{z_{\rm t}^*}$) in the counter-clockwise (clockwise) direction for $Q^{(+)}$ ($Q^{(-)}$).  Therefore, by successively using the connection formula (\ref{e151}), one gets
\begin{align}
	\left\{ \begin{array}{l}
		\displaystyle \begin{pmatrix} \Phi_{+,{\rm R}}^{(+)} \\ \Phi_{-,{\rm R}}^{(+)} \end{pmatrix}
			= \begin{pmatrix} 1 & 0 \\ +{\rm i}  {\rm e}^{ + \sigma_{z_{\rm t}}/\hbar } & 1 \end{pmatrix} \begin{pmatrix} 1 & -{\rm i}  {\rm e}^{ - \sigma_{z_{\rm t}^*}/\hbar }  \\ 0 & 1 \end{pmatrix} 	\begin{pmatrix} \Phi_{+,{\rm L}}^{(+)} \\ \Phi_{-,{\rm L}}^{(+)} \end{pmatrix} \vspace*{4mm}\\
	\displaystyle \begin{pmatrix} \Phi_{+,{\rm R}}^{(-)} \\ \Phi_{-,{\rm R}}^{(-)} \end{pmatrix} 
			= \begin{pmatrix} 1 & -{\rm i}  {\rm e}^{ - \sigma_{z_{\rm t}^*}/\hbar }  \\ 0 & 1 \end{pmatrix} \begin{pmatrix} 1 & 0 \\ +{\rm i}  {\rm e}^{ + \sigma_{z_{\rm t}}/\hbar } & 1 \end{pmatrix} \begin{pmatrix} \Phi_{+,{\rm L}}^{(-)} \\ \Phi_{-,{\rm L}}^{(-)} \end{pmatrix} 
	\end{array} \right.    
 \Leftrightarrow\;
	\begin{pmatrix} \Phi_{+,{\rm R}}^{(\pm)} \\ \Phi_{-,{\rm R}}^{(\pm)} \end{pmatrix}
	= T^{(\pm)} \begin{pmatrix} \Phi_{+,{\rm L}}^{(\pm)} \\ \Phi_{-,{\rm L}}^{(\pm)} \end{pmatrix} , \label{e--192}
\end{align}
where $\Phi^{(-)}_{\pm,{\rm R/L}}$ denotes the Borel sum defined at the right/left Stokes region with respect to the de-degenerated Stokes lines (see Fig.~\ref{fig2}) and 
\begin{align}
	T^{(\pm)}
	\equiv 
	\begin{pmatrix} 
		\displaystyle 1 + \frac{{\rm e}^{-2S_{z_{\rm t}}/\hbar } }{2} \mp \frac{{\rm e}^{-2S_{z_{\rm t}}/\hbar } }{2} 
		&  \displaystyle -{\rm i}\, {\rm e}^{-{\rm i}\,{\rm Im}\,\sigma_{z_{\rm t}}/\hbar } {\rm e}^{-S_{z_{\rm t}}/\hbar }  \\[3.5mm]
		\displaystyle +{\rm i}\,{\rm e}^{+{\rm i}\,{\rm Im}\,\sigma_{z_{\rm t}}/\hbar } {\rm e}^{-S_{z_{\rm t}}/\hbar }
		& \displaystyle 1 + \frac{{\rm e}^{-2S_{z_{\rm t}}/\hbar } }{2} \pm \frac{{\rm e}^{-2S_{z_{\rm t}}/\hbar } }{2}	\end{pmatrix} \label{eq2..18},
\end{align}
with 
\begin{align}
	S_{z_{\rm t}} 
		&\equiv - {\rm Re}\,\sigma_{z_{{\rm t}}} \nonumber \\
		&= +{\rm i}\int^{z_{{\rm t}}^*}_{z_{{\rm t}}} \sqrt{Q(z')}{\rm d}z' 
		> 0,  \label{eq2.19}
\end{align}
where we have used the assumption (ii) that there are no singularities in the complex $z$-plane to get the second line, and the positivity of $S_{z_{\rm t}}$ follows from Eq.~(\ref{eq2.18}).  Equation~(\ref{eq2..18}) indicates that
\begin{align}
	\lim_{\delta \to 0}\,[ T^{(+)} - T^{(-)}] \neq 0
\end{align}
because of the diagonal components.  It means that one cannot obtain $T$ in the naive $\delta \to 0$ limit\footnote{This can be understood as another type of Stokes phenomena that is induced by a Stokes segment.  Namely, the normalization of the Borel sums $ \Phi^{(+)}_{\pm}$ and $\Phi^{(-)}_{\pm}$ becomes discontinuous at $\delta = 0$, which leads to the discontinuity in the connection matrices $T^{(+)}$ and $T^{(-)}$.  In general, the discontinuity can be quantified by the so-called Voros symbol and is studied well for the Weber potential $Q_{\rm Weber}=z^2-c^2$; see, for example, Refs.~\cite{vor83, tak08, she08, AKT2, CNP, DDP1, DDP2, DP}. } of $T^{(\pm)}$.  Nevertheless, if one neglects the exponentially small factors ${\mathcal O}({\rm e}^{-2S_{z_{\rm t}}/\hbar } )$ in the diagonal components, which are negligible compared to the off-diagonal ones in the semi-classical limit, the limit $\delta \to 0$ becomes well-defined and one may obtain $T$ as
\begin{align}
	T^{(\pm)}
	\sim
	\begin{pmatrix} 
		1  
		&  -{\rm i}\, {\rm e}^{-{\rm i}\,{\rm Im}\,\sigma_{z_{\rm t}}/\hbar } {\rm e}^{-S_{z_{\rm t}}/\hbar } \\ 
		+{\rm i}\, {\rm e}^{+{\rm i}\,{\rm Im}\,\sigma_{z_{\rm t}}/\hbar } {\rm e}^{-S_{z_{\rm t}}/\hbar }
		& 1  	\end{pmatrix}
	\xrightarrow{\delta \to 0}{}
	T.  \label{eq3.32}
\end{align}
Below, we use this approximate connection matrix (\ref{eq3.32}) to derive the production number formula for the vacuum pair production.

\subsection{Production number formula in the semi-classical limit}\label{sec34}

We derive the production number formula within the exact WKB analysis in the semi-classical limit (which we shall call {\it semi-classical exact WKB analysis} for brevity).  Let $z_{{\rm t},i}$ ($i=1,2,\cdots,n$) be the $i$-th turning point in the upper half plane such that ${\rm Im}\,z_{{\rm t},i}>0$ and ${\rm Re}\,z_{{\rm t},1} \leq {\rm Re}\,z_{{\rm t},2} \leq \cdots \leq {\rm Re}\,z_{{\rm t},n}$ (see Fig.~\ref{fig1}).  One may identify $T_i$'s in Eq.~(\ref{eq3.19}) as the connection matrices for the Stokes segments emanating from the above $z_{{\rm t},i}$'s.  To proceed, we decompose Eq.~(\ref{eq3.32}) as 
\begin{align}
	T	&= I_{2} 
		+ \begin{pmatrix} 
			0 
			&  -{\rm i}\,{\rm e}^{-{\rm i}\,{\rm Im}\,\sigma_{z_{\rm t}}/\hbar} {\rm e}^{-S_{z_{\rm t}}/\hbar} \\ 
			+{\rm i}\, {\rm e}^{+{\rm i}\,{\rm Im}\,\sigma_{z_{\rm t}}/\hbar} {\rm e}^{-S_{z_{\rm t}}/\hbar}
			& 0 	\end{pmatrix}
			+ {\mathcal O}({\rm e}^{-2S_{z_{{\rm t},i}}/\hbar} ) \nonumber\\
		&\equiv I_2 + \delta T + {\mathcal O}({\rm e}^{-2S_{z_{{\rm t},i}}/\hbar} ), \label{eq3.33}
\end{align}
where $I_2$ is a $2\times 2$ unit matrix.  Plugging this expression (\ref{eq3.33}) into Eq.~(\ref{eq3.19}), we obtain
\begin{align}
	U &= \left[ I_2 + \delta T_n + {\mathcal O}({\rm e}^{-2S_{z_{{\rm t},n}}/\hbar } )  \right] \left[ I_2 + \delta T_{n-1} + {\mathcal O}({\rm e}^{-2S_{z_{{\rm t},{n-1}}}/\hbar } )  \right] \cdots \left[ I_2 + \delta T_1 + {\mathcal O}({\rm e}^{-2S_{z_{{\rm t},1}}/\hbar } )  \right] \nonumber\\
	  &= I_2 + \sum_{i=1}^n \delta T_i + {\mathcal O}({\rm e}^{-(S_{z_{{\rm t},i}}+S_{z_{{\rm t},i'}})/\hbar} ), 
\end{align}
where we have used $\|\delta T \|= {\mathcal O}({\rm e}^{-S_{z_{\rm t}}/\hbar} )$ and neglected terms of the order of ${\mathcal O}({\rm e}^{-(S_{z_{{\rm t},i}}+S_{z_{{\rm t},i'}})/\hbar} )$, so as to be consistent with the semi-classical approximation that we used in Eq.~(\ref{eq3.32}).  The off-diagonal components read
\begin{align}
	U_{12} = U_{21}^* = -{\rm i} \sum_{i=1}^n {\rm e}^{-{\rm i}\,{\rm Im}\,\sigma_{z_{{\rm t},i}}/\hbar} {\rm e}^{-S_{z_{{\rm t},i}}/\hbar} + {\mathcal O}({\rm e}^{-(S_{z_{{\rm t},i}}+S_{z_{{\rm t},i'}})/\hbar} ).  \label{eq3.34} 
\end{align}
Therefore, the phase-space density of the produced particles (\ref{eq3.16}) reads
\begin{align}
	\frac{{\rm d}^6N_{e^-}}{{\rm d}{\bm x}^3{\rm d}{\bm p}^3} 
	= \frac{{\rm d}^6N_{e^+}}{{\rm d}{\bm x}^3{\rm d}{\bm p}^3} 
	= \frac{1}{(2\pi \hbar)^3}\left|  \sum_{i=1}^n {\rm e}^{-{\rm i}\,{\rm Im}\,\sigma_{z_{{\rm t},i}}/\hbar} {\rm e}^{-S_{z_{{\rm t},i}}/\hbar} \right|^2 \left( 1 + {\mathcal O}({\rm e}^{-S_{z_{{\rm t},i}}/\hbar} ) \right). \label{eq3.35}
\end{align}
The same amount of particles and anti-particles are produced, indicating that a particle and an anti-particle are always created as a pair because of the gauge invariance.  The production number (\ref{eq3.35}) is basically controlled by the exponentially small factor ${\rm e}^{-S_{z_{{\rm t},i}}/\hbar}$, and a pair of turning points having the smallest $S_{z_{{\rm t},i}}$ dominates the production.  When there are several pairs of turning points equally contributing to the production, those turning points interfere with each other because of the factor ${\rm e}^{-{\rm i}\,{\rm Im}\,\sigma_{z_{{\rm t},i}}/\hbar}$.  This point was missing in the naive imaginary time formalism developed by Popov \cite{pop05} and was first clarified by Dumlu and Dunne \cite{ces10, ces11}, who demonstrated that the interference is responsible for the characteristic momentum signatures observed in field profiles with subcycle structures \cite{heb09, ort11, chr12, pan15, gre17, gre19, fk, fk2, tay20}.

Equation~(\ref{eq3.35}) agrees with other semi-classical methods such as the steepest descent evaluation of the Bogoilubov coefficients by Brezin and Izykson \cite{bre70} and the worldline instaton method \cite{dun05a, dun06a, dun06b}, as we discuss in detail in Sec.~\ref{sec35}.  The same formula was also used in particle production with other types of an external field in, e.g., Refs.~\cite{li19, has20}, whose derivation is based on Berry's divergent asymptotic series method \cite{ber89}.

\subsubsection{Interplay between the non-perturbative Schwinger mechanism and the perturbative multi-photon pair production process}

The semi-classical exact WKB formula (\ref{eq3.35}) describes the interplay between the non-perturbative Schwinger mechanism for a slow electric field and the perturbative multi-photon pair production process for a fast electric field.  From a view point of the exact WKB analysis, the interplay can be understood in terms of the change of the location of dominant turning points.  We also confirm that the formula (\ref{eq3.35}) is valid in the semi-classical regime and cannot describe processes beyond that regime such as the low-order pair production processes (e.g., one-photon pair production).

To see those points in an analytical manner, let us assume that the electric field has a typical frequency $\Omega>0$ (without assuming some specific field profile).  Then, the Fourier transformation of the electric field $e{\bm {\mathcal E}}$, 
\begin{align}
	e{\bm E}(t)
			&\equiv \int^{+\infty}_{-\infty} \frac{{\rm d}\omega}{2\pi} {\rm e}^{+{\rm i}\omega t} e{\bm {\mathcal E}}(\omega), 
\end{align}
may be peaked sharply at $\omega \sim \pm \Omega$.  Using $e{\bm {\mathcal E}}$, one may approximate the gauge potential $e{\bm A}$ as
\begin{align}
	e{\bm A}(t) 
		&= \int^{+\infty}_{-\infty} \frac{{\rm d}\omega}{2\pi} \frac{{\rm e}^{+{\rm i}\omega t}}{-{\rm i}\omega} e{\bm {\mathcal E}}(\omega) \nonumber\\
		&\sim \frac{{\rm e}^{+{\rm i}\Omega t}}{-{\rm i}\hbar \Omega} \int_{\omega \sim +\Omega} \frac{{\rm d}\omega}{2\pi} \hbar e {\bm {\mathcal E}}(\omega) + \frac{{\rm e}^{-{\rm i}\Omega t}}{+{\rm i} \hbar \Omega} \int_{\omega \sim -\Omega} \frac{{\rm d}\omega}{2\pi} \hbar e{\bm {\mathcal E}}(\omega) \nonumber\\
		&= \frac{{\rm e}^{+{\rm i}\Omega t}}{-{\rm i}\hbar \Omega} \frac{\hbar e\bar{{\bm E}}}{2} + \frac{{\rm e}^{-{\rm i}\Omega t}}{+{\rm i}\hbar\Omega} \frac{\hbar  e\bar{{\bm  E}}^*}{2},
\end{align}
where 
\begin{align}
	\frac{e\bar{{\bm E}}}{2} \equiv \int_{\omega \sim +\Omega} \frac{{\rm d}\omega}{2\pi} e{\bm {\mathcal E}}(\omega) 
\end{align}
characterizes the typical electric field strength [for example, a monochromatic electric field $e{\bm E} = e{\bm E}_0 \cos (\Omega t)$ gives $e\bar{\bm E} = e{\bm E}_0$].  Then, the potential $Q$ in the complex $z$-plane reads
\begin{align}
	Q(z) 	
		&\sim	m^2 + \left( {\bm p} + \frac{{\rm e}^{+{\rm i}\Omega z}}{+{\rm i}\hbar \Omega} \frac{\hbar  e\bar{{\bm  E}}}{2} + \frac{{\rm e}^{-{\rm i}\Omega z}}{-{\rm i}\hbar  \Omega} \frac{ \hbar  e\bar{{\bm  E}}^*}{2} \right)^2.  \label{eqq3.41}
\end{align}
For small $\hbar\Omega$, one may expand the exponentials in Eq.~(\ref{eqq3.41}) as 
\begin{align}
	Q \sim m^2 + \left(  {\bm P} + {\rm Re}\,e\bar{\bm E}  \times z   \right)^2 , \label{eq-3.35}
\end{align}
where 
\begin{align}
	{\bm P} \equiv {\bm p} - e{\bm A}(0) \sim  {\bm p} + \frac{1}{\hbar \Omega} {\rm Im}\,\hbar e\bar{\bm E} \label{eqq339}
\end{align}
is kinetic momentum (at time $t=0$).  We neglected ${\mathcal O}(|\Omega z|^2)$-terms, which is justified if $\gamma \lesssim 1$, with
\begin{align}
	\gamma \equiv \frac{ m \hbar \Omega}{\left|{\rm Re}\,\hbar e\bar{\bm E} \right|} \label{EQ339}
\end{align}
being the so-called Keldysh parameter \cite{kel65, bre70, pop71, tay14}.  Indeed, Eq.~(\ref{eq-3.35}) has only a single pair of turning points $z_{\rm t}$ and $z_{\rm t}^*$ given by
\begin{align}
	z_{\rm t} \sim \frac{-P_{\parallel} + {\rm i} \sqrt{m^2 + |{\bm P}_{\perp}|^2}}{\left|{\rm Re}\,e\bar{\bm E} \right|}, \label{eqq3.36}
\end{align}
where $P_{\parallel}$ and ${\bm P}_{\perp}$ denote kinetic momenta parallel and perpendicular to the electric field $e\bar{\bm E}$, respectively.  Thus, $|\Omega z|^2 \ll 1$ is guaranteed along the integration contour of $S_{z_{\rm t}}$ (\ref{eq2.19}), i.e., $z: z_{\rm t} \to z_{\rm t}^*$, as long as $1 \gg |\Omega z_{\rm t}| \sim \gamma$ is satisfied.  Now, using Eqs.~(\ref{eq-3.35}) and (\ref{eqq3.36}), one can explicitly evaluate $S_{z_{\rm t}}$ (\ref{eq2.19}) as
\begin{align}
	S_{z_{\rm t}} \sim \frac{\pi}{2} \frac{m^2 + |{\bm P}_\perp|^2}{\left|{\rm Re}\,e\bar{\bm E} \right|},  
\end{align}
where we have used $2 \int {\rm d}z \sqrt{1+z^2} = z\sqrt{1+z^2} + {\rm ln}\left[ z + \sqrt{1+z^2}  \right] $.  Therefore, the semi-classical exact WKB formula (\ref{eq3.35}) yields
\begin{align}
	\frac{{\rm d}^6N_{e^{\pm}}}{{\rm d}{\bm x}^3{\rm d}{\bm p}^3}
	\sim \frac{1}{(2\pi \hbar )^3} \exp \left[-\pi \frac{m^2 + |{\bm P}_\perp |^2}{\left|{\rm Re}\,\hbar e\bar{\bm E} \right|}  \right], \label{eq342}
\end{align}
which precisely agrees with Schwinger's result for a constant electric field \cite{sch51}.  On the other hand, for large $\hbar\Omega$ such that $ \gamma \gtrsim 1$, one may approximate $Q$ (\ref{eqq3.41}) by expanding the square as
\begin{align}
		Q \sim m^2 + {\bm p}^2 + \frac{{\rm e}^{+{\rm i}\Omega z}}{+{\rm i}\hbar \Omega} \left( {\bm p} \cdot \hbar e\bar{{\bm  E}}\right) +  \frac{{\rm e}^{-{\rm i}\Omega z}}{-{\rm i} \hbar \Omega} \left( {\bm p}\cdot \hbar e\bar{{\bm  E}} \right)^* .  \label{eqq3.42} 
\end{align}
The corresponding turning points $z_{\rm t}$ and $z_{\rm t}^*$ read
\begin{align}
	z_{\rm t} \sim \frac{1}{-{\rm i}\Omega} {\rm ln} \frac{+{\rm i}\hbar \Omega (m^2 + {\bm p}^2)}{ \left( {\bm p} \cdot \hbar e\bar{{\bm  E}}  \right)^*}.  
\end{align}
Then, we can explicitly evaluate $S_{z_{\rm t}}$ (\ref{eq2.19}) as
\begin{align}
	S_{z_{\rm t}} 
		\sim {\rm i} \int^{z_{\rm t}^*}_{z_{\rm t}} dz \sqrt{m^2 + {\bm p}^2} 
		\sim  \frac{2\sqrt{m^2 + {\bm p}^2}}{\Omega} {\rm ln} \left| \frac{\hbar \Omega (m^2 + {\bm p}^2)}{{\bm p} \cdot \hbar e\bar{{\bm  E}} } \right|.  
\end{align}
Therefore, 
\begin{align}
	\frac{{\rm d}^6N_{e^{\pm}}}{{\rm d}{\bm x}^3{\rm d}{\bm p}^3}
	\sim \frac{1}{(2\pi \hbar )^3} \left| \frac{{\bm p} \cdot \hbar e\bar{{\bm  E}} }{\hbar \Omega (m^2 + {\bm p}^2)} \right|^{2\frac{2{\sqrt{m^2 + {\bm p}^2}}}{\hbar \Omega}},   \label{EQ345}
\end{align}
which describes the multi-photon pair production involving $n=2{\sqrt{m^2 + {\bm p}^2}}/\hbar \Omega \gg 1$ photons.  Equation~(\ref{EQ345}) also indicatess that the low-order perturbative processes such as the one-photon pair production (see Appendix~\ref{appa4}) cannot be described within the semi-classical approximation, for which higher order corrections ${\mathcal O}({\rm e}^{-2S_{z_{\rm t}}/\hbar})$ must be taken into account.  Note that the scalar QED is a derivatively coupled theory, so that $N_{e^{\pm}} \sim 0$ for ${\bm p} \sim {\bm 0}$.

\begin{figure}[!t]
\begin{center}
\hspace*{-16mm}
\includegraphics[clip, width=0.7\textwidth]{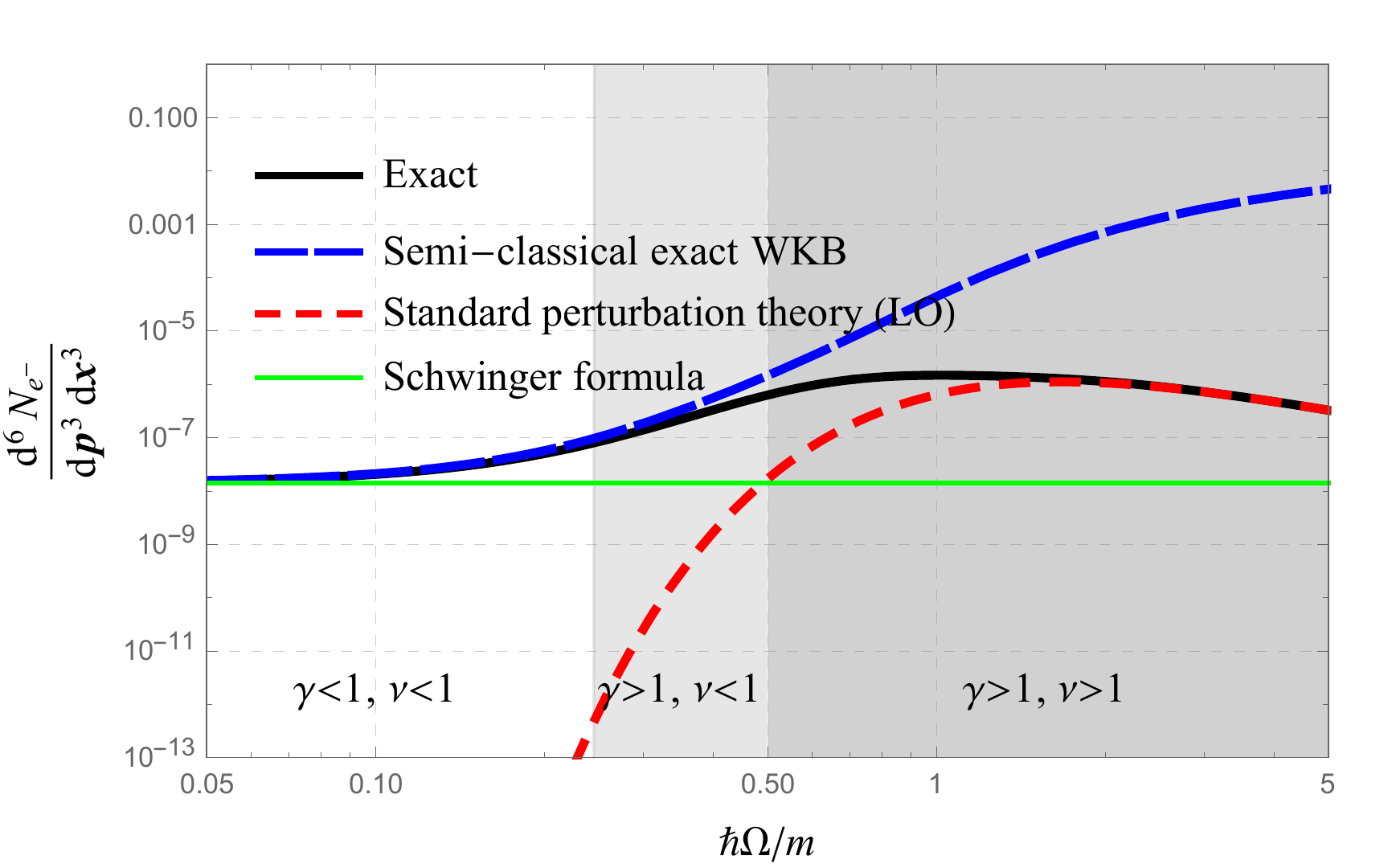} 
\hspace*{-5mm}
\end{center}
\caption{\label{fig-3} (color online) The production number $N_{e^-}$ for the Sauter electric field (\ref{eQ196}) with parameters $|{\bm p}_\perp|/m=0, p_\parallel/m=0.20$, and $\hbar eE_0/m^2=0.25$.  The black line shows the exact result (\ref{eq203}), and colored lines show different theoretical predictions: the semi-classical exact WKB formula [Eq.~(\ref{eq3.35}) or (\ref{eqqa-45})] (long dashed blue line), the lowest order standard perturbation theory [Eq.~(\ref{eqqq89}) or (\ref{eqaaaa48})] (short dashed red line), and the Schwinger formula for a constant electric field (\ref{eq342}) (green line).  The background shading distinguishes values of $\gamma = \frac{m \hbar\omega}{\hbar eE_0}$ (i.e., the Keldysh parameter) and $\nu = \frac{(\hbar \Omega)^2}{\hbar eE_0}$.  }
\end{figure}

We demonstrate the interplay and validity of the semi-classical exact WKB formula (\ref{eq3.35}) by considering, as an example, the so-called Sauter electric field \cite{sau31, tay14}, 
\begin{align}
	{\bm E}(t)  = \frac{E_0}{[\cosh (\Omega t)]^2} \times {\bf e}_\parallel.  \label{eQ196}  
\end{align} 
An advantage of this field configuration is that one can derive an exact formula for the production number by analytically solving the mode equation (\ref{eq3.1}).  The exact result, in comparison with the formula (\ref{eq3.35}) and the standard perturbation theory at the lowest order (see Appendix~\ref{appa4}), is shown in Fig.~\ref{fig-3} (see Appendix~\ref{appa-5} for more details such as the derivation of the analytical formulas and the Stokes graph).  Qualitative features of Fig.~\ref{fig-3} can be understood in terms of two dimensionless parameters \cite{tay14}, i.e., the Keldysh parameter $\gamma$ (\ref{EQ339}) and 
\begin{align}
	\nu \equiv \frac{(\hbar \Omega)^2}{\hbar eE_0}.  
\end{align}
For small $\hbar \Omega$, where the semi-classical approximation is justified, the formula (\ref{eq3.35}) and the semi-classical exact WKB result coincide.  Both results agree with the Schwinger formula for a constant electric field (\ref{eq342}) at around $\hbar \Omega \sim 0$ (or $\gamma \lesssim 1, \nu \lesssim 1$) and deviate from it with increasing $\hbar \Omega$ (or $\gamma \gtrsim 1, \nu \lesssim 1$), implying that the production mechanism becomes dominated by the perturbative multi-photon pair production process.  For larger values of $\hbar \Omega$ (such that $\gamma \gtrsim 1, \nu \gtrsim 1$), the semi-classical approximation is invalid.  In such a parameter region, the low-order perturbative processes such as the one-photon pair production process dominate the production, and the standard perturbation theory works better than the semi-classical approaches.

\subsubsection{The dynamically assisted Schwinger mechanism} \label{sec352}

The production number can be enhanced significantly if one superimposes a weak fast electric field onto a strong slow electric field (the dynamically assisted Schwinger mechanism \cite{sch08, piz09, dun09, mon10a, mon10b}), which is an analog of the Franz-Keldysh effect in semi-conductor physics \cite{fk, fra58, kel58, tah63, cal63}.  The semi-classical exact WKB formula (\ref{eq3.35}) gives a good description of this mechanism in the semi-classical parameter regime.

To get an analytical understanding of the above mechanism within the semi-classical exact WKB formula (\ref{eq3.35}), let us assume that the strong slow electric field is sufficiently slow and treat it as a constant field.  For simplicity, we also assume that the strong and weak fields are pointing in the same direction $\propto {\bf e}_\parallel$.  The gauge potential may be expressed as
\begin{align}
	{\bm A}(t) = - \left(  E_{\rm s} t + \varepsilon \int^t_0 {\rm d}t' E_{\rm w}(t') \right) \times {\bf e}_{\parallel}\ \Rightarrow\ {\bm E}(t) = ( E_{\rm s} + \varepsilon E_{\rm w}(t))  \times  {\bf e}_{\parallel}, 
\end{align}
where we normalized ${\bf e}_{\parallel}$ as $|{\bf e}_\parallel|^2=1$ and set $eE_{\rm s}>0$ and ${\bm A}(0)={\bm 0}$ without losing generality.  $eE_{\rm s}$ and $eE_{\rm w}$ (such that $|eE_{\rm w}| \ll eE_{\rm s}$) denote the strength of the strong and weak fields, respectively.  $\varepsilon$ is a book-keeping parameter, inserted so as to make sure the weakness of the weak field.  Now that $eE_{\rm w}$ is weak, one speculates that the location of a turning point may not change significantly from that for the strong constant electric field alone $eE_{\rm s}$: 
\begin{align}
	z_{\rm t} = z_{\rm t}^{(0)} +  {\mathcal O}(\varepsilon), \label{eq348}
\end{align}
where $z_{\rm t}^{(0)}$ is the turning point (in the upper complex half plane) for $eE_{\rm s}$,
\begin{align}
	z_{\rm t}^{(0)} = \frac{ - p_{\parallel} + {\rm i} \sqrt{m^2 + {\bm p}_{\perp}^2}}{eE_{\rm s}}.  \label{eeq349}
\end{align}
Note that one needs to replace $p_\parallel, {\bm p}_\perp$ with $P_\parallel, {\bm P}_\perp$ as in Eq.~(\ref{eqq339}) if ${\bm A}(0) \neq {\bm 0}$.  Using Eq.~(\ref{eqq339}), we can compute $S_{z_{\rm t}}$ in Eq.~(\ref{eq2.19}) up to the lowest order in $\varepsilon$ as
\begin{align}
	S_{z_{\rm t}} 
		&= \frac{\pi}{2} \frac{m^2 + |{\bm p}_\perp|^2}{eE_{\rm s}}
				- {\rm i} \varepsilon  \int^{z_{\rm t}^{(0)*}}_{z_{\rm t}^{(0)}} \!\!\! {\rm d}z  \sqrt{  m^2 + {\bm p}_\perp^2 + \left(  p_\parallel + eE_{\rm s} z \right)^2 } \frac{eE_{\rm w}(z)}{eE_{\rm s}} 
			 + {\mathcal O}(\varepsilon^2) \nonumber\\
		&= \frac{\pi}{2} \frac{m^2 + |{\bm p}_\perp|^2}{eE_{\rm s}} \left[ 1 - \frac{\varepsilon}{\pi} \int^{+\infty}_{-\infty} {\rm d}\omega\, {\rm e}^{-{\rm i}\frac{\hbar \omega p_\parallel}{\hbar eE_{\rm s}}}  \frac{ I_1 \left( \frac{\hbar \omega \sqrt{m^2 + |{\bm p}_\perp|^2}}{\hbar eE_{\rm s}}  \right)}{\frac{\hbar \omega \sqrt{m^2 + |{\bm p}_\perp|^2}}{\hbar eE_{\rm s}}} \frac{e{\mathcal E}_{\rm w}(\omega)}{eE_{\rm s}}    + {\mathcal O}(\varepsilon^2) \right],   \label{eq352}
\end{align}
where $E_{\rm w} \equiv \int^{+\infty}_{-\infty} \frac{{\rm d}\omega}{2\pi} {\rm e}^{+{\rm i}\omega t}{\mathcal E}_{\rm w}(\omega) $ and we used $\int^{+1}_{-1}{\rm d}z \sqrt{1-z^2} {\rm e}^{+\omega z} = \pi I_1(\omega)/\omega$, with $I_1$ being the Bessel function of the first kind.  Equation~(\ref{eq352}) gives a faithful description as long as the semi-classical approximation is valid and Eq.~(\ref{eq348}) is satisfied, i.e., the weak electric field $E_{\rm w}$ is weak enough that the deviation $z_{\rm t} - z_{\rm t}^{(0)}$ is controlled well only by the book-keeping parameter $\varepsilon$.  To get a better understanding of Eq.~(\ref{eq352}), it may be instructive to expand the Bessel function $I_1$ in terms of $\hbar \omega$ as
\begin{align}
	S_{z_{\rm t}} 
		&=  \frac{\pi}{2} \frac{m^2 + |{\bm p}_\perp|^2}{eE_{\rm s}} \nonumber\\
			&\quad \times \left[ 1 - \frac{\varepsilon}{2\pi} \int^{+\infty}_{-\infty} {\rm d}\omega\, {\rm e}^{-{\rm i}\frac{\hbar \omega p_\parallel}{\hbar eE_{\rm s}}}  \frac{e{\mathcal E}_{\rm w}(\omega)}{eE_{\rm s}}  \left( 1 + \frac{\tilde{\gamma}^2}{8} + {\mathcal O}(\tilde{\gamma}^4) \right)  + {\mathcal O}(\varepsilon^2) \right] \nonumber\\
		&= \frac{\pi}{2} \frac{m^2 + |{\bm p}_\perp|^2}{eE_{\rm s}} \nonumber\\
			&\quad \times \left[ 1 - \varepsilon \left( \frac{eE_{\rm w}(-\hbar p_\parallel/\hbar eE_{\rm s})}{eE_{\rm s}} - \frac{1}{8} \frac{m^2+|{\bm p}_\perp|^2}{|\hbar eE_{\rm s}|^2}  \frac{\hbar^2 eE''_{\rm w}(-\hbar p_\parallel/\hbar eE_{\rm s})}{eE_{\rm s}}   + {\mathcal O}(\tilde{\gamma}^4)  \right)  + {\mathcal O}(\varepsilon^2)  \right], \label{eq351}
\end{align}
where we have used $I_1(x)/x = \frac{1}{2}\left( 1+\frac{x^2}{8}+{\mathcal O}(x^4) \right)$ and introduced the so-called combined Keldysh parameter \cite{sch08}
\begin{align}
	\tilde{\gamma} \equiv \frac{\hbar \omega \sqrt{m^2 + {\bm p}_\perp^2} }{\hbar eE_{\rm s}} .  
\end{align}
In the static limit $E_{\rm w} \to {\rm const.}$, Eq.~(\ref{eq351}) reproduces the Schwinger formula (\ref{eq342}) for the total electric field $E = E_{\rm s} + \varepsilon E_{\rm w}$ up to ${\mathcal O}(\varepsilon^2)$.  The particle production by $E_{\rm s}$ occurs at $t=-\hbar p_{\parallel}/\hbar eE_{\rm s}$, and thus the value of $E_{\rm w}$ at that time becomes relevant.  Physically, this value corresponds to the instant of time when the energy threshold $\sim \sqrt{m^2 + ({\bm p}+e{\bm E}_{\rm s}t)^2}$ becomes the minimum.  From an exact-WKB point of view, it is the location of the crossing between the Stokes segment by the strong constant field $E_{\rm s}$ and the real axis ${\rm Re}\,z_{\rm t}^{(0)} = -\hbar p_{\parallel}/\hbar eE_{\rm s}$, at which the Stokes phenomenon occurs.  Equation~(\ref{eq351}) implies that the production number can be enhanced by the time-dependence of the weak field if $eE''_{\rm w}(-\hbar p_\parallel/\hbar eE_{\rm s})<0$, compared to the naive prediction of the Schwinger formula (\ref{eq342}).  Even though $e E_{\rm w}$ is weak, the enhancement can be significant if $eE_{\rm s}$ is sub-critical $m^2/\hbar eE_{\rm s} \gg 1$, which is the essence of the dynamically assisted Schwinger mechanism.  Note that the expanded result (\ref{eq351}) agrees with the perturbation theory in the Furry picture (see Appendix~\ref{appa-4}) if one neglects high frequency corrections ${\mathcal O}(\tilde{\gamma}^4)$.  The disagreement originates from the ${\mathcal O}({\rm e}^{-2S_{z_{\rm t}}/\hbar})$ terms neglected by the semi-classical approximation.

\begin{figure}[!t]
\begin{center}
\hspace*{-16mm}
\includegraphics[clip, width=0.7\textwidth]{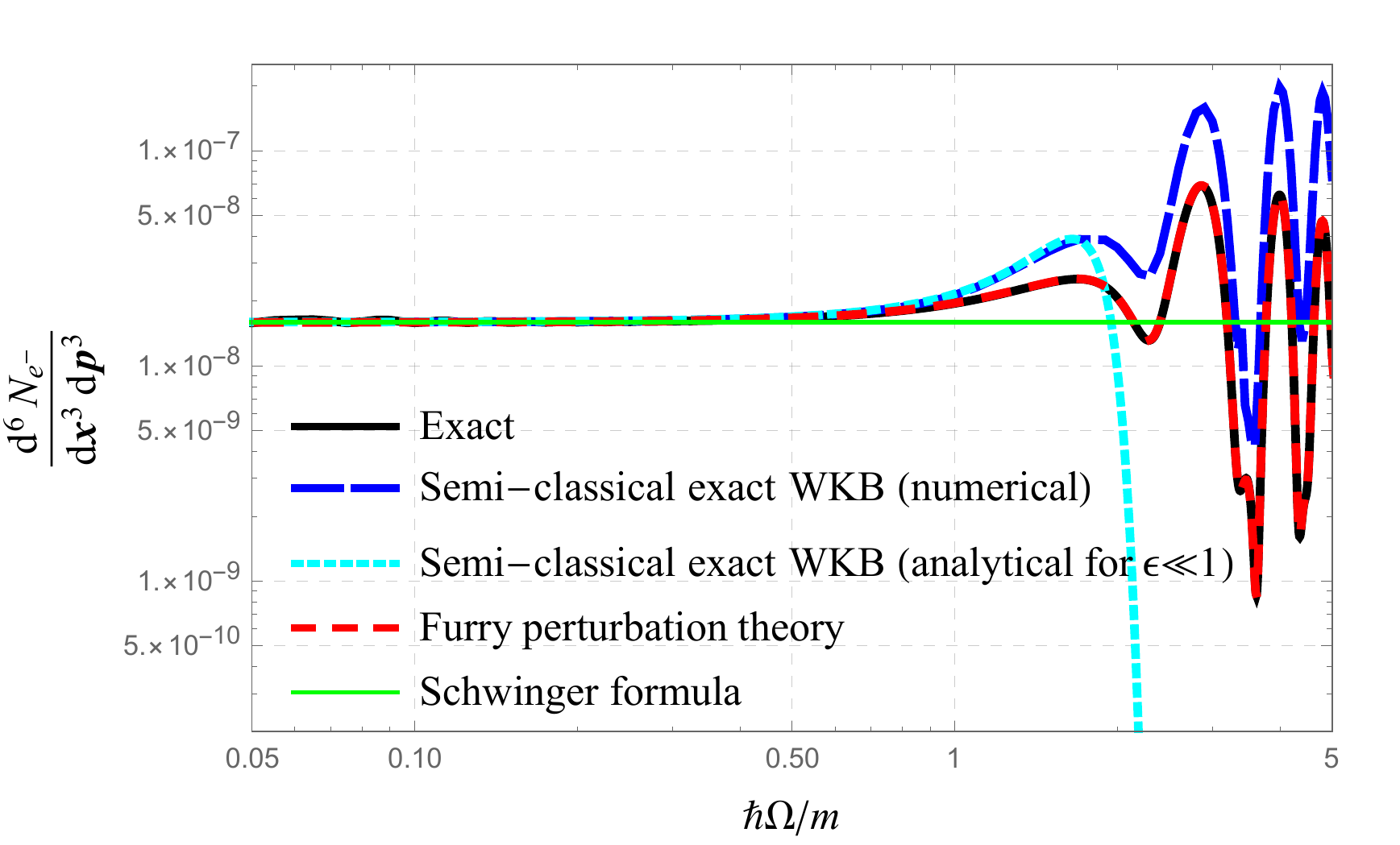} 
\hspace*{-5mm}
\end{center}
\caption{\label{fig-4} (color online) The production number for a constant strong electric field superimposed by a weak monochromatic perturbation (\ref{EQ354}).  The parameters are fixed as $\hbar e E_{{\rm s}}/m^2 = 0.25, E_{{\rm w}}/E_{{\rm s}} = 0.01, |{\bm p}_\perp|/m=0$, and $p_\parallel/m=0.20$.  The black line is an exact result,  obtained by numerically solving the mode equation (\ref{eq3.1}).  The long dashed blue line is obtained by the semi-classical exact WKB formula (\ref{eq3.35}), in which we have numerically determined turning points (see Appendix~\ref{appe} for the resulting Stokes graphs) and computed the corresponding actions $S_{z_{\rm t}}$'s.  The dotted cyan line represents the analytical formula (\ref{eq352}), which is obtained by expanding the exact WKB formula in terms of $|E_{{\rm w}}/E_{{\rm s}}| \ll 1$.  For comparison, results of the perturbation theory in the Furry picture (\ref{eqaaa37}) and the Schwinger formula for a constant electric field (\ref{eq342}) are plotted by the short dashed red line and the green line, respectively.      
}
\end{figure}

As an explicit demonstration, let us consider a situation in which a constant strong electric field is superimposed by a weak monochromatic perturbation, i.e., 
\begin{align}
	e{\bm E} = \left[  eE_{{\rm s}} + eE_{{\rm w}} \cos \Omega t \right] \times {\bf e}_\parallel, \label{EQ354}
\end{align}
where we have assumed $eE_{\rm s} > 0$ and $eE_{\rm s} \gg |eE_{\rm w}|$.    We have computed the production number by numerically solving the mode equation (\ref{eq3.1}), and compared it with the semi-classical WKB formula (\ref{eq3.35}) and the perturbation theory in the Furry picture (see Appendix~\ref{appa-4}).  The semi-classical WKB formula (\ref{eq3.35}) gives a good description for sufficiently small values of $\hbar \Omega $, where the semi-classical approximation is valid.  For large $\hbar \Omega$, it fails to reproduce the exact result quantitatively.  Nevertheless, it qualitatively captures the oscillating behavior of the production number.  The oscillating behavior is an analog of the Franz-Keldysh oscillation in semi-conductor physics, whose origin is the modification to the energy spectrum by the strong electric field $eE_{\rm s}$ \cite{fk}.  Even though the semi-classical approximation cannot fully take into account effects of $eE_{\rm w}$ in the quantum regime where $\hbar \Omega $ is large, it captures those of $eE_{\rm s}$ correctly and thus it shows the qualitative agreement.

\subsection{Relation to other semi-classical approaches}\label{sec35}

We discuss the relationship between our semi-classical exact WKB result (\ref{eq3.35}) and other semi-classical methods to compute the vacuum pair production; in particular, Brezin-Izykson's steepest descent evaluation of the Bogoliubov coefficients \cite{bre70} and the worldline instanton method \cite{dun05a, dun06a, dun06b}.  Although they look different at first sight, we show that they are equivalent in the semi-classical regime, if one neglects unimportant prefactors.

\subsubsection{Brezin-Izykson's steepest descent evaluation}

Brezin and Izykson \cite{bre70} derived an integral representation of the Bogoliubov coefficient $U_{12}$ and evaluated it within the steepest descent method, explicitly for a linearly polarized cosine electric field ${\bm E} \propto \cos \omega t \times {\bf e}_{\parallel} $.  Note that the Dykhne-Davis-Pechukas formula \cite{dyk62, dav76} (see also Ref.~\cite{Fukushima:2019iiq}), which is widely used in analyses of the Landau-Zener transition in the condensed matter community, is obtained in the same manner.  For the sake of simplicity, we here focus on $U_{12}$ only, but one can equally perform the same calculation for $U_{21}$.  We shall see that the steepest descent method and the exact WKB analysis have a close relationship to each other, as already indicated in mathematics \cite{aoki04}, and their leading order results in the semi-classical limit coincide.

First, we derive an integral representation of $U_{12}$.  From Eq.~(\ref{eq3.13}), we have
\begin{align}
	U_{12} = {\rm i}\hbar\, \varphi_{+,{\rm out}} \overset{\leftrightarrow}{\partial}_{t} \varphi_{+, {\rm in}}.  \label{eq3.36}
\end{align}
Since $U_{12}$ is a constant independent of $t$, one can evaluate Eq.~(\ref{eq3.36}) at any convenient time.  Taking $t \to +\infty$, we obtain
\begin{align}
	U_{12} 
		= {\rm i}\hbar \lim_{t \to +\infty} \left( \frac{{\rm e}^{-\frac{\rm i}{\hbar}\int^t_{t_0}\sqrt{Q(t')}{\rm d}t'}}{\sqrt{2Q^{1/2}(t)}} \right) \overset{\leftrightarrow}{\partial}_{t} \varphi_{+,{\rm in}}
		\equiv \lim_{t \to +\infty} u_{12}(t).  
\end{align}
Note that $u_{12}(-\infty)=0$.  To proceed, let us consider to act $\partial_t$ onto $u_{12}$.  We thus get
\begin{align}
	u_{12}' 
		&= -{\rm i}\hbar \left( \frac{5}{16} \frac{{Q'}^2}{Q^2} - \frac{1}{4}\frac{Q''}{Q} \right) \left( \frac{{\rm e}^{-\frac{\rm i}{\hbar}\int^t_{t_0}\sqrt{Q(t')}{\rm d}t'}}{\sqrt{2Q^{1/2}(t)}} \right) \varphi_{+,{\rm in}},
\end{align}
where we have used $[\hbar^2 \partial_t^2 +Q] \varphi_{+,{\rm in}} = 0$. Integrating $u'_{12}$ over $t$, one finds
\begin{align}
	U_{12} 
		&= \lim_{t \to +\infty} \int^{t}_{-\infty} {\rm d}t' u'_{12}(t') \nonumber\\
		&= -{\rm i}\hbar \int^{\infty}_{-\infty} {\rm d}t \left( \frac{5}{16} \frac{{Q'}^2}{Q^2} - \frac{1}{4}\frac{Q''}{Q} \right) \left( \frac{{\rm e}^{-\frac{\rm i}{\hbar}\int^t_{t_0}\sqrt{Q(t')}{\rm d}t'}}{\sqrt{2Q^{1/2}(t)}} \right) \varphi_{+,{\rm in}} . \label{eq3.39}
\end{align}
The production number may be suppressed strongly in the semi-classical limit, so one may simply approximate $\varphi_{+, {\rm in}}$ by the lowest order WKB solution as 
\begin{align}
	\varphi_{+,{\rm in}} \sim \frac{{\rm e}^{-\frac{\rm i}{\hbar}\int^t_{t_0}\sqrt{Q(t')}{\rm d}t'}}{\sqrt{2Q^{1/2}(t)}} . \label{eq3.40}
\end{align}
Note that this approximation is not so accurate and gives an incorrect prefactor \cite{ces11}; Nevertheless, it is sufficient to extract the exponentially small main factor in $U_{12}$ and is the same approximation level considered by Brezin and Izykson \cite{bre70}.  Substituting the approximation (\ref{eq3.40}) into Eq.~(\ref{eq3.39}), one gets an integral representation of $U_{12}$ as\footnote{Equation~(\ref{eq3.41}) looks slightly different from what Brezin and Izykson derived [Eq.~(35) in Ref.~\cite{bre70}] because of the treatment of the first order time derivative of the mode functions (or choice of the ``adiabatic basis" \cite{dab14, dab16}). Within the approximation (\ref{eq3.39}), they give different prefactors, which are, nevertheless, unimportant within the accuracy of the approximation.  }
\begin{align}
	U_{12} 
		\sim -{\rm i}\hbar \int^{\infty}_{-\infty} {\rm d}t \left( \frac{5}{32} \frac{{Q'}^2}{Q^3} - \frac{1}{8}\frac{Q''}{Q^2} \right) {\rm e}^{-2\frac{\rm i}{\hbar}\int^t_{t_0}\sqrt{Q(t')}{\rm d}t'} .  \label{eq3.41}
\end{align}

One may evaluate the integral representation (\ref{eq3.41}) with the steepest descent method, which is a good approximation in the semi-classical regime, in which $\hbar$ is formally regarded as a small quantity.  After complexifing the integration variable $t \in [-\infty, +\infty] \to z \in {\mathbb C}$, we define saddle points $z_{{\rm s},i}$ as
\begin{align}
	0= \left. \partial_z \int^z_{t_0} {\rm d}z' \sqrt{Q(z')} \right|_{z=z_{{\rm s},i}} = \sqrt{Q(z_{{\rm s},i})}.  
\end{align}
Apparently, $z_{{\rm s},i}$ is nothing but the turning point $z_{{\rm t},i}$ in the language of the exact WKB analysis.  We notice that not all the saddle points but only a part of them are relevant in evaluating the integral (this point shall be clarified in more detail later).  By picking up the contributions from the relevant saddle points and noticing Eq.~(\ref{eq3.21}), the integral (\ref{eq3.41}) can be evaluated as
\begin{align}
	U_{12}
		&\sim \sum_{z_{{\rm s},i} \in {\rm relevant\; saddles}} {\rm e}^{-2\frac{{\rm i}}{\hbar} \int^{z_{{\rm s},i}}_{t_0} {\rm d}z' \sqrt{Q(z')}} \nonumber\\
		&\quad\times \frac{-5}{24} \int_{\Gamma}{\rm d}z \frac{ 1 }{ z-z_{{\rm s},i}} {\rm e}^{-\frac{{\rm i}}{\hbar} \frac{2}{3} \sqrt{ 2 \left( {\bm p} + e {\bm A}(z_{{\rm s},i})\right) \cdot e{\bm E}(z_{{\rm s},i}) }  (z-z_{{\rm s},i})^{3/2} } \nonumber\\
		&= \frac{-5 \pi {\rm i}}{18} \sum_{z_{{\rm s},i} \in {\rm relevant\; saddles}} s_i\, {\rm e}^{-2\frac{{\rm i}}{\hbar} \int^{z_{{\rm s},i}}_{t_0} {\rm d}z' \sqrt{Q(z')}}  ,  \label{eq3.43}
\end{align}
\sloppy where $\Gamma$ is an integration contour running from $z=-\infty$ to $+\infty$ during which it wraps each relevant saddle point $z_{{\rm s},i}$ over the angle $4\pi/3$.  The integrand in the first equality is singular at the saddle points, and we have picked up the corresponding residues to get the second equality.  $s_i \equiv +1$ ($-1$) if the contour $\Gamma_{z_{{\rm s},i}}$ wraps the saddle point $z_{{\rm s},i}$ counter-clockwise (clockwise).  One may expect that the contour wraps a saddle point clockwise (counter-clockwise) if that point is located in the lower (upper) half plane, i.e., 
\begin{align}
	s_i = {\rm Im}\,z_{{\rm s},i}.  
\end{align}
Therefore,
\begin{align}
	U_{12}
		&\sim \frac{-5 \pi {\rm i}}{18} \sum_{z_{{\rm s},i} \in {\rm relevant\; saddles}}  {\rm Im}\,z_{{\rm s},i} \; {\rm e}^{-2\frac{{\rm i}}{\hbar} \int^{z_{{\rm s},i}}_{t_0} {\rm d}z' \sqrt{Q(z')}}.  
\end{align}
Essentially, this is Eq.~(42) in Brezin-Izykson's paper \cite{bre70}.

Next, we need to identify which saddle points are relevant.  For this, we need to understand the topology concerning how steepest descent/ascent lines are located in the complex $z$-plane.  This can be achieved by making use of the properties (1)-(5) for the Stokes graph in the exact WKB analysis.  Indeed, in the steepest descent method (or the Lefschetz thimble method in more general), steepest descent/ascent lines ${\mathcal C}_{z_{{\rm s},i}}$ emanating from a saddle point $z_{{\rm s},i}$ are given by a set of points $z \in {\mathbb C}$ satisfying $0 = {\rm Im}\left[ -2\frac{{\rm i}}{\hbar} \int^{z \in {\mathcal C}_{z_{{\rm s},i}}}_{z_{{\rm s},i}} {\rm d}z' \sqrt{Q(z')} \right]$, which are nothing but the Stokes lines (\ref{eq2.13}) in the language of the exact WKB analysis.  Therefore, we can directly use the properties (1)-(5) to determine the structure of steepest descent/ascent lines.  In the language of the steepest descent method, one may rephrase the properties (1)-(5) as: 
\begin{enumerate}
\item[(1')] The locations of saddle points and steepest lines are symmetric in the upper and lower half planes.  

\item[(2')] All the saddle points $z_{{\rm s},i}$'s are of order one if $0 \neq \left( {\bm p} + e {\bm A}(z_{{\rm s},i}) \right) \cdot e{\bm E}(z_{{\rm s},i})$.     

\item[(3')] A saddle point $z_{{\rm s}}$ and its conjugate $z_{{\rm s}}^*$ are connected by a line in which steepest descent and ascent lines are degenerated.  

\item[(4')] The degenerated steepest descent/ascent line connecting  a pair of $z_{{\rm s}}$ and $z_{{\rm s}}^*$ crosses the real axis only once.  

\item[(5')] The other steepest lines emanating from a saddle point, other than the degenerated line, cannot cross the real axis.  
\end{enumerate}
According to the steepest descent method, only saddles whose steepest ascent line crosses the real axis (or the original integration contour of the integral that one wants to evaluate) contribute to the integral.  Therefore, either of $z_{{\rm s},i}$ and $z_{{\rm s},i}^*$ is relevant because of the property (3').  Without loss of generality, we can assume ${\rm Im} \, z_{{\rm s},i}>0$.  Then, by repeating the same argument below Eq.~(\ref{eq2.18}), one can show that the ``action," $-2 \frac{{\rm i}}{\hbar} \int^{z \in {\mathcal C}_{z_{{\rm s},i}}^*}_{z_{{\rm s},i}^*}  {\rm d}z' \sqrt{Q(z')} \in {\mathbb R}$, increases as it goes from $z=z_{{\rm s},i}^*$ to $z_{{\rm s},i}$ along the steepest line ${\mathcal C}_{z_{{\rm s},i}^*}$.  Therefore, the steepest line emanating from $z=z_{{\rm s},i}^*$ is the steepest ascent, and thus $z=z_{{\rm s},i}^*$ is relevant.  Thus, we have 
\begin{align}
	U_{12}
		&\sim \frac{+5 \pi}{18} \times  {\rm i} \sum_{i=1}^n  \exp \left[ -{\rm i}\,{\rm Im} \left[ +2\frac{{\rm i}}{\hbar} \int^{z_{{\rm s},i}}_{t_0} {\rm d}z' \sqrt{Q(z')} \right] \right] \exp\left[-\frac{\rm i}{\hbar}\int^{z_{{\rm s},i}^*}_{z_{{\rm s},i}} {\rm d}z' \sqrt{Q(z')} \right] , \label{eqe105}
\end{align}
where $n$ is the number of saddle points in the upper half plane, and we have used ${\rm Re}[-2\frac{\rm i}{\hbar}\int^{z^*_{{\rm s},i}}_{t_0} {\rm d}z' \sqrt{Q(z')}] = -\frac{\rm i}{\hbar}\int^{z_{{\rm s},i}^*}_{z_{{\rm s},i}} {\rm d}z' \sqrt{Q(z')}$, which is negative definite because of Eq.~(\ref{eq2.19}), and ${\rm Im}[-2\frac{{\rm i}}{\hbar} \int^{z_{{\rm s},i}}_{t_0} {\rm d}z' \sqrt{Q(z')}] = -{\rm Im}[+2\frac{{\rm i}}{\hbar} \int^{z_{{\rm s},i}^*}_{t_0} {\rm d}z' \sqrt{Q(z')}]$.  Equation~(\ref{eqe105}) reproduces the semi-classical exact WKB result (\ref{eq3.34}), putting aside the unimportant prefactor $5\pi/18$, which is inaccurate within the approximation (\ref{eq3.40}).  This coincidence is not an accident, as the Stokes graph in the exact WKB analysis has precisely the same structure as that of steepest ascent/descent lines in the steepest descent method.  In obtaining the exact WKB result (\ref{eq3.34}), we neglected higher order terms of the order of ${\mathcal O}({\rm e}^{-2S_{z_{{\rm t},i}}/\hbar})$ by virtue of the semi-classical approximation.  Similar approximations were implicitly used in the steepest descent method in Eqs.~(\ref{eq3.40}) and (\ref{eq3.43}).

\subsubsection{Worldline instanton method}

The worldline instation method \cite{dun05a, dun06a, dun06b} computes the one-loop QED effective action $\Gamma_{\rm 1\mathchar`-loop}$, whose imaginary part gives the production number, based on the Feynman's worldline path integral representation \cite{fey50, fey51}.  The steepest descent method is applied to evaluate the worldline path integral, and thus the worldline instation method is valid in the semi-classical regime.  We shall see that the semi-classical exact WKB result (\ref{eq3.35}) is equivalent to the worldline instation method by showing that worldline instanton actions are nothing but cycle integrals enclosing turning points of the potential $Q$.

The starting point of the worldline instaton method is Schwinger's proper-time representation of the one-loop effective action $\Gamma_{\rm 1\mathchar`-loop}$ \cite{sch51}: 
\begin{align}
	\Gamma_{\rm 1\mathchar`-loop} 
		&\equiv \hbar\, {\rm tr}\,{\rm ln} \left[  m^2 + \hbar^2 \hat{D}_\mu \hat{D}^\mu \right] \nonumber\\
		&= - \hbar \int_0^\infty \frac{{\rm d}\tau}{\tau} \int {\rm d}^4 x \braket{x| {\rm e}^{-\frac{\rm i}{\hbar} \hat{\mathcal H} \tau} |x}, \label{eqa-1}
\end{align}
where $\tau$ is the so-called proper-time parameter and $\hat{D}_\mu$ is a covariant derivative operator under an external gauge field $A_\mu$ such that $\braket{x|\hat{D}^\mu|y} = \delta^4(x-y) \left[ \partial^\mu_x + \frac{\rm i}{\hbar}eA^\mu(x) \right]$.  We have also assumed an ${\rm i}\epsilon$-prescription $m^2 \to m^2 - {\rm i}0^+$ and used an identity $\hbar\, \hat{\mathcal O}^{-1} = {\rm i}\int^\infty_0 {\rm d}\tau\,{\rm e}^{-\frac{{\rm i}}{\hbar}\hat{\mathcal O}\tau}$ for ${\rm Im}\,{\mathcal O} < 0$ to get the second equality.  One may interpret the last exponential factor as a ``time-translation operator" with respect to the proper-time $\tau$ with a ``Hamiltonian" $\hat{\mathcal H}$, 
\begin{align}
	\hat{\mathcal H} \equiv m^2 + \hbar^2 \hat{D}_\mu \hat{D}^\mu.  
\end{align}
The time-translation operator can be expressed as a path integral.  For a spatially homogeneous gauge potential $A^{\mu}(t,{\bm x}) = (0, {\bm A}(t))$ [see Eq.~(\ref{eq3.1})]\footnote{We here assume a spatially homogeneous gauge potential in order to discuss the momentum distribution and to directly see the relationship with our exact WKB analysis at the semi-classical level.  In general, the worldline instanton method can be applied to inhomogeneous gauge potentials depending on several coordinate variables \cite{dun06b}.  }, the Hamiltonian $\hat{\mathcal H}$ does not depend on ${\bm x}$ and one can explicitly carry out the spatial part of the path integration \cite{dum-11}.  The resulting expression is the so-called worldline representation of the one-loop effective action $\Gamma_{\rm 1\mathchar`-loop}$: 
\begin{align}
	\Gamma_{\rm 1\mathchar`-loop} 
		= - \hbar \int {\rm d}^4 x \int \frac{ {\rm d}^3{\bm p}}{(2\pi \hbar)^3} \int_0^\infty \frac{{\rm d}\tau}{\tau}  \int_{z(0)=x^0}^{z(\tau)=x^0} {\mathcal D}z(u) {\rm e}^{+\frac{{\rm i}}{\hbar}S[z(u)]},  \label{eqa-3}
\end{align}
where ${\bm p}$ is canonical momentum, which is an eigenvector of the spatial derivative $-{\rm i}\hbar{\bm \partial}$, and we have introduced an ``action" $S$ as
\begin{align}
	S[z] 
		\equiv - \int^\tau_{0}{\rm d}u \left[ \left( \frac{{\rm d}z}{{\rm d}u} \right)^2 + Q[z(u)]  \right] , \label{eq-4}
\end{align}
with $Q[z] = m^2 + \left( {\bm p} - e{\bm A}[z] \right)^2$ as before.  Note that all the integration variables $z, u$, and $\tau$ are real at this stage.

The path- and proper-time $\tau$-integrations in Eq.~(\ref{eqa-3}) are not analytically doable, except for a few special cases.  In the worldline instanton method, those integrations are evaluated by the steepest descent method, which is justified in the semi-classical regime.  To be concrete, we first expand the path-integration variable $z$ and the action $S$ as\footnote{We here carry out the path integral first and then the $\tau$-integration.  Alternatively, one may carry out the $\tau$-integration first and then the path-integral.  For this case, the resulting classical equation of motion as well as the classical action $S_{\rm cl}$ might look a bit different, but they are essentially the same and one can discuss the coincidence of our semi-classical exact WKB formula in a similar manner that we discuss below.  }
\begin{align}
	&z \equiv z_{\rm cl} + \delta z, \nonumber\\
	&S[z] = \underbrace{S[z_{\rm cl}]}_{\equiv S_{\rm cl}} + \underbrace{S[z] - S[z_{\rm cl}]}_{\equiv \delta S},
\end{align}
where $z_{\rm cl}$ (or ``instanton" on the worldline) is a solution of the classical equation of motion for $S$, i.e., 
\begin{align}
	0 = \frac{{\rm d}^2z_{\rm cl}}{{\rm d}u^2} - \frac{1}{2} \frac{{\rm d}Q}{{\rm d}z_{\rm cl}}
	\ \ {\rm with}\ \ x^0 = z_{\rm cl}(0)= z_{\rm cl}(\tau).   \label{eqa-6}
\end{align}
Here, we complexified the integration variable $z \in {\mathbb R} \to {\mathbb C}$ by virtue of the steepest descent method, and thus the classical solution $z_{\rm cl}$ is in general complex-valued.  The one-loop effective action (\ref{eqa-1}) now reads
\begin{align}
	\Gamma_{\rm 1\mathchar`-loop} 
		&= \hbar \int {\rm d}^4 x \int \frac{ {\rm d}^3{\bm p}}{(2\pi \hbar)^3} \int_0^\infty \frac{{\rm d}\tau}{\tau} \, C  {\rm e}^{+\frac{\rm i}{\hbar}S_{\rm cl}(\tau)} , 
\end{align}
where the prefactor $C$ comes from the path-integration of the fluctuation $\delta S$.  The remaining $\tau$-integration may be evaluated by the steepest descent method, which is again valid in the semi-classical regime.  Complexifying the variables $\tau, u \in {\mathbb R} \to {\mathbb C}$ and assuming that the integral is dominated around $\tau \sim \tau_{\rm st}$ at which the classical action $S_{\rm cl}$ becomes stationary, 
\begin{align}
	0 	= \left. \frac{{\rm d} S_{\rm cl}}{{\rm d}\tau} \right|_{\tau = \tau_{\rm st}},   \label{eqa-8}
\end{align}
we find
\begin{align}
	\Gamma_{\rm 1\mathchar`-loop} 
		&= \hbar \int {\rm d}^3 {\bm x} \int \frac{ {\rm d}^3{\bm p}}{(2\pi \hbar)^3} \, C' {\rm e}^{+\frac{\rm i}{\hbar}S_{\rm cl}(\tau_{\rm st})} \nonumber\\
		&\sim \hbar \int {\rm d}^3 {\bm x} \int \frac{ {\rm d}^3{\bm p}}{(2\pi \hbar)^3} \, {\rm e}^{+\frac{\rm i}{\hbar}S_{\rm cl}(\tau_{\rm st})} , \label{eqa--9}
\end{align}
where the prefactor $C$ is replaced by $C'$ after absorbing contributions from the fluctuations around the stationary point $\int_{\Gamma} \frac{{\rm d}\tau}{\tau} {\rm e}^{+\frac{\rm i}{\hbar} \left( S_{\rm cl}(\tau) - S_{\rm cl}(\tau_{\rm st}) \right)}$, with $\Gamma$ being the steepest descent path, and from the time integral $\int {\rm d}x^0 = \int {\rm d}u \frac{{\rm d}z_{\rm cl}(0)}{{\rm d}u} = {\rm finite}$ \cite{dun06a}.  In the second equality of Eq.~(\ref{eqa--9}), we have assumed $C' = {\mathcal O}(1)$ for simplicity.  In principle, the prefactor $C'$ is calculable numerically/analytically \cite{gie05, dun06a}, but it is quite complicated and not essential in our discussion below.  In fact, the dominant factor of the production number comes from the exponential ${\rm e}^{+\frac{\rm i}{\hbar}S_{\rm cl}}$, compared to which the prefactor $C'$ is not so important.  Note that the stationary condition (\ref{eqa-8}) is just a necessary condition for a {\it relevant} stationary point, which shall be discussed later.  So far, we have implicitly assumed that there is only one classical solution $z_{\rm cl}$ with a single relevant stationary point $\tau_{\rm st}$.  Generally, one can have several classical solutions $z_{{\rm cl},i}$ and/or stationary points $\tau_{{\rm st},ij}$.  In such a case, the effective action $\Gamma_{\rm 1\mathchar`-loop}$ shall be given by a sum of all the relevant classical solutions $z_{{\rm cl},i}$ and/or stationary points $\tau_{{\rm st},ij}$ as
\begin{align}
	\Gamma_{\rm 1\mathchar`-loop} 
		\sim \hbar \sum_{i,j} \int {\rm d}^3 {\bm x} \int \frac{ {\rm d}^3{\bm p}}{(2\pi \hbar)^3} \, {\rm e}^{+\frac{\rm i}{\hbar}S_{{\rm cl},i}(\tau_{{\rm st},ij})} .   \label{eqa--10}
\end{align}

Having obtained the effective action (\ref{eqa--10}), we turn to compute the production number.  The effective action (\ref{eqa--10}) is related to the vacuum persistence probability $P$ as
\begin{align}
	P	&\equiv | \braket{{\rm vac;in}|{\rm vac;out} } |^2  \nonumber\\
		&= {\rm e}^{-2\,{\rm Im}\,\Gamma} \nonumber\\
		&\sim 1 - 2\,{\rm Im}\,\Gamma .  
\end{align} 
Since the vacuum decay occurs due to the particle production, it is natural to assume\footnote{Precisely speaking, $1-P$ and the production number $N_{e^-}+N_{e^+}$ are different quantities and coincide only when the production number is sufficiently small \cite{coh08}.  The worldline instanton method cannot compute the production number directly.  }
\begin{align}
	P  &\sim 1 - (N_{e^-}+N_{e^+}) \nonumber\\
		&= 1 -2N_{e^-},  
\end{align}
where $N_{e^-}=N_{e^+}$ is assumed because of the gauge invariance (no spontaneous charge production).  Then, using Eq.~(\ref{eqa--10}), we arrive at
\begin{align}
	N_{e^\pm}	
		\sim {\rm Im}\,\Gamma 
	\ \Rightarrow\ 
	\frac{{\rm d}^3 N_{e^\pm}}{{\rm d}{\bm p}^3 {\rm d}{\bm x}^3}
		\sim \frac{ 1}{(2\pi \hbar)^3}\sum_{i,j} {\rm e}^{+\frac{\rm i}{\hbar}S_{{\rm cl},i}(\tau_{{\rm st},ij})}.  \label{eqaa13}
\end{align}

\begin{figure}[!t]
\begin{center}
\hspace*{3mm}
\includegraphics[clip, width=0.5\textwidth]{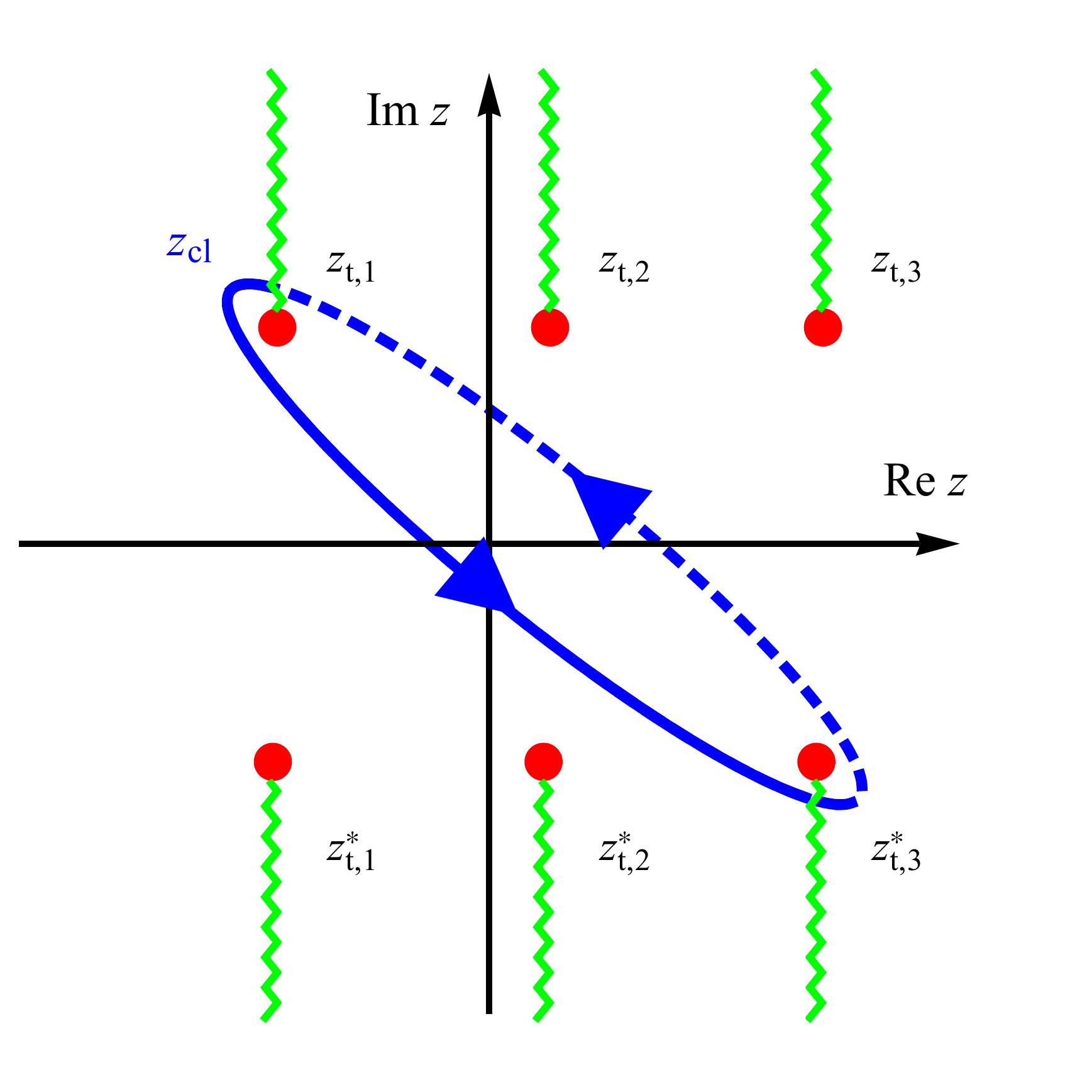} 
\vspace*{-5mm}
\end{center}
\caption{\label{fig-5} (color online) A non-trivial classical path $z_{\rm cl}$ (blue line) enclosing a single pair of turning points $z_{{\rm t},i}$ and $z_{{\rm t},j}^*$ (red dots).  The wavy green lines represent branch cuts.  The dashing of the blue line distinguishes Riemann sheets; thick (dashed) for the first (second) Riemann sheet such that ${\rm d}z_{\rm cl}/{\rm d}u = + \sqrt{Q[z_{\rm cl}]}$ $(-\sqrt{Q[z_{\rm cl}]})$.  }
\end{figure}

We show that the worldline instanton method (\ref{eqaa13}) agrees with our semi-classical exact WKB formula (\ref{eq3.35}) (see also Ref.~\cite{dum-11}).  To this end, we first integrate the classical equation of motion (\ref{eqa-6}) over $u$ to find 
\begin{align}
	\left( \frac{{\rm d}z_{\rm cl}}{{\rm d}u} \right)^2 - Q[z_{\rm cl}(u)] = a\ 
	\Rightarrow\ 
	\frac{{\rm d}z_{\rm cl}}{{\rm d}u} = s \sqrt{Q[z_{\rm cl}]+a},   \label{eqa-11}
\end{align}
where $s \equiv \pm 1$ is fixed after choosing a Riemann sheet associated with the square root in Eq.~(\ref{eqa-11}).  The integration constant $a$ is independent of $u$ but can be dependent on $\tau$.  The arbitrariness of $a$ is essentially related to the number of classical solutions.  For our steepest descent evaluation (\ref{eqa--9}), it is sufficient to know the value of $a$ at $\tau = \tau_{\rm st}$, which is uniquely fixed by the stationary condition (\ref{eqa-8}) and thus we just have to consider a single classical solution characterized by the unique value $a(\tau_{\rm st})$.  Noticing
\begin{align}
	\tau
	&= \int^{\tau}_0 {\rm d}u \nonumber\\
	&= \oint {\rm d}z_{\rm cl} \left( \frac{{\rm d}z_{\rm cl}}{{\rm d}u} \right)^{-1} \nonumber\\
	&= \oint \frac{ {\rm d}z_{\rm cl} }{s \sqrt{Q[z_{\rm cl}]+a(\tau)}} , \label{eqacy}
\end{align}
the stationary condition (\ref{eqa-8}) yields\footnote{$a=0$ corresponds to ``Bogomolnyi-Prasad-Sommerfield (BPS) instantons \cite{bog76, pra75}" in the language of instanton calculus, and one may say that {\it BPS worldline instantons} dominate the particle production.  }
\begin{align}
	0 	&= \left. \frac{{\rm d} S_{\rm cl}}{{\rm d}\tau} \right|_{\tau = \tau_{\rm st}} \nonumber\\
		&= \left. \frac{{\rm d}}{{\rm d}\tau} \left[ a(\tau)\tau - 2 \oint {\rm d}z_{\rm cl}\,s \sqrt{Q[z_{\rm cl}]+a(\tau)} \right] \right|_{\tau = \tau_{\rm st}} \nonumber\\
		&= a(\tau_{\rm st}).   \label{eqa-13}
\end{align}
Using Eq.~(\ref{eqa-13}), one can evaluate the classical action $\frac{{\rm i}}{\hbar}S_{\rm cl}$ at a relevant stationary point $\tau = \tau_{\rm st}$ as
\begin{align}
	+\frac{{\rm i}}{\hbar}S_{\rm cl}(\tau_{\rm st}) 
		&= -2\frac{{\rm i}}{\hbar} \int_0^{\tau_{\rm st}} {\rm d}u \frac{ {\rm d}z_{\rm cl} }{{\rm d}u} s \sqrt{Q[z_{\rm cl}(u)]}  \nonumber\\
		&= -2\frac{{\rm i}}{\hbar} \oint {\rm d}z_{\rm cl}\,s \sqrt{Q[z_{\rm cl}]}.   \label{eqa-14}
\end{align}
Equation~(\ref{eqa-14}) implies that the classical action $S_{\rm cl}$ can have non-trivial values only when a classical path $z_{\rm cl}$ encloses some singularities of the integrand $\sqrt{Q}$ as shown in Fig.~\ref{fig-5}.  One can always construct such a classical path by properly choosing a path for the complexified time variable $u$.  Topologically distinct paths $u$ give distinct stationary points $\tau_{\rm st}$, whose contributions should be summed up as in Eq.~(\ref{eqa--10}).  In principle, one may consider multi-loops and/or loops containing more than two pairs of turning points, whose contributions are exponentially suppressed in the semi-classical regime by the worldline instanton action ${\mathcal O}({\rm e}^{+\frac{2{\rm i}}{\hbar}S_{\rm cl}})$ [i.e., ${\mathcal O}({\rm e}^{-\frac{2}{\hbar} S_{z_{\rm t}}})$-contributions in the notation of the exact WKB analysis] and are negligible compared to loops containing a single pair of turning points shown in Fig.~\ref{fig-5}.  For such a classical path enclosing only a single pair of turning points $z_{{\rm t},i}$ (such that ${\rm Im}\,z_{{\rm t},i}>0$) and $z_{{\rm t},j}^*$, we have
\begin{align}
	+\frac{{\rm i}}{\hbar} S_{{\rm cl},ij}
		&= -\frac{{\rm i}}{\hbar} \int^{z_{{\rm t},i}^*}_{z_{{\rm t},j}} {\rm d}z (+1) \sqrt{Q[z]} -\frac{{\rm i}}{\hbar} \int^{z_{{\rm t},i}}_{z_{{\rm t},j}^*} {\rm d}z (-1) \sqrt{Q[z]}  \nonumber\\
		&= -2\frac{{\rm i}}{\hbar} \int^{z_{{\rm t},i}^*}_{z_{{\rm t},j}} {\rm d}z \sqrt{Q[z]} ,   \label{eqqaa18}
\end{align}
in which, without loss of generality, the path is assumed to be on the first (second) Riemann sheet ${\rm d}z_{\rm cl}/{\rm d}u = + \sqrt{Q[z_{\rm cl}]}$ $(-\sqrt{Q[z_{\rm cl}]})$ when going from $z_{{\rm t},i}$ to $z_{{\rm t},j}^*$ (coming back from $z_{{\rm t},j}^*$ to $z_{{\rm t},i}^*$).  Before proceeding, we remark that the classical path must be in the counter-clockwise direction, so that ${\rm Im}\,\tau_{\rm st} < 0$.  To show this, we notice that if $\tau_{\rm st} $ is a stationary point, its conjugate $\tau_{\rm st}^*$ must also be a stationary point because $0 = \left[  \left.  {\rm d} S_{\rm cl}/{\rm d}\tau \right|_{\tau = \tau_{\rm st}} \right]^* = \left. {\rm d} S_{\rm cl}/{\rm d}\tau \right|_{\tau = \tau_{\rm st}^*}$, where we have used $[z_{\rm cl}(u)]^*=z_{\rm cl}(u^*)$ and $[Q[z]]^* = Q[z^*]$.  Repeating a similar discussion that we presented in Sec.~\ref{sec35}, one can show that, in the complex $\tau$-plane, the pair of the stationary points $\tau_{\rm st}$  and $\tau_{\rm st}^*$ is connected by steepest descent and ascent lines and that the lines cross the real axis.  One can also identify that  the steepest line emanating from the point with negative imaginary part is ascent.  Thus, a stationary point which has negative imaginary part is relevant, and its conjugate pair, having positive imaginary part, is always irrelevant in the steepest descent evaluation of the $\tau$-integration.  The value of $\tau_{\rm st}$ is determined by the cycle integral (\ref{eqacy}).  For a classical path in the counter-clockwise direction as shown in Fig.~\ref{fig-5}, it can be evaluated as
\begin{align}
	\tau_{\rm st}
	&= \int_{z_{{\rm t},j}}^{z_{{\rm t},i}^*}\frac{ {\rm d}z_{\rm cl} }{+\sqrt{Q[z_{\rm cl}]}} + \int^{z_{{\rm t},j}}_{z_{{\rm t},i}^*}\frac{ {\rm d}z_{\rm cl} }{-\sqrt{Q[z_{\rm cl}]}} \nonumber\\
	&= \underbrace{ 2 \int_{x_j}^{x_i}\frac{ {\rm d}z }{+\sqrt{Q[z]}} }_{\in {\mathbb R}}
		+ \underbrace{ 2 \int_{z_{{\rm t},j}}^{x_j}\frac{ {\rm d}z }{+\sqrt{Q[z]}}
		+ 2 \int_{x_i}^{z_{{\rm t},i}^*}\frac{ {\rm d}z }{+\sqrt{Q[z]}} }_{\in -{\rm i}\times {\mathbb R}^+}, \label{eqqa19}
\end{align}
where we have deformed the integration contour in the second line and $x_{\alpha} \in {\mathbb R}$ $(\alpha=i,j)$ is a crossing between the real axis and a Stokes line ${\mathcal C}$ connecting the pair of turning points $z_{{\rm t},\alpha}$ and $z_{{\rm t},\alpha}^*$ on which ${\rm Im}\left[ +{\rm i}\int^{z \in {\mathcal C}}_{z_{{\rm t},\alpha}}  {\rm d}z /\sqrt{Q[z]} \right] = 0$.  In a similar manner as in Eq.~(\ref{eq2.18}), one can show $ +{\rm i}\int^{z \in {\mathcal C}}_{z_{{\rm t},\alpha}} {\rm d}z /\sqrt{Q[z]} > 0$, and thus Eq.~(\ref{eqqa19}) surely has negative imaginary part, as we wanted.  Note that one can show in a similar manner that ${\rm Re}\left[ +\frac{{\rm i}}{\hbar} S_{{\rm cl},ij} \right] < 0$ for the counter-clockwise classical path, and thus the production number (\ref{eqaa13}) never becomes exponentially large but is always suppressed exponentially by the worldline instanton action.  We also remark that classical paths enclosing $(z_{{\rm t},i},z_{{\rm t},j})$ or $(z_{{\rm t},i}^*,z_{{\rm t},j}^*)$ do not contribute because they never cross the real axis, and hence the boundary condition $z_{\rm cl}(0) = z_{\rm cl}(\tau) = x^0 \in {\mathbb R}$ cannot be satisfied.  Having explained that only counter-clockwise classical paths enclosing a pair of $z_{{\rm t},i}$ and $z_{{\rm t},j}^*$ are relevant at the leading order in the worldline instatnton action, we substitute Eq.~(\ref{eqqaa18}) into Eq.~(\ref{eqaa13}) to arrive at
\begin{align}
	\frac{{\rm d}^6 N_{e^{\pm}}}{{\rm d}{\bm p}^3 {\rm d}{\bm x}^3}
		&\sim \frac{ 1}{(2\pi \hbar)^3} \sum_{i,j} \exp \left[ -2\frac{{\rm i}}{\hbar} \int^{z_{{\rm t},i}^*}_{z_{{\rm t},j}} {\rm d}z \sqrt{Q[z]}  \right] \nonumber\\
		&= \frac{ 1}{(2\pi \hbar)^3}\left| \sum_{i} \exp \left[ -2\frac{{\rm i}}{\hbar} \int^{z_{{\rm t},i}^*}_{t_0} {\rm d}z \sqrt{Q[z]}  \right] \right|^2, \label{eqaa19}
\end{align} 
with $t_0$ being an arbitrary point on ${\mathbb R}$.  Equation~(\ref{eqaa19}) agrees with our semi-classical exact WKB formula (\ref{eq3.35}).

\section{Summary and discussion} \label{sec4}

We have studied the vacuum pair production by a time-dependent strong electric field on the basis of the exact WKB analysis under the semi-classical approximation.  First, we have explained that the vacuum pair production can be formulated in terms of a Bogoliubov transformation, which can be regarded as a connection matrix that describes a Stokes phenomenon of WKB solutions at the asymptotic times $t=\pm\infty$.  To apply the exact WKB analysis, we have identified the generic structure of a Stokes graph for the vacuum pair production (see Fig.~\ref{fig1}), assuming that the potential is adiabatic at the infinite times and is an analytic function on the entire complex plane.  Then, we have shown that the total connection matrix is given by a product of that for a Stokes segment connecting a pair of turning points, $z_{{\rm t}}$ and $z_{{\rm t}}^*$, which we have evaluated in the semi-classical limit [Eq.~(\ref{eq3.33})].  From the connection matrix, we have obtained the production number formula [Eq.~(\ref{eq3.35})], which is given by a sum of exponential factors ${\rm e}^{-S_{z_{\rm t}}/\hbar}$, controlling the magnitude of the production, as well as imaginary factors ${\rm e}^{-\frac{\rm i}{\hbar}\,{\rm Im}\,\sigma_{z_{\rm t}}}$, responsible for interference between different pairs of turning points.  The obtained formula is equivalent to other semi-classical approaches such as the steepest descent evaluation by Brezin and Izykson \cite{bre70} and the worldline instanton method \cite{dun05a, dun06a, dun06b}, and generalizes the divergent asymptotic series method by Berry \cite{ber89}.  The time-dependent effects such as the interplay between the perturbative multi-photon pair production and non-peturbative Schwinger mechanism and the interference effects including the dynamically assisted Schwinger mechanism have also been discussed within the obtained formula, and we have found a good agreement with the exact results in the semi-classical regime.

As a future work, it is desirable to include the higher order ${\mathcal O}({\rm e}^{-2S_{z_{\rm t}}/\hbar})$ corrections, which are important for the production in the quantum regime where $\hbar$ cannot be regarded as a small quantity and to describe smooth interplay between the multi-photon pair production processes and the low-order ones.  For this, it is essential to improve the connection matrix $T$ for a Stokes segment [Eq.~(\ref{eq3.33})].  To the best of our knowledge, there does not exist such a mathematical formula applicable in generic situations.  For some special cases, one may transform a generic potential $Q$ into a simple potential that can be analyzed exactly.  For example, in the case of the so-called merging-turning-points (MTP) equation, it is rigorously proved that the corresponding potential $Q$ can be locally transformed into the Weber potential \cite{AKT2}.  The monodromy structure of the Weber potential (or the resulting parabolic cylinder function) is well-known, so that one can compute $T$ without resorting to any approximations.  In a physics sense, this mathematical transformation amounts to mapping a generic electric field configuration into a superposition of constant electric fields.  It is interesting to study such special cases as a first step toward a complete formula for the vacuum pair production by a time-dependent electric field.

Another interesting direction is to extend our formulation to the vacuum pair production by other kinds of fields/forces other than a strong electric field or to analogous processes such as the Landau-Zener transition in materials.  The formulation presented in Sec.~\ref{sec3} is quite general and is not limited to a strong electric field, as we have just assumed that the potential $Q$ is adiabatic at the infinite times and is an analytic function on the entire complex plane.   Our formulation gives a powerful and general framework to discuss the time-dependent effects (or similar effects may occur for spatial variation) in the particle production.  One of the most interesting examples is the Hawking radiation.  One can use the similar Bogoliubov transformation technique to the Hawking radiation.  The problem is then reduced to solving a field equation under a potential determined by the background gravitational field (e.g., for a Klein-Gordon equation under the Kerr-Newman background, the potential is given by that for a confluent Heun equation \cite{Vieira:2014waa}), and thus our exact WKB analysis can be applied straightforwardly.  It is also reassuring that the worldline instanton approach has been applied recently to the Hawking radiation \cite{Dumlu:2017kfp}, which implies the applicability of our semi-classical WKB analysis because of the equivalence between the two approaches that we have shown in this paper.  The spacetime dependent effects, which can be conveniently captured with our framework, could play an important role in the Hawking radiation, e.g., by giving rise to non-thermal corrections.

{\it Note added in proof:} While preparing the final draft of our paper, a preprint \cite{Enomoto:2020xlf} has appeared in arXiv, which applied the exact WKB analysis to cosmological particle production (particularly for $Q(t)\propto t^2+g t^4$-type potential) and has some overlap with our paper.

\section*{Acknowledgments}
The authors thank Tomohiro~Matsuda for useful comments and participants of the international molecule-type workshop at Yukawa Institute for Theoretical Physics (YITP) ``Potential Toolkit to Attack Nonperturbative Aspects of QFT -- Resurgence and related topics -- (YITP-T-20-03)'' for fruitful discussions.  This work is supported by the Japan Society for the Promotion of Science (JSPS) Grant-in-Aid for Scientific Research (KAKENHI) Grant Number (18H01217).  This work is also supported in part by JSPS KAKENHI Grant Numbers 18K03627 (T.~F.) and 19K03817 (T.~M.).

\appendix

\section{Exact WKB analysis of the Airy equation}\label{sec2.2}

We discuss the exact WKB analysis of the Airy equation.  The Airy potential is defined as a potential having one simple turning point at some point $z=z_{\rm t}$, 
\begin{align}
	Q(z) = \xi (z-z_{\rm t}),  \label{eqa107}
\end{align}
where $\xi, z, z_{\rm t} \in {\mathbb C}$.  For the Airy potential (\ref{eqa107}), one can analytically compute the coefficients $\psi_{\pm,n}$ in Eq.~\eqref{qqq86}.  The solution reads
\begin{align}
	\psi_{\pm,n} 
		= d_n \times (\pm {\rm i})^n \xi^{-1/4-n/2}(z-z_{\rm t})^{-1/4-3n/2},
\end{align}
with
\begin{align}
	d_n \equiv \frac{1}{\sqrt{2}}\frac{1}{2\pi} \left( \frac{3}{4} \right)^{n} \frac{\Gamma(n+1/6)\Gamma(n+5/6)}{n!}.  
\end{align}
Notice that $d_n$ is factorially divergent, so is $\psi_{\pm,n}$.  In turn, one can explicitly compute the Borel transformation $\tilde{\psi}_{\pm}$ (\ref{eqa113}) as
\begin{align}
	\tilde{\psi}_{\pm} (z; \eta) 
		= \frac{1}{\sqrt{2}} \xi^{-1/4} (z-z_{\rm t})^{-1/4} {}_2 F_1 \left( \frac{1}{6},\frac{5}{6}; 1 ;\pm {\rm i} \frac{3}{4} \eta \xi^{-1/2} (z-z_{\rm t})^{-3/2} \right),
\end{align}
where ${}_2 F_1(a,b;c;z)$ is the hypergeometric function, 
\begin{align}
	{}_2 F_1(a,b;c;z) \equiv \sum_{n=0}^{\infty} \frac{(a)_n(b)_n}{(c)_n} \frac{z^n}{n!} = \sum_{n=0}^{\infty} \frac{\Gamma(a+n)\Gamma(b+n)\Gamma(c)  }{\Gamma(a)\Gamma(b)\Gamma(c+n)} \frac{z^n}{n!}.  
\end{align}
Therefore, the Borel sum (\ref{eq1a15}) reads
\begin{align}
	\Psi_{\pm} (z; \hbar) = \frac{1}{\sqrt{2}} \xi^{-1/4} (z-z_{\rm t})^{-1/4} \int^{\infty}_0 \frac{{\rm d}\eta}{\hbar} {\rm e}^{-\eta/\hbar}   {}_2 F_1 \left( \frac{1}{6},\frac{5}{6}; 1 ;\pm {\rm i} \frac{3}{4} \eta \xi^{-1/2} (z-z_{\rm t})^{-3/2} \right) .  
\end{align} 
Since the hypergeometric function ${}_2 F_1(a,b;c;z)$ has a branch cut on $z \in [1, \infty]$, the integrand becomes singular on 
\begin{align}
	\pm {\rm i} \frac{3}{4} \eta \xi^{-1/2} (z-z_{\rm t})^{-3/2} \in [1,\infty]\ \Leftrightarrow\ \eta \in \mp {\rm i} \frac{4}{3} \xi^{1/2} (z-z_{\rm t})^{3/2} \times [1,\infty].   \label{eqa.7}
\end{align} 
Therefore, the integration contour of the Laplace transformation $\eta: 0 \to \infty$ can hit the singularity if $0 = {\rm Im} \left[ {\rm i}   \xi^{1/2} (z-z_{\rm t})^{3/2} \right] = {\rm Im} \left[ {\rm i}  \int^z_{z_{\rm t}} {\rm d}z' \sqrt{Q(z')} \right]$, which is nothing but the Stokes line (\ref{eq2.13}).

\begin{figure}[t]
\begin{center}
\includegraphics[clip, width=0.4\textwidth]{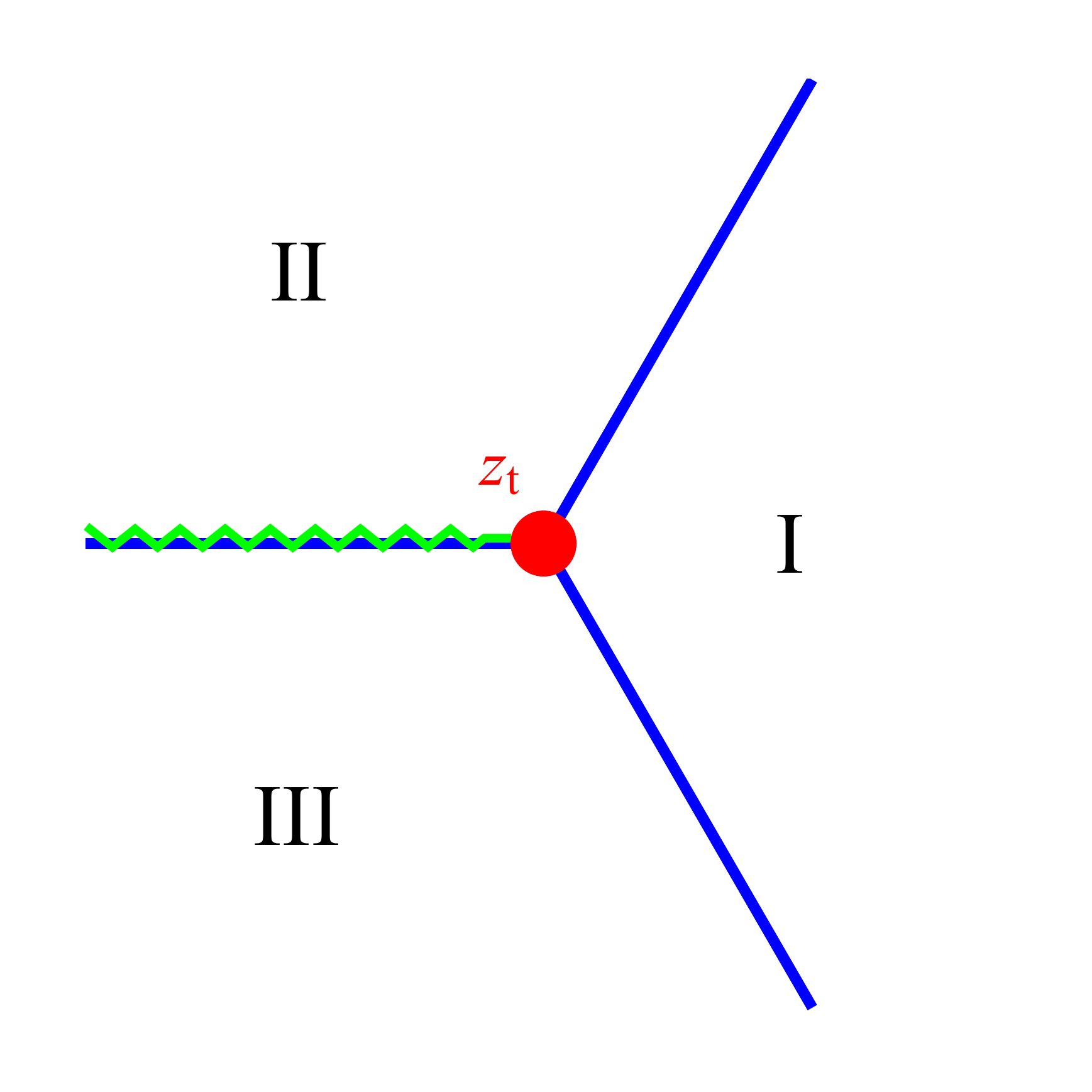}
\end{center}
\caption{Stokes graph for Airy potential with the case ${\rm arg}[\xi] = 0$.}
\end{figure}

For the Airy potential (\ref{eqa107}), there exist three Stokes lines emanating from the turning point $z_{\rm t}$.  The Stokes lines are straight lines emanating with angle $\arg (z-z_{\rm t}) + (1/3)\arg \xi = -\pi/3, +\pi/3, \pi \ ({\rm mod}\, 2\pi)$ and eventually flow into the infinity.  For convenience, let us insert a cut on $\arg(z-z_{\rm t})+\frac{1}{3} \arg\xi = \pi\; ({\rm mod}\;2\pi)$ and restrict ourselves on the first Riemann sheet such that $-\pi < \arg(z-z_{\rm t})+\frac{1}{3} \arg\xi < \pi$.  Then, we have three Stokes regions $A={\rm I},{\rm II},{\rm III}$ in the $z$-plane that are separated by the Stokes lines as
\begin{align}
	A=\left\{ \begin{array}{cl} {\rm I} & \displaystyle {\rm for}\ -\frac{\pi}{3} < \arg(z-z_{\rm t})+\frac{1}{3} \arg\xi < +\frac{\pi}{3} \vspace*{3mm}\\ {\rm II} &\displaystyle {\rm for}\ +\frac{\pi}{3} < \arg(z-z_{\rm t})+\frac{1}{3} \arg \xi< +\pi \vspace*{3mm}\\ {\rm III}&\displaystyle {\rm for}\ -\pi < \arg(z-z_{\rm t})+\frac{1}{3} \arg\xi < -\frac{\pi}{3}  \end{array} \right. .
\end{align}
In each Stokes region $A={\rm I}, {\rm II}, {\rm III}$ the Borel sum $\Phi_{\pm,A}$ is well-defined, but the Borel sums in the different Stokes regions are not necessarily identical $\Phi_{\pm,A} \neq \Phi_{\pm,B}$ if $A\neq B$ because of the Stokes phenomenon of WKB solutions.  Namely, suppose $z$ is initially located in a Stokes region $A$ and smoothly moved to another region $B$ by crossing the Stoke line separating the regions $A$ and $B$.  In the $\eta$-plane, the singularity can hit the integration contour of the Laplace transformation as changing $z$.  Then, the Borel sum in the region $B$ gets an additional contribution from the singularity compared to that in the region $A$ as
\begin{align}
	\Phi_{\pm,B} 
		&= \Phi_{\pm,A} + \exp \left[ \mp \frac{\rm i}{\hbar} \int^z_{t_0} {\rm d}z' \sqrt{Q(z')}  \right] \int_{\Gamma} \frac{{\rm d}\eta}{\hbar} {\rm e}^{-\eta/\hbar} \tilde{\psi}_{\pm}(z; \eta) , \label{eqa-135}
\end{align}
where the second term describes the contribution for the singularity and $\Gamma$ is a path that wraps the singularity of the Borel transformation $\tilde{\psi}_{\pm}$.  The second term can be evaluated with the help of 
\begin{align}
	&{}_2F_1(a,b;c;z+{\rm i}0^+) - {}_2F_1(a,b;c;z-{\rm i}0^+)  \nonumber\\
	&= \frac{2\pi{\rm i}}{\Gamma(a)\Gamma(b)} (z-1)^{c-a-b}{}_2F_1(c-a,c-b;c-a-b+1;1-z)\ \ {\rm for}\ 1<z \in {\mathbb R} ,
\end{align}
and the result reads 
\begin{align}
\left\{\begin{array}{l}
	\displaystyle \begin{pmatrix} \Phi_{+,{\rm II}} \\ \Phi_{-,{\rm II}} \end{pmatrix}
	= \begin{pmatrix} 1 & +{\rm i}{\rm e}^{ -2\frac{\rm i}{\hbar}  \int^{z_{\rm t}}_{t_0} {\rm d}z' \sqrt{Q(z')} } \\ 0 & 1 \end{pmatrix}\begin{pmatrix} \Phi_{+,{\rm I}} \\ \Phi_{-,{\rm I}} \end{pmatrix} \vspace*{3mm} \\
	\displaystyle \begin{pmatrix} \Phi_{+,{\rm I}} \\ \Phi_{-,{\rm I}} \end{pmatrix}
	= \begin{pmatrix} 1 & 0 \\ +{\rm i}  {\rm e}^{ +2\frac{\rm i}{\hbar}  \int^{z_{\rm t}}_{t_0} {\rm d}z' \sqrt{Q(z')} } & 1 \end{pmatrix}\begin{pmatrix} \Phi_{+,{\rm III}} \\ \Phi_{-,{\rm III}} \end{pmatrix} 
\end{array}\right. . 
\end{align} 
This is nothing but Eq.~(\ref{e151}), as ${\rm Re} \left[ {\rm i} \int^z_{z_{\rm t}} {\rm d}z' \sqrt{Q(z')} \right] < 0$ on the Stokes line between I and II and $>0$ between III and I.

\section{Standard perturbation theory (at the lowest order)} \label{appa4}

The standard perturbative treatment with respect to the applied electric field $e{\bm A}$ gives a faithful description when the field varies very fast compared to the mass scale; or, physically, when the particle production mechanism is dominated by the one-photon (or a few-photon) pair production process $\gamma \to e^+ e^-$ \cite{tay14}.  In such a regime, the electric field interacts with particles incoherently rather than coherently, for which the semi-classical treatment is not justifiable.  Thus, the standard perturbation theory is applicable to different parameter regimes that the semi-classical exact WKB analysis as well as other semi-classical approaches covers, and vice versa.

For later convenience, we here use the Green function technique to derive the production number formula within the standard perturbation theory.  We first rewrite the mode equation [i.e., the Klein-Gordon equation (\ref{eq3.1}) in terms of the mode function $\varphi_{\pm,{\rm as}}$] as \begin{align}
	\left[ \hbar^2 \partial_t^2  + Q_0 \right] \varphi_{\pm,{\rm as}}(t)
	= V(t) \varphi_{\pm,{\rm as}}(t), \label{eq----45}
\end{align}
where 
\begin{align}
	Q_0 \equiv m^2 + {\bm p}^2,\ 
	V \equiv Q - Q_0 = 2 {\bm p}\cdot e{\bm A}  - |e{\bm A}|^2.
\end{align}
Below, we assume smallness of $e{\bm A}$, or $V$, which is justified when the applied electric field is very fast\footnote{To be precise, the lowest order perturbation works for $\nu \equiv  (\hbar \omega)^2/ | \hbar e{\bm E}| \gg 1$ and $\gamma \equiv m  \hbar\omega/| \hbar e{\bm E}| \gg 1$ \cite{tay14}.  Notice that the largeness of the Keldysh parameter alone $\gamma \gg 1$ is not enough to justify the lowest order perturbation theory.  }.  Indeed, for an electric field with a monochromatic frequency $\omega$, i.e., ${\bm E} = {\bm E}(\omega t)$, the corresponding gauge potential decreases linearly with $\omega$ as $|{\bm A}| \propto \omega^{-1}$.  Then, one may solve Eq.~(\ref{eq----45}) perturbatively using the Green function technique to obtain 
\begin{align}
	\varphi_{\pm,{\rm as}}(t) 
		&= \varphi_{\pm,(0)}(t) + \int {\rm d}t'\, G_{(0)} (t,t') V(t') \varphi_{\pm,{\rm as}}(t') \nonumber\\
		&=  \varphi_{\pm,(0)}(t) + \int {\rm d}t'\, G_{(0)} (t,t') V(t') \varphi_{\pm,(0)}(t') + {\mathcal O}(|V|^2), \label{eq---75}
\end{align}
where $G_{(0)}$ is a retarded Green function satisfying
\begin{align}
	0 = G_{(0)}(t - t' < 0), \ \ 
	\delta(t - t') = \left[ \hbar^2 \partial_t^2  + Q_0 \right] G_{(0)}(t, t'),  \label{eqa--23}
\end{align}
$\varphi_{\pm,(0)}$ is a plane wave solution of $0 = \left[ \hbar^2  \partial_t^2 + Q_0 \right] \varphi_{\pm,(0)}$, i.e.,
\begin{align}
	\varphi_{\pm,(0)} 
		= \frac{1}{\sqrt{2Q_0^{1/2}(t)}} {\rm e}^{\mp \frac{\rm i}{\hbar} Q_0^{1/2}t} 
		= \lim_{|e{\bm A}| \to 0} \frac{1}{\sqrt{2Q^{1/2}(t)}} \exp\left[ \mp \frac{\rm i}{\hbar} \int^{t}{\rm d}t' \, Q^{1/2}(t') \right], \label{eqa--24}
\end{align}
and we imposed a boundary condition onto $\varphi_{\pm,{\rm as}}$ as [the same as Eq.~(\ref{eq3.10})]
\begin{align}
	0 = \lim_{t \to -\infty} \left[ \varphi_{\pm,{\rm in}}  - \varphi_{\pm,(0)} \right], \ 
	0 = \lim_{t \to +\infty} \left[ \varphi_{\pm,{\rm out}}  - \varphi_{\pm,(0)} \right].  
\end{align}
Using
\begin{align}
	G_{(0)}(t, t') 
		= - \frac{\rm i}{\hbar} \theta(t - t') \left[  [\varphi_{+,(0)}(t)]^* \varphi_{+,(0)}(t')  - [\varphi_{-,(0)}(t)]^* \varphi_{-,(0)}(t') \right], 
\end{align}
one can express Eq.~(\ref{eq---75}) as
\begin{align}
	\varphi_{\pm,{\rm in}}(t) 
		&= \varphi_{\pm,(0)}(t) \times \left[ 1 \pm \frac{\rm i}{\hbar}  \int^{t}_{-\infty} {\rm d}t' [\varphi_{\pm,(0)}(t')]^*  V(t') \varphi_{\pm,(0)}(t')  + {\mathcal O}(|V|^2) \right] \nonumber\\
			&\quad + \varphi_{\mp,(0)}(t)  \times \left[ 0 \mp \frac{\rm i}{\hbar}  \int^{t}_{-\infty} {\rm d}t' [\varphi_{\mp,(0)}(t')]^*  V(t') \varphi_{\pm,(0)}(t')  + {\mathcal O}(|V|^2) \right] .  
\end{align}
Taking $t \to \infty$ and comparing with Eq.~(\ref{eq3.13}), one understands 
\begin{subequations}
\begin{align}
	U_{11}	 = [U_{22}]^* 
				&= 1 + \frac{\rm i}{\hbar}  \int^{+\infty}_{-\infty} {\rm d}t' [\varphi_{-,(0)}(t')]^*  V(t') \varphi_{-,(0)}(t')  + {\mathcal O}(|V|^2) , \\
	U_{12}	 = [U_{21}]^* 
				&= 0 + \frac{\rm i}{\hbar}  \int^{+\infty}_{-\infty} {\rm d}t' [\varphi_{-,(0)}(t')]^*  V(t') \varphi_{+,(0)}(t')  + {\mathcal O}(|V|^2) . \label{eqaa28b}
\end{align}
\end{subequations}
Using the analytical expression for the plane wave $\varphi_{\pm,(0)} $ (\ref{eqa--24}), one can explicitly evaluate the off-diagonal component of $U$ (\ref{eqaa28b}) and arrives at
\begin{align}
	\frac{{\rm d}^6 N_{e^{\pm}}}{{\rm d}{\bm x}^3{\rm d}{\bm p}^3} 
	&= \frac{1}{(2\pi \hbar)^3} |U_{12}|^2 \nonumber\\
	&= \frac{1}{(2\pi \hbar)^3} \frac{1}{4 \hbar^2 Q_0^2 }\left|  {\bm p} \cdot \hbar e{\bm {\mathcal E}}(2Q_0^{1/2}/\hbar) \right|^2 + {\mathcal O}(|\hbar e{\bm E}|^3).   \label{eqqq89}
\end{align}
where ${\bm {\mathcal E}}(\omega) \equiv \int^{+\infty}_{-\infty}{\rm d}t\,{\rm e}^{- {\rm i}\omega t} {\bm E}(t)$ is the Fourier transformation of ${\bm E}(t)$, as in the main text.  The Fourier transformation ${\bm {\mathcal E}}(\omega)$ may be peaked at some characteristic frequency $\omega \sim \omega_{\rm c}$ of the applied field ${\bm E}$.  Then, Eq.~(\ref{eqqq89}) implies that the production occurs only when $ \hbar \omega_{\rm c} \sim 2Q_0^{1/2} = 2\sqrt{m^2 + {\bm p}^2}$, reflecting the threshold nature of the one-photon pair production.

\section{Perturbation theory in the Furry picture} \label{appa-4}

The perturbation theory in the Furry picture is an improved version of the standard perturbation theory (see Appendix~\ref{appa4}) by using a dressed propagator and wavefunction instead of the bare ones $G_{\rm (0)} (\ref{eqa--23})$ and $\varphi_{\pm,(0)}$ (\ref{eqa--24}).  This theory works quite well when there is a clear scale separation in the applied electric field as 
\begin{align}
	{\bm E} = {\bm E}_{\rm s} + {\bm E}_{\rm w}\ \ {\rm with}\ |{\bm E}_{\rm s}| \gg |{\bm E}_{\rm w}|,
\end{align} 
where ${\bm E}_{\rm w}$ is some perturbation on top of a strong field ${\bm E}_{\rm s}$.  In particular, when the strong field ${\bm E}_{\rm s}$ is sufficiently slow such that it is well approximated by a constant electric field (such a situation may be realized in the dynamically assisted Schwinger mechanism), one can carry out all the calculations exactly, finding an analytical production number formula for ${\bm E}_{\rm w}$ with arbitrary time-dependence.

The derivation of the formula can be done in a parallel manner as the standard perturbation theory (we resort details of the derivation to Ref.~\cite{fk}).  The only difference is that the bare propagator and wavefunction, $G_{\rm (0)}$ and $\varphi_{\pm,(0)}$, respectively, are replaced by the dressed ones $G_{\rm (0)} \to G_{\rm (d)}$ and $\varphi_{\pm,(0)} \to \varphi_{\pm,{\rm as},{\rm (d)}}$ under the strong field ${\bm E}_{\rm s}$ such that
\begin{align}
	&0 = G_{\rm (d)}(t - t' < 0), \ \ 
	\delta(t - t') = \left[ \hbar^2 \partial_t^2  + m^2 + \left( {\bm p} - e{\bm A}_{\rm s}(t) \right)^2 \right] G_{\rm (d)}(t, t'), \nonumber\\
	&0 = \left[ \hbar^2 \partial_t^2  + m^2 + \left( {\bm p} - e{\bm A}_{\rm s}(t) \right)^2 \right] \varphi_{\pm,{\rm as},{\rm (d)}}(t),
\end{align}
where ${\bm A}_{\rm s}$ is the gauge potential for the strong field ${\bm E}_{\rm s}$ and we required that $\varphi_{\pm,{\rm as},{\rm (d)}}$ asymptotes a plane wave at $t \to \pm \infty$ as in Eq.~(\ref{eq3.10}).  Treating the remaining field ${\bm E}_{\rm w}$ as a perturbation and repeating a similar calculation that we explained in Appendix~\ref{appa4}, one can obtain
\begin{align}
	\frac{{\rm d}^6 N_{e^{\pm}}}{{\rm d}{\bm x}^3{\rm d}{\bm p}^3}
	=  \frac{1}{(2\pi \hbar)^3}\biggl| ( \varphi_{-,{\rm out},{\rm (d)}} | \varphi_{+,{\rm in},{\rm (d)}} ) + \frac{\rm i}{\hbar} \int^{+\infty}_{-\infty} {\rm d}t\, [\varphi_{-,{\rm out},{\rm (d)}}]^* V \varphi_{+,{\rm in},{\rm (d)}} + {\mathcal O} \left( |V|^2 \right)   \biggl|^2 , \label{eqaa36a}
\end{align}
where the Klein-Gordon inner product $(A|B) \equiv {\rm i} A^* \overset{\leftrightarrow}{\partial}_t B$ is a conserved quantity and $V$ denotes
\begin{align}
	V	&\equiv Q - \left[ m^2 + \left( {\bm p} - e{\bm A}_{\rm s} \right)^2 \right] \nonumber\\ 
		&=  2 \left( {\bm p} - e{\bm A}_{\rm s}\right) \cdot e{\bm {\bm A}}_{\rm w} + |e{\bm A}_{\rm w} |^2, 
\end{align}
with ${\bm A}_{\rm w}$ being the gauge potential for the perturbation ${\bm E}_{\rm w}$.

One can explicitly evaluate Eq.~(\ref{eqaa36a}) for a constant strong electric field \cite{fk, tay20b}, which is quite powerful to discuss the dynamically assisted Schwinger mechanism.  Namely, we consider 
\begin{align}
	{\bm E}_{\rm s} = E_{\rm s} \times {\bf e}_\parallel
	\ {\rm with}\ eE_{\rm s}>0.  
\end{align}
For this case, one can derive an analytical expression for the dressed wavefunction $\varphi_{\pm,{\rm as},{\rm (d)}}$: 
\begin{align}
	\varphi_{+,{\rm out},{\rm (d)}}(t) 
	&= [\varphi_{-,{\rm out},{\rm (d)}}(t)]^*
	= \frac{{\rm e}^{- \frac{\pi}{4} \frac{m^2 + {\bm p}_{\perp}^2 }{2\hbar eE_{\rm s}}}}{(2\hbar eE_{\rm s})^{1/4}} \left[ D_{- {\rm i}\frac{m^2 + {\bm p}_{\perp}^2 }{2\hbar eE_{\rm s}}-1/2}\left( -{\rm e}^{{\rm i}\pi/4}\sqrt{\frac{2}{\hbar eE_{\rm s}}} (eE_{\rm s} t + p_\parallel) \right) \right]^*, \nonumber\\
	\varphi_{+,{\rm in},{\rm (d)}}(t) 
	&= [\varphi_{-,{\rm in},{\rm (d)}}(t)]^*
	= \frac{{\rm e}^{- \frac{\pi}{4} \frac{m^2 + {\bm p}_{\perp}^2 }{2\hbar eE_{\rm s}}}}{(2\hbar eE_{\rm s})^{1/4}}  D_{- {\rm i}\frac{m^2 + {\bm p}_{\perp}^2 }{2\hbar eE_{\rm s}}-1/2}\left( +{\rm e}^{{\rm i}\pi/4}\sqrt{\frac{2}{\hbar eE_{\rm s}}} (eE_{\rm s} t + p_\parallel) \right)  , 
\end{align}
with $D_\nu(z)$ being the parabolic cylinder function.  Inserting this expression into Eq.~(\ref{eqaa36a}), one obtains
\begin{align}
	\frac{{\rm d}^6 N_{e^{\pm}}}{{\rm d}{\bm x}^3{\rm d}{\bm p}^3}
		&= \frac{1}{(2\pi \hbar)^3} \exp\left[ -\pi \frac{m^2 + {\bm p}_{\perp}^2}{\hbar eE_{\rm s}}  \right] \left|  1 +  \int_0^{\infty} d\omega \, {\rm e}^{-{\rm i}\frac{\omega p_{\parallel}}{\hbar eE_{\rm s}} }  \frac{\hbar e{\mathcal E}_{\rm w}(\omega)}{\hbar eE_{\rm s}}  I(\omega) + {\mathcal O}(|\hbar eE_{\rm w}|^2) \right|^2,  \label{eqaaa37}
\end{align}
where 
\begin{align}
	I(\omega) 
		&\equiv \frac{\hbar eE_{\rm s}}{\omega} \partial_\omega \left[ {\rm e}^{ +{\rm i} \frac{\omega^2}{4\hbar eE_{\rm s}} } {}_1F_1 \left( \frac{1}{2}+\frac{\rm i}{2}\frac{m^2 + {\bm p}_{\perp}^2}{\hbar eE_{\rm s}}; 1; -\frac{\rm i}{2} \frac{\omega^2}{\hbar eE_{\rm s}} \right) \right] \nonumber\\
		&= \frac{m^2+{\bm p}_\perp^2}{2\hbar eE_{\rm s}} \left[ 1 + \frac{1}{8} \left( 1 - \frac{|\hbar eE_{\rm s}|^2}{(m^2+{\bm p}_\perp^2)^2} \right) \left| \frac{\omega \sqrt{m^2+{\bm p}_\perp^2}}{\hbar eE_{\rm s}} \right|^2 + {\mathcal O}(\omega^4) \right].
\end{align}

\section{Analysis of the Sauter electric field} \label{appa-5}

As a supplement to Fig.~\ref{fig-3}, we here describe the details of the analysis of the Sauter electric field (\ref{eQ196}).  

\subsection{Exact result}

Under the Sauter electric field (\ref{eQ196}), one can cast the mode equation (\ref{eq3.1}) into Gauss's hypergeometric differential equation.  Namely, we introduce
\begin{align}
	u \equiv \frac{1}{2} \left[ 1 + \tanh \Omega t \right], \ 
	m_\perp \equiv \sqrt{m^2+{\bm p}_\perp^2},\ 
	P_\pm \equiv p_\parallel \pm \frac{eE_0}{\Omega} ,
\end{align}
and decompose $\phi$ as
\begin{align}
	\phi = u^{-\frac{{\rm i}}{2\hbar\Omega} \sqrt{ m_\perp^2 + P_-^2} } (1-u)^{+\frac{{\rm i}}{2\hbar\Omega} \sqrt{ m_\perp^2 + P_+^2} } f.  
\end{align}
Then, we can rewrite the mode equation (\ref{eq3.1}) as
\begin{align}
	0 = \left[ u(1-u) \frac{{\rm d}^2}{{\rm d}u^2} + \left\{ c - (a+b+1)u  \right\} \frac{{\rm d}}{{\rm d}u} - ab \right] f, \label{eq199}
\end{align}
where
\begin{subequations}
\begin{align}
	a &\equiv \frac{1}{2} - \frac{1}{2} \sqrt{1 - \left( \frac{2eE_0}{\hbar\Omega^2} \right)^2 } - \frac{1}{2}\frac{{\rm i}}{\hbar\Omega} \sqrt{ m_\perp^2 + P_-^2} + \frac{1}{2}\frac{{\rm i}}{\hbar\Omega} \sqrt{ m_\perp^2 + P_+^2}, 
\\
	b &\equiv \frac{1}{2} + \frac{1}{2} \sqrt{1 - \left( \frac{2eE_0}{\hbar\Omega^2} \right)^2 } - \frac{1}{2}\frac{{\rm i}}{\hbar\Omega} \sqrt{ m_\perp^2 + P_-^2} + \frac{1}{2}\frac{{\rm i}}{\hbar\Omega} \sqrt{ m_\perp^2 + P_+^2}, \\
	c &\equiv 1 - \frac{{\rm i}}{\hbar\Omega} \sqrt{ m_\perp^2 + P_-^2}.  
\end{align}
\end{subequations}
Requiring the boundary condition (\ref{eq3.10}), one can readily solve Eq.~(\ref{eq199}) and derive the Bogoliubov coefficients $U_{12}$ and $U_{21}$, whose square gives the production number.  The result is 
\begin{align}
	\frac{{\rm d}^6 N_{e^{\pm}}}{{\rm d}{\bm p}^3 {\rm d}{\bm x}^3}
	&= \frac{ 1}{(2\pi \hbar)^3} \frac{1}{{ \sinh \frac{\pi \sqrt{ m_\perp^2 + P_-^2}}{ \hbar \Omega} \sinh \frac{\pi \sqrt{ m_\perp^2 + P_+^2}}{ \hbar \Omega} }}\nonumber\\
	&\quad \times \cosh\left[ \pi \left( + \frac{{\rm i}}{2} \sqrt{1 -\left( \frac{2 \hbar eE_0}{ (\hbar \Omega)^2} \right)^2} +  \frac{\sqrt{ m_\perp^2 + P_-^2}-\sqrt{ m_\perp^2 + P_+^2}}{2 \hbar \Omega} \right)   \right] \nonumber\\
	&\quad \times \cosh\left[ \pi \left( - \frac{{\rm i}}{2} \sqrt{1 -\left( \frac{2 \hbar eE_0}{ (\hbar \Omega)^2} \right)^2} +  \frac{\sqrt{ m_\perp^2 + P_-^2 }-\sqrt{ m_\perp^2 + P_+^2}}{2 \hbar \Omega} \right)   \right] . \label{eq203}  
\end{align}

For later use, we expand the exact result (\ref{eq203}) with $\hbar \Omega \to 0$ and $\infty$: 
\begin{align}
	&\lim_{\hbar \Omega \to 0} (2\pi \hbar)^3 \frac{{\rm d}^6 N_{e^{\pm}}}{{\rm d}{\bm p}^3 {\rm d}{\bm x}^3} \nonumber\\ 
		&= \exp\Biggl[ -2\pi \Biggl( \frac{m_\perp^2}{2 \hbar eE_0} + \frac{ (\hbar eE_0)^2 - m_\perp^4 + 4m_\perp^2 p_\parallel^2}{8(\hbar eE_0)^3} (\hbar \Omega)^2 + \frac{m_\perp^6 - 12 m_\perp^4 p_\parallel^2 + 8m_\perp^2 p_\parallel^4}{16(\hbar eE_0)^5} (\hbar \Omega)^4 \nonumber\\
		&\quad\quad\quad\quad\quad\quad +\frac{  (\hbar eE_0)^4 -5 m_\perp^8+120 m_\perp^6 p_\parallel^2-240 m_\perp^4 p_\parallel^4+64 m_\perp^2 p_\parallel^6}{128 (\hbar eE_0)^7} (\hbar \Omega)^6 + {\mathcal O}((\hbar \Omega)^{8}) \Biggl) \Biggl],    \label{EQQ170}
\end{align}
and
\begin{align}
	&\lim_{\hbar \Omega \to \infty} (2\pi \hbar)^3 \frac{{\rm d}^6 N_{e^{\pm}}}{{\rm d}{\bm p}^3 {\rm d}{\bm x}^3} \nonumber\\
		&= \left[ \frac{ \frac{ \pi^2 p_\parallel^2 (\hbar eE_0)^2}{m_\perp^2+p_\parallel^2} (\hbar \Omega)^{-4}  }{ \left( \sinh \frac{\pi\sqrt{m_\perp^2 + p_\parallel^2}}{\hbar \Omega} \right)^2 } \right] \times  \frac{1 - \frac{(\hbar eE_0)^2m_\perp^2 }{(m_\perp^2 + p_\parallel^2)^2} (\hbar \Omega)^{-2} + {\mathcal O}((\hbar \Omega)^{-4})}{  1 + \frac{(\hbar eE_0)^2(m_\perp^2-p_\parallel^2)}{(m_\perp^2+p_\parallel^2)^2}  (\hbar \Omega)^{-2} + {\mathcal O}((\hbar \Omega)^{-4}) } \nonumber\\
		&=  \frac{ (\hbar eE_0)^2 p_\parallel^2 }{ \left( m_\perp^2 + p_\parallel^2 \right)^2} (\hbar \Omega)^{-2} + \frac{ (\hbar eE_0)^2 p_\parallel^2 \left(3 (\hbar eE_0)^2 \left( p_\parallel^2 - 2 m_\perp^2 \right) - \pi^2 \left( m_\perp^2 + p_\parallel^2 \right)^3 \right)}{ 3 \left( m_\perp^2 + p_\parallel^2 \right)^4 } (\hbar \Omega)^{-4} \nonumber\\
		&\quad + {\mathcal O}((\hbar \Omega)^{-6}).  \label{qqq183}
\end{align}

\subsection{The exact WKB analysis in the semi-classical limit} \label{app344}

\begin{figure}[!t]
\begin{center}
\hspace*{-10mm}
\includegraphics[clip, width=0.5\textwidth]{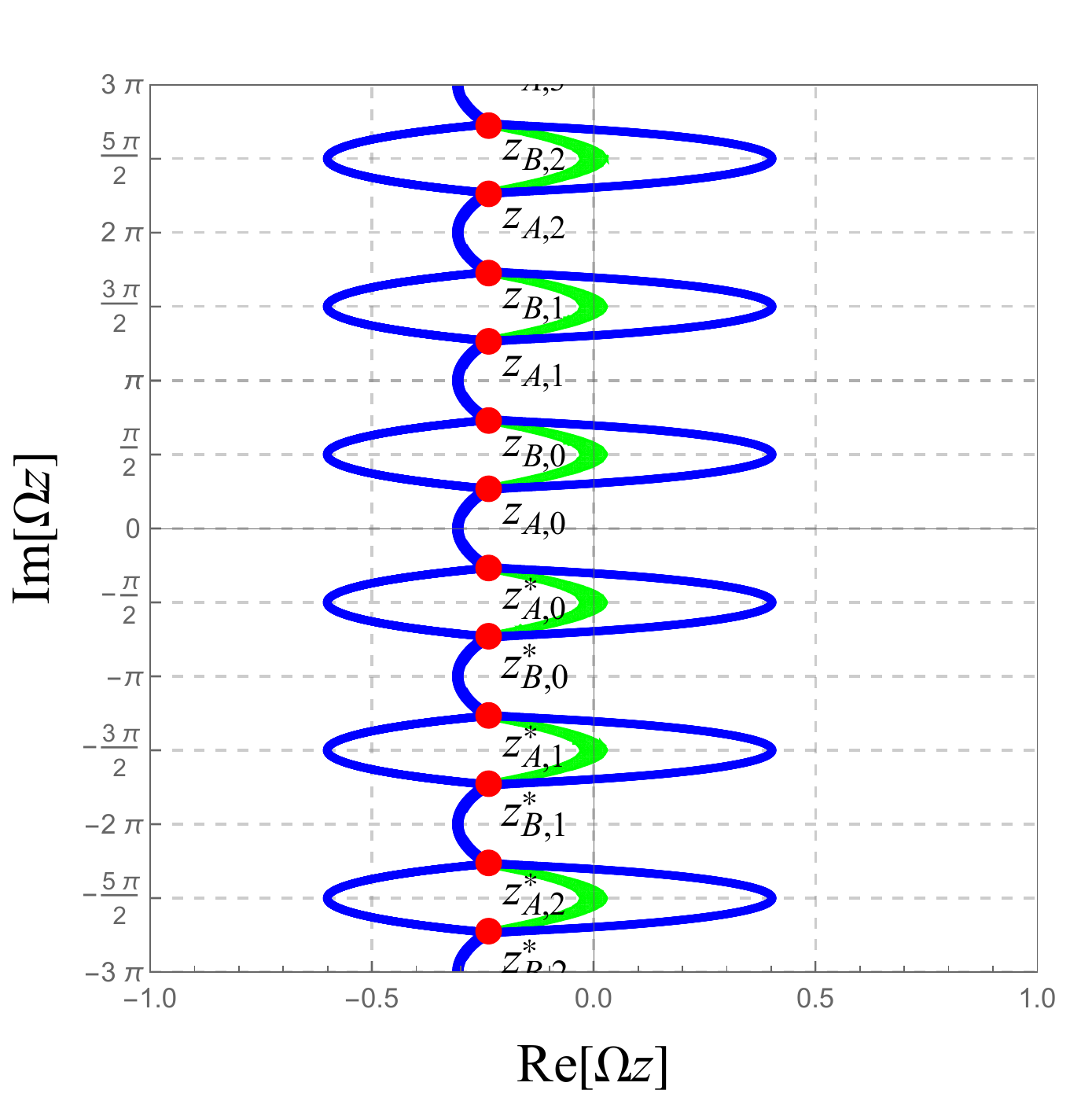}
\end{center}
\caption{\label{figa2} (color online) A Stokes graph for the Sauter electric field (\ref{eQ196}).  The red points, the blue lines, and the green lines are the turning points (\ref{eqtur}), Stokes lines, and branch cuts, respectively.  Parameters are taken as $m_\perp/\Omega=1, eE_0/\Omega^2=1, p_\parallel/\Omega=0.5$.  
}
\end{figure}

For the Sauter electric field (\ref{eQ196}), there exist an infinite number of pairs of turning points $(z_{A,n}, z_{A,n}^{*})$ and $(z_{B,n}, z_{B,n}^{*})$: 
\begin{subequations}
\label{eqtur}
\begin{align}
	z_{A,n} &\equiv +\frac{\rm i}{\Omega} \arctan \left( \hbar\Omega \frac{m_\perp + {\rm i} p_\parallel}{\hbar eE_0} \right) + \frac{{\rm i} n \pi}{\Omega} , \\
	z_{B,n} &\equiv -\frac{\rm i}{\Omega} \arctan \left( \hbar\Omega \frac{m_\perp - {\rm i} p_\parallel}{\hbar eE_0} \right) + \frac{{\rm i} (n+1) \pi}{\Omega},  \label{q-186}
\end{align}
\end{subequations}
with $n=0,1,\cdots \in {\mathbb N}$.  Note that $0<{\rm Im}\,\Omega z_{A,0}<\pi/2<{\rm Im}\,\Omega z_{B,0}<\pi<{\rm Im}\,\Omega z_{A,1}<3\pi/2<{\rm Im}\,\Omega z_{B,1}<\cdots$.  The Stokes lines emanating from the turning points form Stokes segments as shown in Fig.~\ref{figa2}.  The Stokes segment crossing the real axis is infinitely degenerated with Stokes lines emanating from all the turning points $z_{\rm t} = z_{A,n}, z_{A,n}^{*}, z_{B,n}, z_{B,n}^{*}$.  The connection matrix for such a multiply degenerated Stokes segment is given by Eq.~(\ref{eq-330}), and the resulting production number formula in the semi-classical limit takes exactly the same form as that for a doubly degenerated Stokes segment (\ref{eq3.35}), as we remarked in footnote~\ref{foot3}.  One can explicitly evaluate relevant actions $S_{z_{\rm t}}$'s, whose integrals are enclosing pairs of turning points $(z_{\rm t},z_{\rm t}^*)$ with $z_{\rm t}=z_{A,n}, z_{B,n}$, as
\begin{subequations}
\label{eqd9}
\begin{align}
	S_{z_{A,n}} 
		&= +{\rm i} \int_{z_{A,n}}^{z^*_{A,n}} {\rm d}z \sqrt{Q(z)} 
		= (2n+1) S_\alpha + 2n S_\beta , \label{eqd9a}\\
	S_{z_{B,n}} 
		&= +{\rm i} \int_{z_{B,n}}^{z^*_{B,n}} {\rm d}z \sqrt{Q(z)} 
		= (2n+1) S_\alpha + (2n+2) S_\beta , 
		\label{eqd9b}
\end{align}
\end{subequations}
where we introduced
\begin{subequations}
\begin{align}
	S_{\alpha} 
		&\equiv {\rm Re}\left[ +{\rm i} \int_{z_{A,0}}^{z_{A,0}^*} {\rm d}z \sqrt{Q(z)} \right] 
		= \frac{\pi}{\Omega} \left( \sum_{\pm}\frac{\sqrt{m_\perp^2 + P_\pm^2}}{2 } -  \frac{\hbar eE_0}{\hbar \Omega} \right) ,\\
	S_{\beta} &\equiv  {\rm Re} \left[ +{\rm i} \int_{z_{B,0}}^{z_{A,0}} {\rm d}z \sqrt{Q(z)}  \right]
		= \frac{\pi}{\Omega} \left( {\rm sgn}\,p_\parallel \sum_{\pm} (\mp 1)\frac{\sqrt{m_\perp^2 + P_\pm^2}}{2 }  + \frac{\hbar eE_0}{\hbar \Omega} \right) , 
\end{align}
\end{subequations}
and used the periodicity of the Sauter electric field $Q(z)=Q(z+{\rm i}\pi/\Omega)$ to get the second equalities of Eqs.~(\ref{eqd9a}) and (\ref{eqd9b}).  Note that $S_\alpha, S_\beta>0$, which guarantees $S_{A,n}, S_{B,n}>0$.  Inserting $	{\rm Im}\,\sigma_{z_{A,n}} = {\rm Im}\,\sigma_{z_{B,n}} = {\rm Im}\left[ +2{\rm i}\int_{t_0}^{z_{A,0}} {\rm d}z \sqrt{Q(z)} \right]$ and Eq.~(\ref{eqd9}) into the production number formula (\ref{eq3.35}), one arrives at
\begin{align}
	\frac{{\rm d}^6 N_{e^{\pm}}}{{\rm d}{\bm p}^3 {\rm d}{\bm x}^3}
		&= \frac{ 1}{(2\pi \hbar)^3}  \left| \sum_{n=0}^\infty \left( {\rm e}^{-S_{z_{A,n}}/\hbar} + {\rm e}^{-S_{z_{B,n}}/\hbar} \right) \right|^2 \nonumber\\
		&= \frac{ 1}{(2\pi \hbar)^3} {\rm e}^{-2S_\alpha/\hbar} \left| \frac{1+{\rm e}^{-2S_\beta/\hbar}}{1-{\rm e}^{-2(S_\alpha+S_\beta)/\hbar}}  \right|^2 \nonumber\\
		&\sim \frac{ 1}{(2\pi \hbar)^3} {\rm e}^{-2S_\alpha/\hbar} \left|  1+{\rm e}^{-2S_\beta/\hbar} \right|^2. \label{eqqa-45}  
\end{align}
In the last line, we neglected ${\mathcal O}(|{\rm e}^{-2S_\alpha/\hbar}|^2)$ terms, so as to be consistent with the semi-classical approximation.

In the limit of $\hbar \Omega \to 0$, the production number formula (\ref{eqqa-45}) behaves as
\begin{align}
	&\lim_{\hbar \Omega \to 0} (2\pi \hbar)^3 \frac{{\rm d}^6 N_{e^{\pm}}}{{\rm d}{\bm p}^3 {\rm d}{\bm x}^3} \nonumber\\
	&= \exp\Biggl[ -2\pi \Biggl( \frac{m_\perp^2}{2 \hbar eE_0} + \frac{ \hbox{ \sout{$(\hbar eE_0)^2$}}  - m_\perp^4 + 4m_\perp^2 p_\parallel^2}{8(\hbar eE_0)^3} (\hbar \Omega)^2 + \frac{m_\perp^6 - 12 m_\perp^4 p_\parallel^2 + 8m_\perp^2 p_\parallel^4}{16(\hbar eE_0)^5} (\hbar \Omega)^4 \nonumber\\
		&\quad\quad\quad\quad\quad\quad +\frac{ \hbox{ \sout{$(\hbar eE_0)^4$}} -5 m_\perp^8+120 m_\perp^6 p_\parallel^2-240 m_\perp^4 p_\parallel^4+64 m_\perp^2 p_\parallel^6}{128 (\hbar eE_0)^7} (\hbar \Omega)^6 + {\mathcal O}((\hbar \Omega)^{8}) \Biggl) \Biggl],
\end{align}
where the slashed parts $\sim \left( \frac{(\hbar \Omega)^2}{\hbar eE_0} \right)^{2n+1} = \nu^{2n+1}$ are absent compared to the exact formula (\ref{EQQ170}).  Thus, the semi-classical exact WKB formula (\ref{eq3.35}) works well in the semi-classical regime, or when the applied field is slow enough compared to the electric field strength, such that $\nu \lesssim 1$.

\subsection{Standard perturbation theory at the lowest order} 

The Fourier transformation of the Sauter electric field (\ref{eQ196}) is given by
\begin{align}
	e{\mathcal E}(\omega) = {\rm i}\pi \frac{ \hbar eE_0}{(\hbar \Omega)^2} \frac{ \hbar \omega }{ \sinh \frac{\pi}{2} \frac{ \hbar \omega}{\hbar \Omega} }.  
\end{align}
Thus, the formula (\ref{eqqq89}) is expressed as
\begin{align}
(2\pi \hbar)^3 \frac{{\rm d}^6 N_{e^{\pm}}}{{\rm d}{\bm p}^3 {\rm d}{\bm x}^3}
	= \frac{ \frac{ \pi^2 p_\parallel^2 (\hbar eE_0)^2}{m_\perp^2+p_\parallel^2} (\hbar \Omega)^{-4}  }{ \left( \sinh \frac{\pi\sqrt{m_\perp^2 + p_\parallel^2}}{\hbar \Omega} \right)^2 }, \label{eqaaaa48}
\end{align}
which agrees with the square bracket in the first equality in Eq.~(\ref{qqq183}).  In the limit of $\hbar \Omega \to \infty$, Eq.~(\ref{eqaaaa48}) is expanded as 
\begin{align}
	&\lim_{\hbar \Omega \to \infty} (2\pi \hbar)^3 \frac{{\rm d}^6 N_{e^{\pm}}}{{\rm d}{\bm p}^3 {\rm d}{\bm x}^3} \nonumber\\
	&=  \frac{ (\hbar eE_0)^2 p_\parallel^2 }{ \left( m_\perp^2 + p_\parallel^2 \right)^2} (\hbar \Omega)^{-2} + \frac{ (\hbar eE_0)^2 p_\parallel^2 \left(  \hbox{ \sout{$3 (\hbar eE_0)^2 \left( p_\parallel^2 - 2 m_\perp^2 \right) $}} - \pi^2 \left( m_\perp^2 + p_\parallel^2 \right)^3 \right)}{ 3 \left( m_\perp^2 + p_\parallel^2 \right)^4 } (\hbar \Omega)^{-4} \nonumber\\
		&\quad + {\mathcal O}((\hbar \Omega)^{-6}) .  
\end{align}
Compared with the exact formula (\ref{qqq183}), the slashed part is $\sim \left( \frac{(\hbar \Omega)^2}{\hbar eE_0} \right)^{-4} = \nu^{-4}$ absent in the above.  This implies that the standard perturbation theory is valid outside of the semi-classical regime, or when the applied field is fast enough compared to the electric field strength, such that $\nu \gtrsim 1$.

\section{Stokes graphs in the dynamically assisted Schwinger mechanism} \label{appe}

\begin{figure}[!t]
\begin{center}
\includegraphics[clip, width=0.33\textwidth]{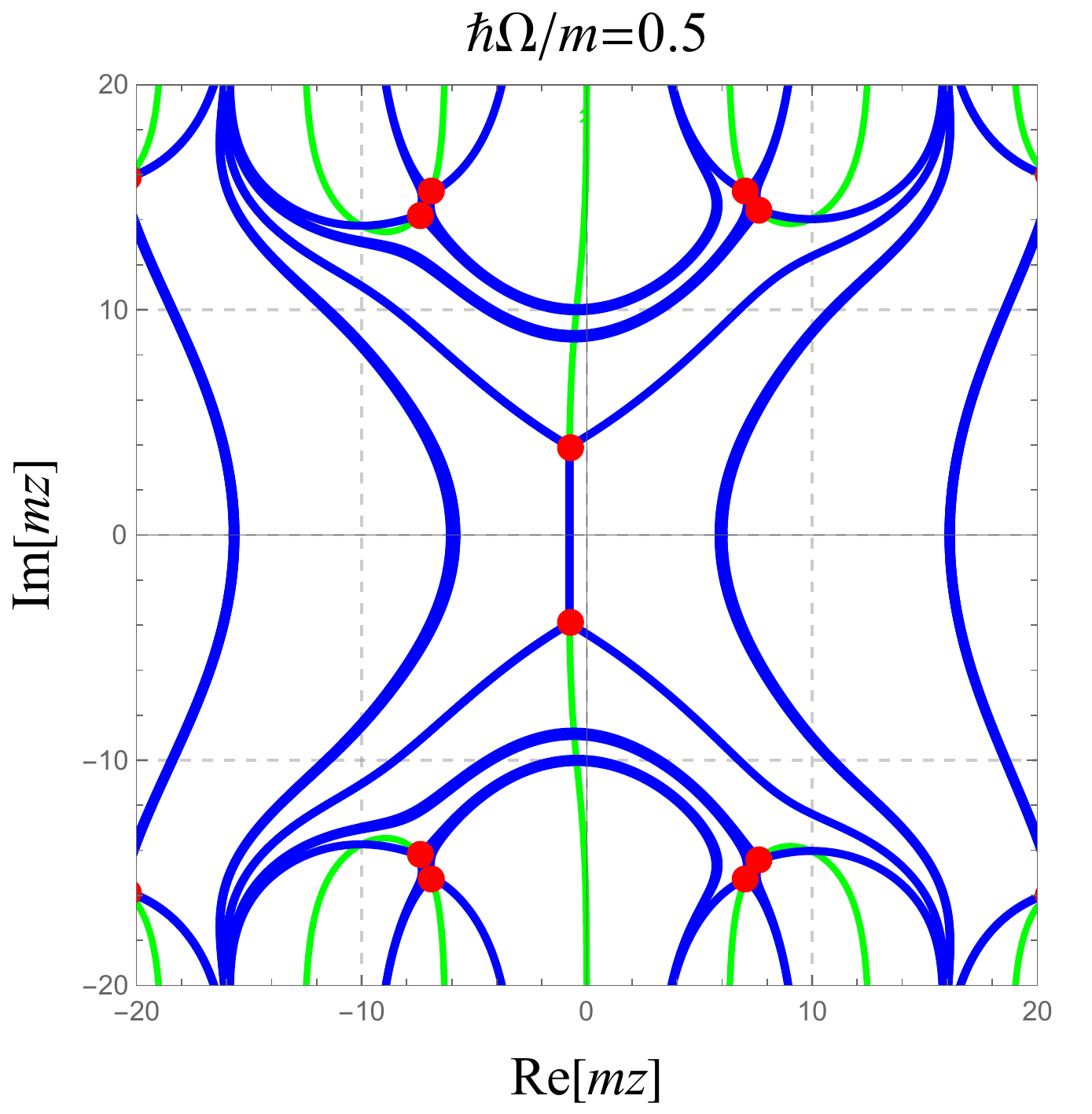}
\hspace*{-2mm}
\includegraphics[clip, width=0.33\textwidth]{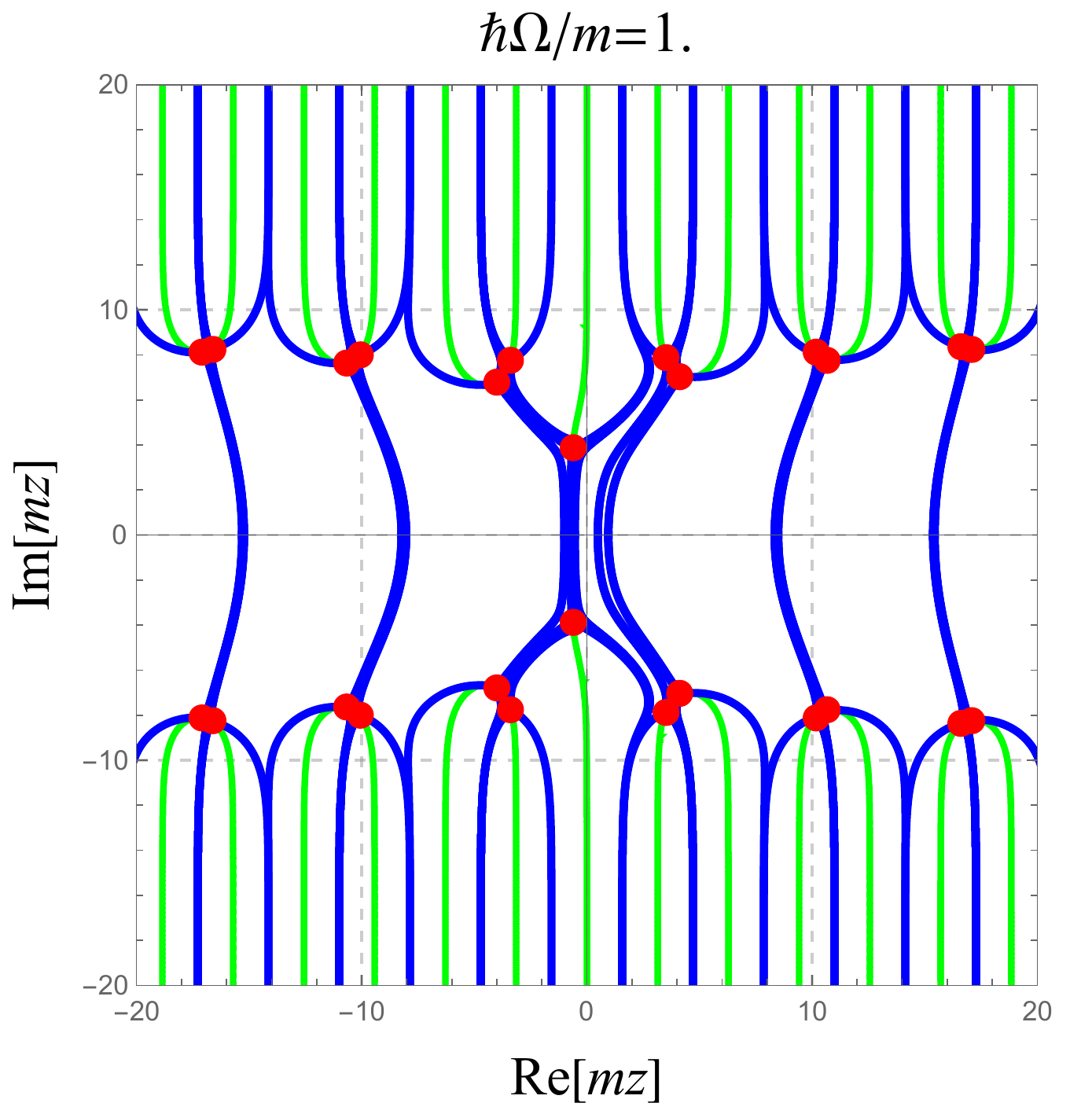}
\hspace*{-2mm}
\includegraphics[clip, width=0.33\textwidth]{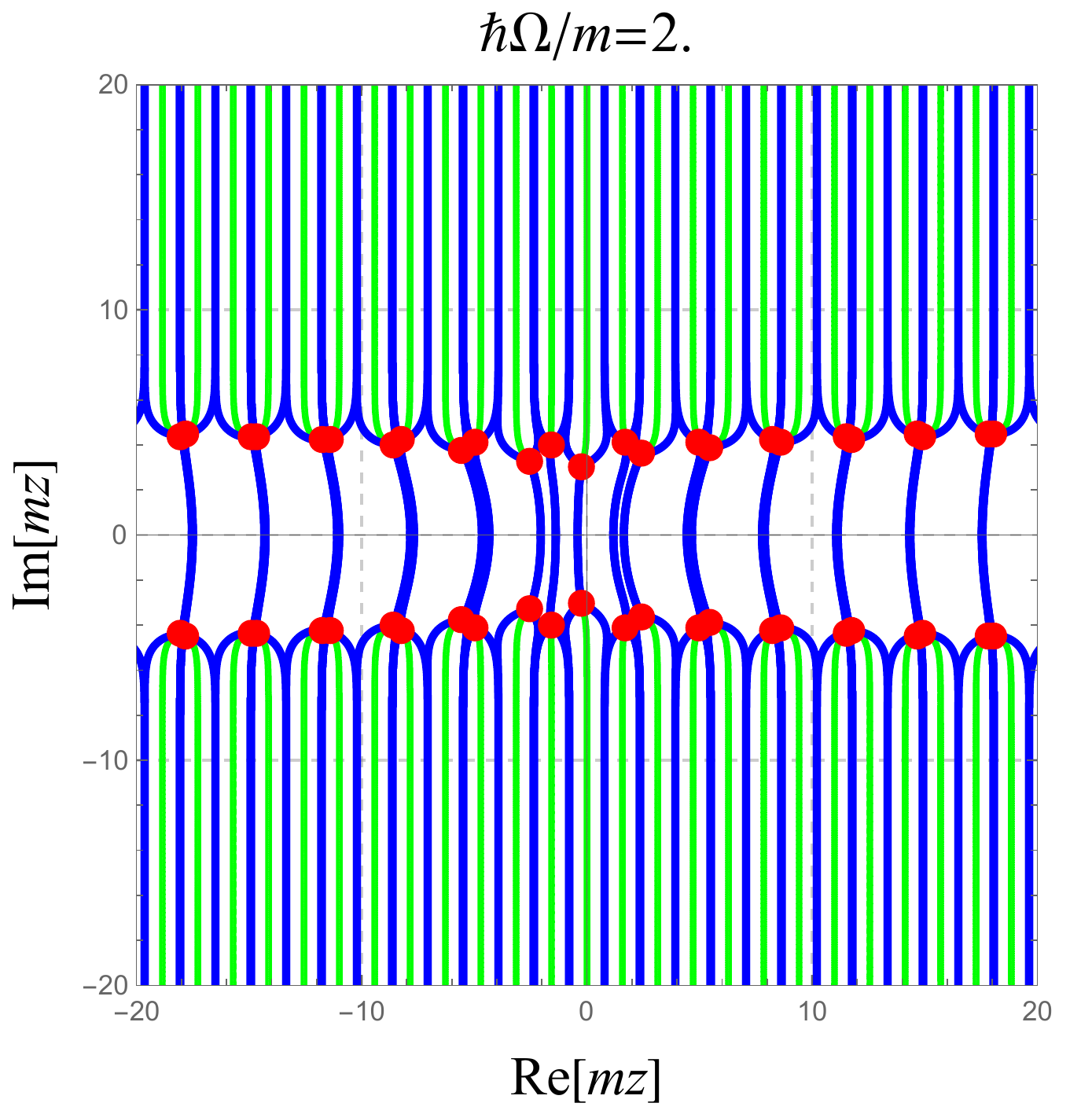}
\end{center}
\caption{\label{figa3} (color online) Stokes graphs for a constant strong electric field superimposed by a weak cosine perturbation (\ref{EQ354}).  The parameters are common to the three panels $\hbar e E_{{\rm s}}/m^2 = 0.25, E_{{\rm w}}/E_{{\rm s}} = 0.01, |{\bm p}_\perp|/m=0$, and $p_\parallel/m=0.20$, except for the frequency $\hbar \Omega/m=0.5, 1, 2$ from the left to right, respectively.  The red points, the blue lines, and the green lines are representing turning points, Stokes lines, and branch cuts, respectively.   
}
\end{figure}

As a supplement to Sec.~\ref{sec352}, i.e., the analysis of the dynamically assisted Schwinger mechanism with a constant strong electric field superimposed by a weak cosine perturbation (\ref{EQ354}), we here discuss the structure of Stokes graphs.  We numerically solved $Q(z_{\rm t})=0$ and the condition (\ref{eq2.13}) to get turning points and Stokes lines, respectively.  Figure~\ref{figa3} displays the obtained Stokes graphs for various frequencies of the perturbation $\hbar \Omega$.

For small $\hbar \Omega$ (the leftmost panel in Fig.~\ref{figa3}), the particle production is dominated by the pair of turning points closest to the real axis around the origin of the plot.  The location of the dominant pair $(z_{{\rm t},{\rm dom}}, z_{{\rm t},{\rm dom}}^*)$ is approximated well by that for a constant electric field (\ref{eeq349}) as $(z_{{\rm t},{\rm dom}}, z_{{\rm t},{\rm dom}}^*) \sim (z^{(0)}_{\rm t}, z_{\rm t}^{(0)*})$, and the Stoke graph around the origin has essentially the same structure as that for a constant electric field, i.e., the Weber potential.  This pair $(z_{{\rm t},{\rm dom}}, z_{{\rm t},{\rm dom}}^*) \sim (z^{(0)}_{\rm t}, z_{\rm t}^{(0)*})$ gives the production number consistent with the Schwinger formula for a constant electric field alone.  Other turning points $z_{\rm t}$'s in Fig.~\ref{figa3} have considerably larger imaginary parts than that the dominant one $z_{{\rm t},{\rm dom}}$ has.  Thus, the corresponding actions $S_{z_{\rm t}} \sim 2 m \, {\rm Im}\,z_{\rm t}$ become large, and their contributions are negligible in the semi-classical limit.  Note that some Stokes segments (e.g., the four Stokes segments emanating from the four turning points ${\rm Im}\, z_{\rm t} \sim 15$ in the leftmost panel) go to the infinity first and then come back to the real axis, which is in contrast to the dominant Stokes segment that goes to the real axis directly without passing the infinity.

As increasing $\hbar \Omega$, the dominant production mechanism smoothly changes from the non-perturbative Schwinger mechanism to the perturbative multi-photon pair production processes.  During this interplay, all the turning points approach the real axis, and not only $z_{{\rm t},{\rm dom}} \sim z^{(0)}_{\rm t}$ but also other turning points, which were negligible for small values of $\hbar \Omega$, start contributing.  Those additional contributions make the production number deviates from the naive Schwinger formula.  It is interesting to point out that the topology of the Stokes graphs changes at intermediate values of $\hbar \Omega$ (the middle panel of Fig.~\ref{figa3}).  Namely, we had some Stokes segments passing the infinity before crossing the real axis for small $\hbar \Omega$ as in the leftmost panel, but those Stokes segments change their topology as increasing $\hbar \Omega$ and eventually cease to passing the infinity.  For example, the four turning points ${\rm Im}\, z_{\rm t} \sim 15$ in the leftmost panel go down to the real axis as increasing $\hbar \Omega$.  At some point they cross a Stokes line from the dominant turning point $z_{{\rm t},{\rm dom}}$, after which the Stokes segments emanating from those turning points cease to passing the infinity.  This is intuitively because any Stokes lines cannot cross each other except at a turning point or a pole for finite $|z|<\infty$, and thus if a Stokes segment eventually crossing the real axis is (not) separated from the real axis by some Stokes lines, it must (needs not) pass the infinity.

For sufficiently large values of $\hbar \Omega$ (the rightmost panel of Fig.~\ref{figa3}), the above-mentioned change in the topology of the Stokes graphs finishes, and all the Stokes segments cross the real axis without passing the infinity.  Contributions from each Stokes segment are roughly equal, as the distances between the turning points and the real axis are almost the same and so are the actions $S_{z_{\rm t}}$'s.  The structure of the Stokes graph is essentially the same as that for the cosine perturbation (\ref{EQ354}) alone and is almost unaffected by the presence of the strong constant electric field.  This is reasonable since the particle production in this parameter regime is dominated by the perturbative multi-photon pair production processes by the perturbation.

\end{document}